\documentclass[final]{amsart}
\usepackage{amssymb,amsmath,amsfonts,dsfont}
\usepackage{graphicx}
\usepackage{epstopdf}
\usepackage[comma, authoryear]{natbib}
\usepackage{subcaption}
\usepackage{mathabx}

\usepackage{multirow}

\usepackage{float}
\restylefloat{figure}
\restylefloat{table}

\newcommand{\dR}{\ensuremath{\mathbb{R}}}

\newcommand{\cB}{\ensuremath{\mathcal{B}}}

\newcommand{\cF}{\ensuremath{\mathcal{F}}}

\newcommand{\cL}{\ensuremath{\mathcal{L}}}

\newcommand{\cQ}{\ensuremath{\mathcal{Q}}}

\newcommand{\cS}{\ensuremath{\mathcal{S}}}

\newcommand{\fa}{\ensuremath{\mathfrak{a}}}

\newcommand{\fb}{\ensuremath{\mathfrak{b}}}

\newcommand{\I}{\mathbf{1}}

\newtheorem{thm}{Theorem}

\newtheorem{prop}{Proposition}

\newtheorem{rem}{Remark}

\usepackage[color, notref, notcite]{showkeys}
\usepackage{enumerate}
\definecolor{labelkey}{rgb}{0,0,1}
\usepackage{tabularx,colortbl}
\definecolor{Gray}{gray}{0.9}

\begin{document}
\title[Option Pricing and Hedging for Discrete Time ARHMM]{Option Pricing and Hedging for
	   Discrete Time Autoregressive Hidden Markov Model}

\author{Massimo Caccia}
\address{Department of Decision Sciences, HEC Montr\'eal, 3000 chemin de la C\^ote Sainte-Catherine, Montr\'eal (Qu\'ebec), Canada H3T 2A7}
\email{massimo.caccia@hec.ca}

\author{Bruno R\'emillard}
\address{Department of Decision Sciences, HEC Montr\'eal, 3000 chemin de la C\^ote Sainte-Catherine, Montr\'eal (Qu\'ebec), Canada H3T 2A7}
\email{bruno.remillard@hec.ca}
\thanks{This
		work is supported in part by the Fonds pour la formation de
		chercheurs et l'aide \`a la recherche du Gouvernement du Qu\'ebec and by the Natural Sciences and Engineering Research Council of
		Canada. We would like to thank Genevi\`eve Gauthier (HEC Montr\'eal) and Fr\'ed\'eric Godin (Concordia University) for their  helpful comments and suggestions.}


\maketitle 

\begin{abstract}
		In this paper we solve the discrete time mean-variance hedging problem when asset returns follow a multivariate autoregressive hidden Markov model.
		Time dependent volatility and serial dependence are well established properties of financial time series and our model covers both.
		To illustrate the relevance of our proposed methodology, we first compare the proposed model with the well-known hidden Markov model via likelihood ratio tests and a novel goodness-of-fit test on the S\&P 500 daily returns. Secondly, we present out-of-sample hedging results on S\&P 500 vanilla options as well as a trading strategy based on theoretical prices, which we compare to simpler models including the classical Black-Scholes delta-hedging approach.
		\end{abstract}

\section{Introduction}

	The quest for the perfect option pricing model is clearly an important topic in the mathematical finance literature.  	\cite{Cox/Ross:1976} provided the following observation: if a claim is priced by arbitrage in a world with one asset and one bond, then its value can be found by first adapting the model so that the asset earns the risk-free rate, and then computing the expected value of the claim. The idea of finding a self-financing optimal investment strategy that replicates the terminal payoff of the claim is now known as dynamic hedging.
	
	One can model the underlying asset's returns with the geometric Brownian motion and retrieve a tractable and intuitive way of pricing and replicating options. This is precisely what \cite{Black/Scholes:1973} proposed. Unfortunately, financial markets are far too complex for a model as simple as this one and this hedging protocol can lead to large hedging errors, as it will be shown later in this paper. The main drawback of this framework is the constant volatility's assumption.  Indeed, volatility seems to vary over time \citep{Schwert:1989}, \citep{Hamilton/Lin:1996}, mainly for macroeconomics reason. Furthermore, this model assumes serial independence for the returns, which is also an hypothesis that is violated in general.
	
	Optimal hedging was later introduced, which consists in minimizing the quadratic error of replication. The solutions were derived in continuous time \citep{Schweizer:1992} and later in discrete time \citep{Schweizer:1995a}. This methodology can be applied to geometric Brownian motion, or more interestingly to stochastic volatility models. \\

	Hidden Markov models \cite{Hamilton:1989}, \cite{Hamilton:1990} were proven to be extremely useful for modeling economic and financial time series. They are robust to time-varying volatility, serial correlation and higher-order moments, which are all well-established stylized facts of asset returns. The premise for these models is that identifiable events can quickly change the characteristics of an asset's returns. This should be taken into account when pricing a derivative. These events could be on a long horizon - fundamental changes in monetary, fiscal or income policies - or on a shorter horizon - news related to the underlying stock or changes in the target band for the Federal funds rate. However, the classical implementation of an HMM can't account for multiple horizons.\\

	Elliot's work on energy finance and interest rate modeling, where mean-reversion is a widely accepted feature, addresses this problem. \cite{Wu/Elliott:2005} introduced a way to parameterize a regime-switching mean-reverting model with jumps. They found the calibration of the model to be difficult because of the small amounts of jumps in the time series exhibited. \citet{Elliott/Miao/Wu:2011} later introduced a similar model  with no jumps, and where it is the volatility that is subject to mean-reverting regime-switches. His basis was that volatility, being driven by macroeconomic forces, was not to be modeled by price movements. Hence the need to model it by a hidden Markov chain. Finally, \citet{Elliott/Chan/Siu:2013} investigates the valuation of European and American options under another model where the volatility is subject to regime-switches, but this time the Markov chain being observable. The paper suggest that it would be interesting to develop some methods and their corresponding criteria to determine the optimal number of states for the hidden Markov chain in their setting. This is precisely one of the contributions of our paper. \\
	
	In light of all the above, we decided to generalize the work of \cite{Remillard/Hocquard/Lamarre/Papageorgiou:2017}: we combine the regime-switching model with an autoregressive parameter to account for trends and mean-reversions \citep{Fama/French:1988} without having to change regime. Autoregressive hidden Markov models (ARHMM) have been applied to financial engineering and have shown promising results \citep{Shi/Weigend:1997}. Still, this model has never been used in conjunction with optimal hedging. We derive the solution of the hedging strategy and obtain derivatives prices under this class of models. It is also noteworthy to add that we will use semi-exact techniques to compute expectations necessary for the optimal hedging, instead of Monte Carlo techniques, which will greatly speed up computations. For parameterization, we will implement the EM algorithm \citep{Dempster/Laird/Rubin:1977} to the ARHMM. This method is widely used in unsupervised machine learning in order to find hidden structures, in our case, regimes. In order to choose the optimal number of regimes and to assess the suitability of the model, we propose a new goodness-of-fit test based on the work of \cite{Genest/Quessy/Remillard:2006a} and \cite{Remillard:2011a}.  It is based on the Rosenblatt transform and on parametric bootstrap. Compared to Elliot's work, our model can exhibit mean-reversion, but is not restricted to it. It could thus be more adequate for the modeling of a wider variety of financial assets. \\

In his famous study \cite{Fama:1965}, Fama presented strong and voluminous evidence in favor of the random walk hypothesis. He although suggests that other tests - statistical or profit generating strategies - could either confirm or contradict his findings. In this paper, we will explore both avenues. We will statistically show that the ARHMM is an adequate model for financial modelling using the goodness-of-fit test as well as likelihood ratio tests, and we will show that it is possible to generate money by buying/selling options and replicating them until maturity. To support our approach, we will compare the trading strategy's returns with different methodologies: Black-Scholes delta-hedging and optimal hedging when assets follow a geometric random walk. We will also compare the hedging results with the delta-hedging using the market's implicit volatility. \\

First, we present likelihood ratio test results confirming the ARHMM is a better fit than the classical HMM on S\&P daily returns, in particular, because our model has the capacity for mean-reversion. Secondly, empirical pricing and hedging results suggest that our methodology is superior to its counterparts by achieving the best mean-squared error in six out of eight cases, as well as being the most profitable strategy.

The rest of the paper is organized as follows. Section 2 describes the model and implements the EM algorithm for parameter estimation. In addition, we will introduce the goodness-of-fit test and study its suitability in the financial markets. Then, in Section 3, we will state the optimal dynamic discrete time hedging model when assets follow a ARHMM. The results of the implementation of the dynamic hedging strategies will be presented in Section 4. Section 5 concludes.

\section{Regime-Switching Autoregressive models}\label{sec:model}

	The proposed  models are quite intuitive. The regime process $\tau$
	is a homogeneous Markov chain on $\{1,\ldots,l\}$, with transition
	matrix $Q$. At period $t-1$, if $\tau_{t-1}=i$, and the return
	$Y_{t-1}$ has value  $y_{t-1}$, then at time $t$, $\tau_t=j$ with probability $Q_{ij}$, and the return $Y_t$ has conditional
	distribution $f_j(y_t;y_{t-1})$; here lower case letters $y_1,\ldots, y_n$ are used to denote  a realization of $ Y_1, \ldots,Y_n$. It follows from this construction  that $(Y_t,\tau_t)$ is a Markov process.

For example, for $j\in
	\{1,\ldots,l\}$, one could take a Gaussian AR(1) model meaning that given $Y_{t-1}=y_{t-1}$ and $\tau_{t}=j$, $Y_t = \mu_j+ \Phi_j(y_{t-1}-\mu_j)+\varepsilon_t$, with $\varepsilon_t \sim N(0,A_j)$;  more precisely, the conditional density of $Y_t$ at $y_{t}\in \dR^d$ is	
	\begin{equation}\label{eq:VAR1}
	f_j(y_t|y_{t-1}) =
	\frac{e^{-\frac{1}{2}\left\{y_{t}-\mu_j-\Phi_j(y_{t-1}-\mu_j)\right\}^\top
			A_j^{-1}\left\{y_t-\mu_j-\Phi_j(y_{t-1}-\mu_j)\right\} }}{(2\pi)^{d/2}
		|A_j|^{1/2}},
	\end{equation}
	where $\mu_j\in \dR^d$, $\Phi_j$ is a $d\times d$ matrix such that
	$\Phi_j^n\to 0$ as $n\to
	\infty$\footnote{This condition ensures that for any $j\in\{1,\ldots,l\}$, the matrix $B_j
		= \sum_{k=0}^\infty \Phi_j^k A_j \left(\Phi_j^k\right)^\top$ is well
		defined and satisfies $B_j = \Phi_j B_j\Phi_j^\top +A_j$.}, and $A_j$ is a $d\times d$ non degenerate covariance
	matrix. The matrices $\Phi_1,\ldots, \Phi_l$ are mean-reversion parameters. Let $\cB_d$ be the set of $d\times d$ matrices $B$ such
	$B^n\to 0$ as $n\to \infty$ and let $S_d^+$ be the set of symmetric
	positive definite $d\times d$ matrices. Note that $\cB_d$ is the set
	of $d\times d$ matrices with spectral radius smaller than $1$,
	meaning that the eigenvalues are all in the unit complex ball of
	radius $1$; in particular, $I-B$ is invertible for any $B\in \cB_d$.
    Note that the the so-called Hidden Markov  Model is obtained by setting $\Phi_1 = \cdots = \Phi_l = 0$.

	\subsection{Regime prediction}\label{ssec:prediction}
	
	Since the regimes are not observable, we have to find a
	way to predict them. This will be of utmost importance for pricing
	and hedging derivatives.
	
	In many applications, one has to predict an non-observable signal
	by using observations $Y_1,\ldots, Y_t$ linked in a certain way to the signal. This
	is known as a filtering problem \citep{Remillard:2013}. In our case, we need to find the most likely regime at time $t$, in other words $\eta_t(i) = P(\tau_t = i|Y_1=y_1,\ldots, Y_t=y_t)$. It is
	remarkable that for the present model, one can compute exactly this
	conditional distribution, given a starting distribution $\eta_0$.
	For more details, see an extension of the Baum-Welch algorithm in Appendix \ref{app:BWA}.

	\subsubsection{Filtering algorithm}
	
		Choose an a priori probability distribution $\eta_0$ for the
		regimes.
		Equivalently, one can choose a positive vector
		$q_0$ and set $\eta_0(i)= q_0(i)/Z_0$, where $Z_0 = \sum_{j=1}^l
		q_0(j)$. The choice of $q_0$ or $\eta_0$ is not critical since its
		impact on predictions decays in time and have virtually no impact on
		terminal regime probabilities for any reasonable time series length.
		For simplicity, we assume a uniform distribution, i.e. $q_0 \equiv
		1/l$.
		
		For $t=1,\ldots,n$, define $ q_t(i) = E\left [\I(\tau_t=i)
		\prod_{k=1}^t f_{\tau_k}(y_k|y_{k-1})\right ]$, $i\in
		\{1,\ldots,l\}$, and $Z_t = \sum_{j=1}^l q_t(j)$.  \footnote{The first observation of the sequence is burn-in in order to compute $f_{\tau_1}(y_1 | y_0)$}
		Hence, $Z_t$ is the joint density of $Y_1,\ldots,Y_t$ because
		\begin{equation*}
			\sum_{j=1}^l q_t(j) = E\left [ \prod_{k=1}^t f_{\tau_k}(y_k|y_{k-1})\right ] = f_{1:t}(y_1,\ldots,y_t)
		\end{equation*}
		Then if $q_0 = \eta_0$, then for any $i\in\{1,\ldots,l\}$, and any $t\ge 1$,
		\begin{eqnarray}
			q_{t}(i) &=& E\left [\I(\tau_{t}=i) \prod_{k=1}^{t} f_{\tau_k}(y_k|y_{k-1})\right ] \nonumber \\
			&=& f_{i}(y_{t}|y_{t-1})\sum_{j=1}^l E\left [\I(\tau_{t}=i) \I(\tau_{t-1}=j) \prod_{k=1}^{t-1} f_{\tau_k}(y_k|y_{k-1})\right ] \nonumber \\
			&=& f_{i}(y_{t}|y_{t-1})\sum_{j=1}^l Q_{ji} E\left [\I(\tau_{t-1}=j) \prod_{k=1}^{t-1} f_{\tau_k}(y_k|y_{k-1})\right ] \nonumber \\
			&=& f_{i}(y_{t}|y_{t-1})\sum_{j=1}^l Q_{ji} q_{t-1}(j).
		\end{eqnarray}
		and
		\begin{equation}\label{eq:etak}
		\eta_{t}(i) = P(\tau_{t}=i | Y_1,\ldots,Y_{t}) = \frac{q_{t}(i)}{Z_{t}}.
		\end{equation}
		Having computed the conditional probabilities, $\eta_{t}(i)$, one can finally estimate $\tau_{t}$ by
		\begin{equation}\label{eq:most_prob_reg}
		\tau_{t} = \arg\max_{i} 	\eta_{t}(i) ,
		\end{equation}
		i.e. as the most probable regime.
		
		In view of applications, it is preferable to rewrite (\ref{eq:etak}) only in terms of $\eta$, i.e.,
			\begin{equation}
			\eta_t(i) = \frac{f_i(y_t|y_{t-1})}{Z_{t|t-1}}\sum_{j=1}^l
			\eta_{t-1}(j) Q_{ji},
			\end{equation}
		where
			\begin{equation*}
				Z_{t|t-1} = \frac{Z_t}{Z_{t-1}} = \sum_{j=1}^{l} \sum_{i=1}^{l} f_i(y_t|y_{t-1}) \eta_{t-1}(j) Q_{ji}.
			\end{equation*}
As a result, $Z_{t|t-1}$ is the conditional density of $Y_t$ given $Y-1,\ldots, Y_{t-1}$, evaluated at $y_1,\ldots, y_t$.

	\subsubsection{Conditional distribution }\label{sssec:cond_dist}
		
		From the results of the previous section,  the joint  density $f_{1:t}$
		of $Y_1,\ldots, Y_t$ is $Z_t$. Also,
		for any $t\ge 2$, the conditional density $f_{t|t-1}=Z_{t|t-1} $ of  $Y_t$ given
		$Y_1,\ldots, Y_{t-1}$, can be expressed as a mixture, viz.
		\begin{equation}\label{eq:cond_dens}
		f_{t|t-1}(y_t|y_1,\ldots,y_{t-1}) = \sum_{i=1}^l f_i(y_t|y_{t-1})
		\sum_{j=1}^l \eta_{t-1}(j)Q_{ji}  = \sum_{i=1}^l f_i(y_t|y_{t-1}) W_{t-1}(i)
		\end{equation}
		where
		\begin{equation}\label{eq:cond_weights}
		W_{t-1}(i) = \sum_{j=1}^l \eta_{t-1}(j)Q_{ji}, \quad i \in
		\{1,\ldots,l\}.
		\end{equation}
		Note that  for all $t>1$, $W_{t-1}(i) = P(\tau_t = i|Y_{t-1}=y_{t-1},\ldots,
		Y_1=y_1)$.  As a result, it follows that
\begin{equation}\label{eq:pred-regime}
P(\tau_{t+k} = i|Y_{t}=y_{t},\ldots,
		Y_1=y_1) = \sum_{j=1}^l (Q^k)_{ji}\eta_t(j),\quad i\in \{1,\ldots,l\},
\end{equation}
so the conditional law of $Y_{t+1}$ given
		$Y_1,\ldots, Y_t$ has density
\begin{equation}\label{eq:fore20}
		f_{t+1|t}(y_{t+1}|y_1,\ldots,y_t) = \sum_{i=1}^l  f_i(y_{t+1}|y_t)W_t(i).
		\end{equation}
Next, it is easy to check that the conditional law of $Y_{t+1}, \ldots, Y_{t+m}$ given
		$Y_1,\ldots, Y_t$ has density
		\begin{eqnarray}\label{eq:fore2}
		\quad f_{t+m|t}(y_{t+1}, \ldots, y_{t+m}|y_1,\ldots,y_t) &= &\sum_{i_0=1}^l\sum_{i_1=1}^l \cdots \sum_{i_m=1}^l \eta_t(i_0)  \\
&& \qquad \quad \times \prod_{k=1}^m  Q_{i_{k-1}i_k} f_{i_k}(y_{t+k}|y_{t+k-1}).\nonumber
		\end{eqnarray}

	\subsubsection{Stationary distribution in the Gaussian case}\label{sssec:sta_dist}
	
Suppose that the model specified by \eqref{eq:VAR1} holds, ergo the innovations are Gaussian. If $Y_n$ converges in law to a stationary distribution, for any given starting point $y_0$, then this distribution must be Gaussian, with mean $\mu$ and covariance matrix $A$.  Suppose the Markov chain is ergodic with stationary distribution $\nu$. Then with probability $\nu_i$, $i\in \{1,\ldots,l\}$, $Y_1 = (I-\Phi_i)\mu_i+ \Phi_i Y_0 + \epsilon_i$, where $\epsilon\sim N(0,A_i)$ is independent of $Y_0 \sim N(\mu,A)$.
It then follows that
$$
\mu = \left\{\sum_{i=1}^l \nu_i (I-\Phi_i) \right\}^{-1} \left\{\sum_{i=1}^l  \nu_i \{ (I-\Phi_i)\mu_i \right\}.
$$
Similarly, $A$ must satisfies $A = T(A)$, where
\begin{equation}\label{eq:A}
T(A)=  B + \sum_{i=1}^l  \nu_i \Phi_i A \Phi_i^\top,
\end{equation}
with
$B =  -\mu \mu^\top + \sum_{i=1}^l  \nu_i \left[ (I-\Phi_i)\mu_i+\Phi_i \mu\}\{(I-\Phi_i)\mu_i+\Phi_i \mu\} ^\top + A_i\right]$.
From the conditions on $\Phi_1,\ldots,\Phi_l$, there is a
		norm $\|\cdot\|$ on the space of matrices such that $\|\Phi_i\| < 1$ for every $i\in \{1,\ldots,l\}$.\footnote{Recall that all norms are equivalent.} The operator $T$ is then a contraction since for any two matrices $A_0,A_1$, $\|T(A_1)-T(A_0)\| \le \|A_1-A_0\|\sum_{i=1}^l \nu_i \|\Phi_i\|^2 \le c\|A_1-A_0\|$,
		with $c = \max_{1\le i\le l}\|\Phi_i\|^2 < 1$. Also, since $T(A)$ is a covariance matrix whenever $A$ is one, and $B$ is positive definite, it follows that there is a unique fixed point $A$ of $T$, meaning that $A=T(A)$,  and this unique fixed point $A$ is a positive definite covariance matrix. If fact, $A$ is the limit of any sequence $A_n = T(A_{n-1})$, with $A_0$ a non-negative definite covariance matrix. For example, one could take even take $A_0=0$. This provides a way to approximate the limiting covariance $A$ by setting $A\approx A_n$ for $n$ large enough.
		

	\subsection{Estimation of parameters}\label{sec:est}

		The EM algorithm \citep{Dempster/Laird/Rubin:1977} is a quite efficient estimation procedure for incomplete datasets. Under hidden Markov models, observations are partial since $\tau$ is unobservable. The algorithm proceeds iteratively to converge to the maximum likelihood estimation of parameters \citep{Dempster/Laird/Rubin:1977}. We derived its implementation for ARHMM and the details are in Appendix \ref{app:em}.  It seems  that starting the  parameter's estimation of the ARHMM with the HMM parameters' estimate (obtained by setting  $\Phi_1 = \cdots = \Phi_d=0$)  was slightly more stable. The optimal number of regimes must be known a priori, an issue we will discuss next.
		\\
	\subsection{Goodness-of-fit test and selection of the number of regimes}\label{sec:gof}
	
		To select the optimal number or regimes, one must test the adequacy of fitted models with different number of regimes. This is generally done by using a test based on likelihoods. However, goodness-of-fit tests based on likelihoods are not recommended for regime-switching models \citep{Cappe/Moulines/Ryden:2005}. We opt for a simpler approach based on a parametric bootstrapping. It was shown to work on a large number of dynamic models, including hidden Markov models. The test was built on the work of \cite{Genest/Remillard:2008} and its implementation is in Appendix \ref{app:gof}.
				
\subsubsection{Selecting the number of regimes}\label{sssec:selreg}		Choosing the optimal number of regimes. The goodness-of-fit test methodology described in Appendix B produces P-value from Cram\'er-von Mises type statistics, for a given number of regimes $\ell$. As suggested in \cite{Papageorgiou/Remillard/Hocquard:2008}, it make sense to choose the optimal number of regimes, $\ell^\star$, as the first $\ell$ for which the P-value is larger than 5\%. An illustration of the proposed methodology is given in Section 2.4.

	\subsection{Application to S\&P 500 daily returns}	
	
 To assess the relevance of our model on real data, we estimated the parameters on the close-to-close log-returns of the daily price series of the S\&P 500 Total Return. To find stationary estimation windows, we used a nonparametric change\-point test for a univariate series  using a  Kolmogorov-Smirnov type statistic \citep{Remillard:2013}. We focused on recent data, i.e. from early 2000 to today. We found two stationary estimation window: from 05/01/2004 to 02/01/2008 and from 05/01/2010 to 20/01/2017. We can refer to the former as the 2000's recovery and 2010's recovery for the latter. Results of the tests are presented is Table \ref{tab:changePointTest}. We will also study the interesting period in between, the 2008-2009 Financial Crisis, even though  the null hypothesis of stationarity has a $P$-value of $0.4\%$.
			
			\begin{table}[t]
				\centering
				\caption{P-values (in percentage) for the nonparametric change point test using the Kolmogorov-Smirnov statistic with N=10000 bootstrap samples.}
				\label{tab:changePointTest}
				\begin{tabular}{|l|c|}
				\hline
					 Period &  $P$-value\\
					\hline
					2000's recovery & 39.8 \\
					2008-2009 Financial Crisis & 0.4\\
					2010's recovery & 9.7	\\
				\hline	
				\end{tabular}
			\end{table}

			Next, we perform the goodness-of-fit test (GoF for short) described in Appendix \ref{app:gof} for the ARHMM (AR(1)) as well as for the HMM (AR(0)), as a mean of comparison. The results are presented in Tables \ref{tab:gof_lrt_2000Bull}, \ref{tab:gof_lrt_2008Crisis} and \ref{tab:gof_lrt_2010Bull}. According to the selection methods described in Section \ref{sssec:selreg}, we optimally select a three-regime model for the 2000's recovery, since $3$ is the smallest number of regimes for which the $P$-value is larger than 5\%. This is also true for the HMM model. Likewise, we choose a three-regime model for the 2008-2009 Financial Crisis, and a four-regime model for the 2010's recovery. Note that in the case of the 2010's Bull markets, a four regime model for the HMM was not enough to get a $P$-value $>5\%$.

			 Furthermore, to measure the significance of ARHMM  over HMM, we perform a  likelihood ratio test. This is possible because the HMM is a special case of the ARHMM corresponding to $\Phi_1 = \cdots = \Phi_l=0$. The corresponding statistic is computed as follow:
				$$
				D = -2 \log\bigg(\dfrac{L(\hat \theta_0 | x)}{L(\hat \theta_1 | x)}\bigg) = -2 \log\bigg(\dfrac{f_{1:n}(y_1,\ldots,y_n |\hat \theta_0)}{f_{1:n}(y_1,\ldots,y_n | \hat\theta_1)} \bigg)
				$$
			where $\hat \theta_0$ are the model's parameters estimated under the null hypothesis, i.e. $\Phi_1 = \cdots = \Phi_\ell=0$, so the returns follow a Gaussian hidden Markov model, and $\hat \theta_1$ are the model's  parameters estimated under the alternative, i.e. returns follow an autoregressive hidden Markov model. Under the null hypothesis, this statistic is distributed as a chi-square distribution with the number of degrees of freedom equal to the number of extra parameters in the alternative model. In our case, we have one extra parameter per regime, i.e. $\Phi_i$, so the number of degrees of freedom is $\ell$. Hence, under the null hypothesis, $D\sim \chi^2(\ell)$.  The log-likelihoods of both models, the statistical test $D$ and the \raisebox{2pt}{$\chi^2$} critical value at a significance level of 5\% are  also presented in Tables \ref{tab:gof_lrt_2000Bull}, \ref{tab:gof_lrt_2008Crisis} and \ref{tab:gof_lrt_2010Bull}. We clearly reject the null hypothesis for all models, proving we should favor ARHMM over HMM for each dataset.
  \\

			The estimated parameters for the three periods are presented in Tables \ref{tab:calib_2000Bull}, \ref{tab:calib_2008Crisis}, and \ref{tab:calib_2010Bull}, where the mean and standard deviation of each AR(1) and AR(0) Gaussian regime density  $f_i$  are respectively denoted by $\mu_i$ and $\sigma_i$, and are presented as annualized percentages values. The tables further contains the long-term, i.e. stationary, regime probabilities $\nu$, together with the estimated transition matrix, $Q$.
			
			Regimes are ordered by increasing volatility $\sigma_i$, and incidentally by decreasing mean $\mu_i$, which is in line with what we typically observe on the markets.

\begin{table}
	\caption{P-values (in percentage) for the proposed goodness-of-fit test using N=10000 bootstrap samples on the S\&P 500 daily returns for the 2000's recovery, along with the log-likelihood of the models and the
P-values (in percentage) of the likelihood ratio test statistic D.}
	\label{tab:gof_lrt_2000Bull}
\begin{center}
	\begin{tabular}{|l|ccc|}
\hline
& \multicolumn{3}{c|}{Number of regimes}\\
		 &  1 & 2 & 3 \\
		\hline
		GoF $P$-value (ARHMM)& 0 &	0&	26.51 \\
		GoF $P$-value (HMM)   & 0 & 0 & 25.12     \\
		Log-likelihood (ARHMM)&  3479 &	3542 &	  3559  \\
		Log-likelihood  (HMM)& 3475 & 3539 & 3552     \\
		$P$-value (D) &0.43 & 3.63 & 0.18\\
		\hline
	\end{tabular}
\end{center}
\end{table}

\begin{table}
	\caption{Parameters estimation for the three-regime models on the S\&P 500 Total Return daily returns for the 2000's recovery. $\mu$ and $\sigma$ are presented as annualized percentage.}
	\label{tab:calib_2000Bull}
\begin{center}
	\begin{tabular}{|c|ccc|ccc|}
\hline
		& \multicolumn{3}{c|}{AR(0)} & \multicolumn{3}{c|}{AR(1)}\\
		Parameter & \multicolumn{3}{c|}{Regime} & \multicolumn{3}{c|}{Regime} \\
		& 1         & 2       & 3             & 1         & 2       & 3                \\
\hline
		$\mu$     &   31.41 &  13.88 &  -17.23  & 34.89 &  6.99  &    -21.60                       \\
		$\sigma$  &    2.18 & 10.09 &    18.02 &     3.34 &11.03 &   18.95                         \\
		$\Phi$     & 0 & 0 & 0 &      -0.14 & 0.03 & -0.19							\\
$\nu$     & 0.11 &    0.65 &   0.24 &    0.19  &    0.63 &    0.18         \\
		\hline
		& 0    & 0.92 &    0.08 &       0  &  0.96 &    0.04  \\
		Q           & 0.17  &  0.83   &      0 &    0.32 &    0.68 &         0 \\
		& 0    & 0.03 &  0.97 &     0   & 0.04 &    0.96\\
\hline
	\end{tabular}
\end{center}
\end{table}

\begin{figure}
\begin{center}
	\includegraphics[width=10cm]{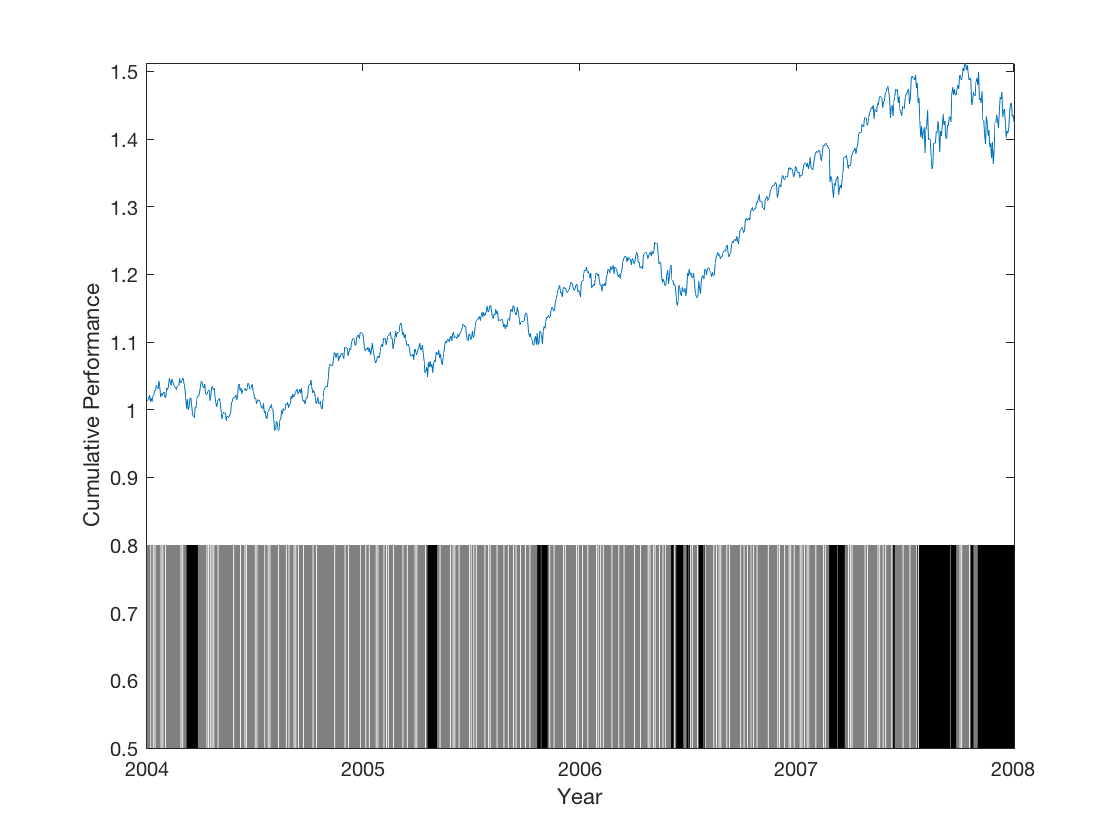}
\end{center}
	\caption{Most probable regimes for the three-regime AR(1) model fitted on the S\&P 500
Total Return index from 05/01/2004 to 02/01/2008 together with the cumulative performance
of the index. Darker areas represent higher volatility states.}\label{fig:2000r}
\end{figure}

\begin{figure}
\begin{center}
	\includegraphics[width=10cm]{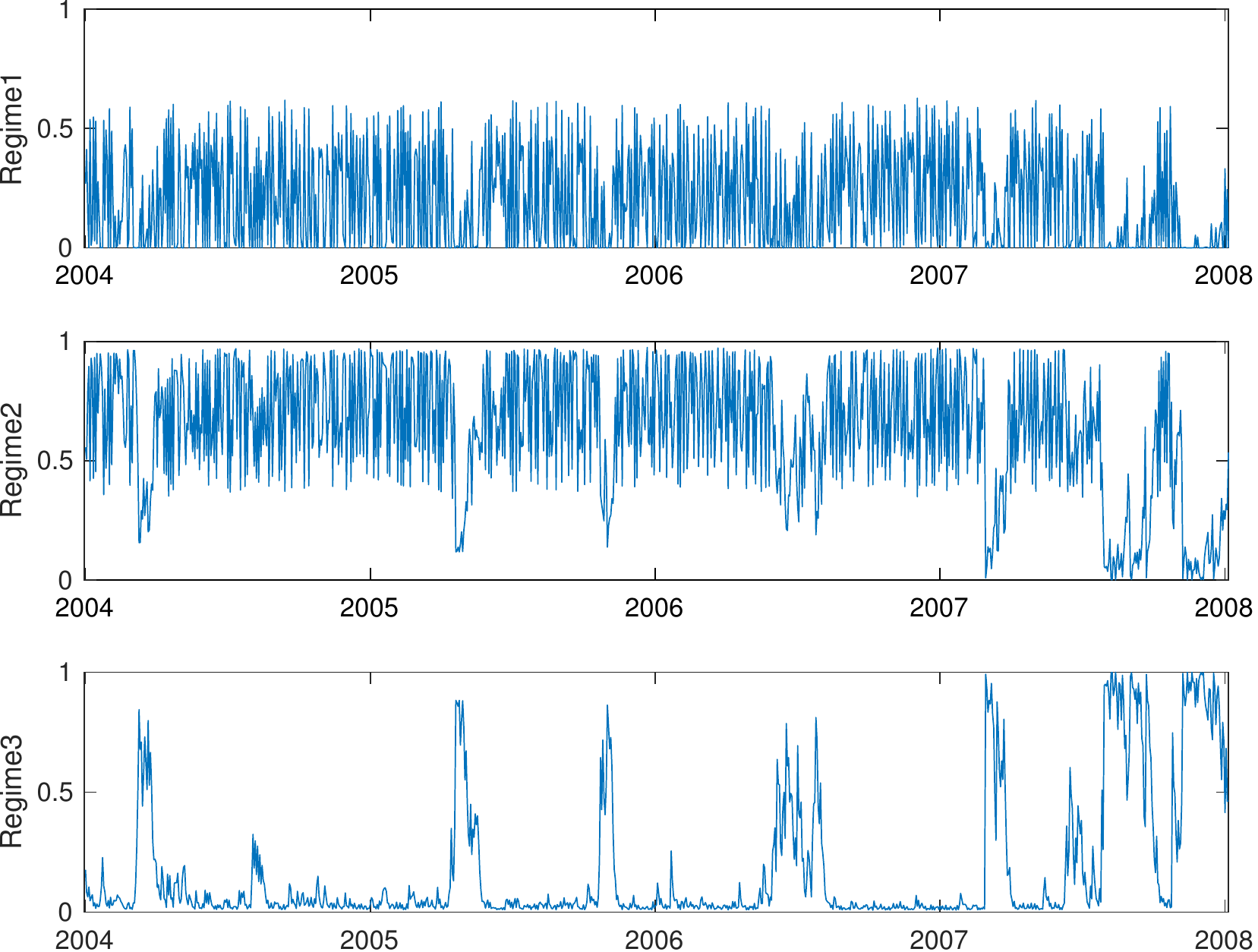}
\end{center}
	\caption{Probability of the regimes, i.e. $\eta_t$, for the three-regime AR(1) model fitted on the S\&P 500
Total Return index from 05/01/2004 to 02/01/2008.}\label{fig:2000r_eta}
\end{figure}

In the case of the 2000's recovery, Regime 1 is associated to bull markets, which are characterized by strong positive premium and low risk ($\mu_1 = 35.89$ and $\sigma_i=3.34$). It seems that this state is intermittent in the sense that the Markov chain does not stay or has a very small probability of staying in regime 1 since $Q_{11}\approx 0$. However, this state is not due to outliers since the percentage of time the Markov chain is in this state is 11\% for the HMM and 19\% for the ARHMM.

Regime 2 is an intermediate state. Lastly, Regime 3 is associated with bear markets or corrections, as highlighted by the negative premium of -21.60 and the volatility of 18.95. The regimes are less distinct in the HMM case. Also, the likelihood ratio test statistic $D=15.04$ informs us that ARHMM is a much better fit for this data. Indeed, we observe strong mean-reversion in regime 1 and 3 ($\Phi_1=-0.14$ and $\Phi_3=-0.19$). This could explain why the HMM is blurring everything to together. Figure \ref{fig:2000r} displays the filtered most probable regimes (see Section \ref{ssec:prediction} for the filtering procedure) for the whole time series. The regimes are depicted by different shades of grey, ranging from dark for the high volatility regime to white for the low volatility regime. The probability of the regimes, i.e. $\eta_t$, are presented in Figure \ref{fig:2000r_eta}.  Interestingly enough, the crisis in the subprime mortgage market is adequately captured by the high risk regime.
			
\begin{table}
	\caption{P-values (in percentage) for the proposed goodness-of-fit test using N=10000 bootstrap samples on the S\&P 500 daily returns for the 2008-2009 Financial Crisis, along with the log-likelihood of the models and the P-values (in percentage) of the likelihood ratio test statistic D.}
	\label{tab:gof_lrt_2008Crisis}
\begin{center}
	\begin{tabular}{|l|ccc|}
\hline
& \multicolumn{3}{c|}{Number of regimes}\\
		 &  1 & 2 & 3 \\
		\hline
		GoF $P$-value (ARHMM)& 0 &	0&	59.59  \\
		GoF $P$-value (HMM)     & 0 & 0 & 72     \\
		Log-likelihood (ARHMM)& 1,209 &	 1,323 &1,334 \\
		Log-likelihood  (HMM)  &  1,214 &  1,318 &   1.329    \\
		$P$-value (D) & 0.19 & 0.95 & 2.59 \\
\hline	
	\end{tabular}
\end{center}
\end{table}

\begin{table}
	\caption{Parameters estimation for the three-regime models on the S\&P 500 Total Return daily returns for the 2008-2009 Financial Crisis. $\mu$ and $\sigma$ are presented as annualized percentage.}
	\label{tab:calib_2008Crisis}
\begin{center}
	\begin{tabular}{|c|ccc|ccc|}
	\hline
		& \multicolumn{3}{c|}{AR(0)} & \multicolumn{3}{c|}{AR(1)}\\
		Parameter & \multicolumn{3}{c|}{Regime} & \multicolumn{3}{c|}{Regime} \\
		& 1         & 2       & 3             & 1         & 2       & 3                \\
		\hline
		$\mu$     &  74.97 &   -5.28 &  -66.87 &   72.22 &   -0.42 &  -64.73                       \\
		$\sigma$  &   5.52 &   23.17 &   57.80 &    5.12 &  22.56 &   55.68                         \\
		$\Phi$     & 0 & 0 & 0 &   -0.03 &   -0.16 &  -0.15  \\
$\nu$     &   0.15 &   0.58 &   0.27 &    0.14 &     0.58 &    0.28          \\
		\hline
		&      0  &   0.97 &    0.03 &         0    & 0.98 &    0.02 \\
		Q &	0.25   &  0.75    &       0  &  0.23 &    0.77 &         0 \\
		& 0   &  0.01 &    0.99 &      0  &  0.01 &    0.99 \\
		\hline		
	\end{tabular}
\end{center}                        	
\end{table}

\begin{figure}
\begin{center}
	\includegraphics[width=10cm]{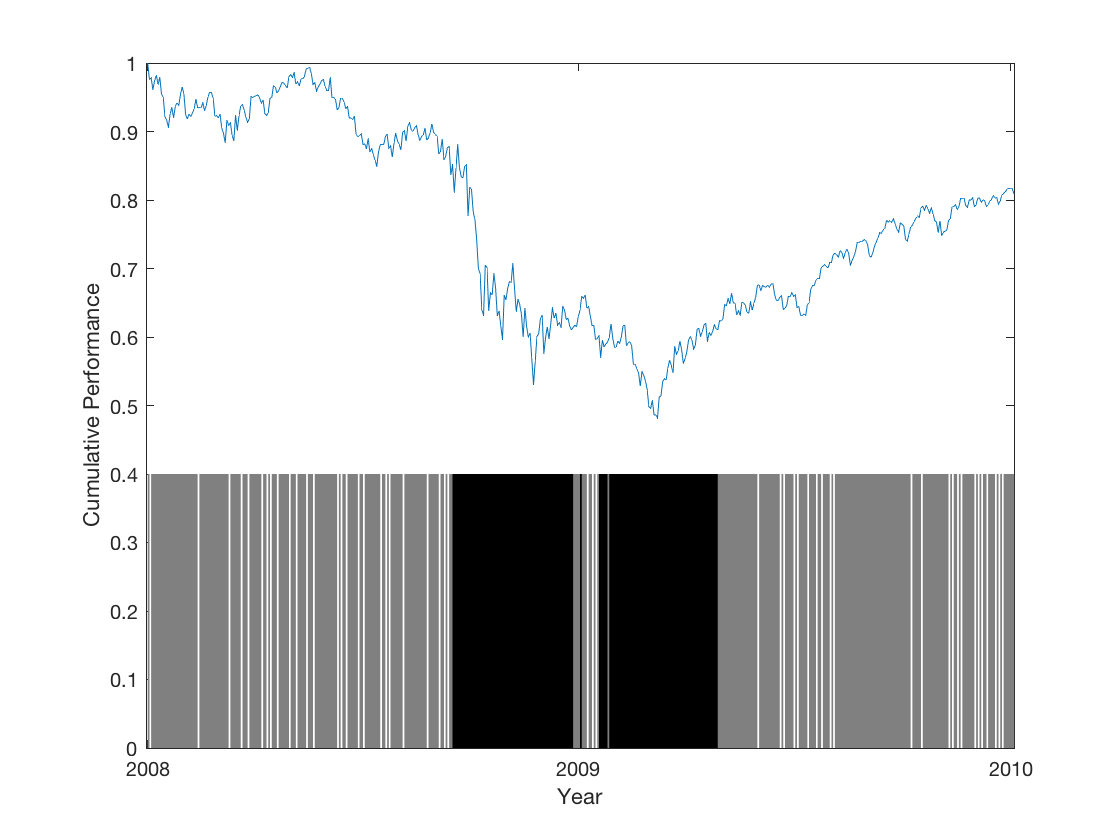}
\end{center}
	\caption{Most probable regimes for the three-regime AR(1) model fitted on the S\&P 500
Total Return index from 03/01/2008 to 04/01/2010, together with the cumulative performance
of the index. Darker areas represent higher volatility states.}\label{fig:2008c}
\end{figure}

\begin{figure}
\begin{center}
	\includegraphics[width=10cm]{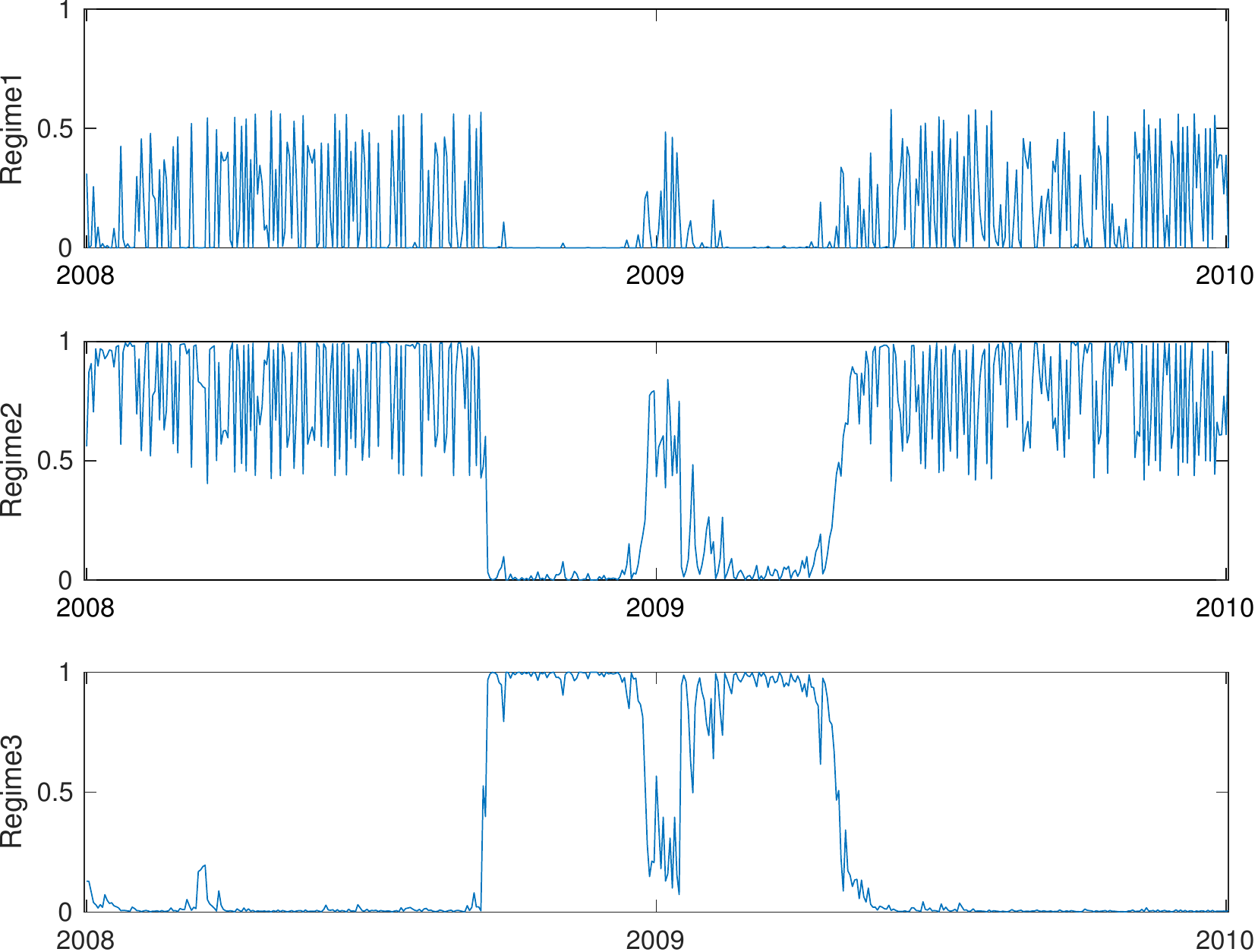}
\end{center}
	\caption{Probability of the regimes, i.e. $\eta_t$, for the three-regime AR(1) model fitted on the S\&P 500
Total Return index from 03/01/2008 to 04/01/2010.}\label{fig:2008c_eta}
\end{figure}


The second period studied is pretty interesting. For the 2008-2009 Financial Crisis, regimes are extremely polarized, with expected returns ranging from 72.22 to -64.73. The bear markets regime, i.e. Regime 3, is exceptionally persistent and volatile, as highlighted by $Q_{3,3}=0.99$ and $\sigma_3=55.68$. Once more, we find two regimes exhibiting mean-reversion, i.e. $\Phi_2 = -0.16$ and $\Phi_3 = -0.15$.

Figure \ref{fig:2008c} and \ref{fig:2008c_eta} are analogous to Figure \ref{fig:2000r} and \ref{fig:2000r_eta} respectively. We can see that the Markov chain switched to the high risk regime right after the collapse of the investment bank Lehman Brothers. Remarkably, it stayed in that regime throughout almost all the Banking Crisis, even though we observe numerous small upwards trends, meaning many thought we hit the bottom. \\

\begin{table}
	\caption{P-values (in percentage) for the proposed goodness-of-fit test using N=10000 bootstrap samples on the S\&P 500 daily returns for the 2010's recovery, along with the log-likelihood of the models and the the P-values (in percentage) of the likelihood ratio test statistic D.}
	\label{tab:gof_lrt_2010Bull}
\begin{center}
	\begin{tabular}{|l|cccc|}
\hline
& \multicolumn{4}{c|}{Number of regimes}\\
		 &  1 & 2 & 3 & 4 \\
		\hline
		GoF $P$-value (ARHMM)& 0 &	0&	0 & 1.56 \\
		GoF $P$-value (HMM)    & 0 & 0 & 0 & 5.83     \\
		Log-likelihood (ARHMM) &  5,696 &	 5,936 &  5,985    & 6,012 \\
		Log-likelihood  (HMM) &   5,694 &     5,931 &     5,981 &   6,006   \\
		$P$-value (D) &  4.19 & 1.52 & 4.46 & 3.02	 \\
		\hline
	\end{tabular}
\end{center}
\end{table}

\begin{table}
	\caption{Parameters estimation for the four-regime models on the S\&P 500 Total Return daily returns for the 2010's recovery. $\mu$ and $\sigma$ are presented as annualized percentage.}
	\label{tab:calib_2010Bull}

\begin{center}
	\begin{tabular}{|c|cccc|cccc|}
	\hline
		& \multicolumn{4}{c|}{AR(0)} & \multicolumn{4}{c|}{AR(1)}\\
		Parameter & \multicolumn{4}{c|}{Regime} & \multicolumn{4}{c|}{Regime} \\
		& 1         & 2       & 3         & 4     & 1         & 2       & 3         & 4                \\ \hline
		$\mu$     &29.41 &  303.07 &  -68.77 &  -28.25 &   32.04 &  365.83 & -68.82 &  -27.50                       \\
		$\sigma$  &   6.70  &  8.13 &  13.50  &  29.05 &    6.73 &    7.56 & 13.39 &   28.50                        \\
		$\Phi$     & 0 & 0 & 0 & 0           &  -0.04 &    0.14 &  -0.09 &   -0.08    \\
$\nu$     &   0.44 &    0.09 &     0.33 &    0.14 &    0.45 &    0.09 &  0.32 &   0.14 \\
		\hline
		&	0.91  &       0 &   0.09    &     0    &0.90  &       0  & 0.10 &        0 \\
		Q &		0.47 &    0.07 &    0.45 &         0   & 0.53 &    0.06 &  0.41&       0 \\
		&	0 &   0.23 &   0.76 &    0.01 &         0  &  0.23 &  0.76 &    0.01  \\
		&	0   & 0.03 &         0 &   0.97 &         0   & 0.03 &        0  &  0.97 \\
		\hline
	\end{tabular}
	\end{center}
\end{table}

\begin{figure}
	\begin{center}
	\includegraphics[width=10cm]{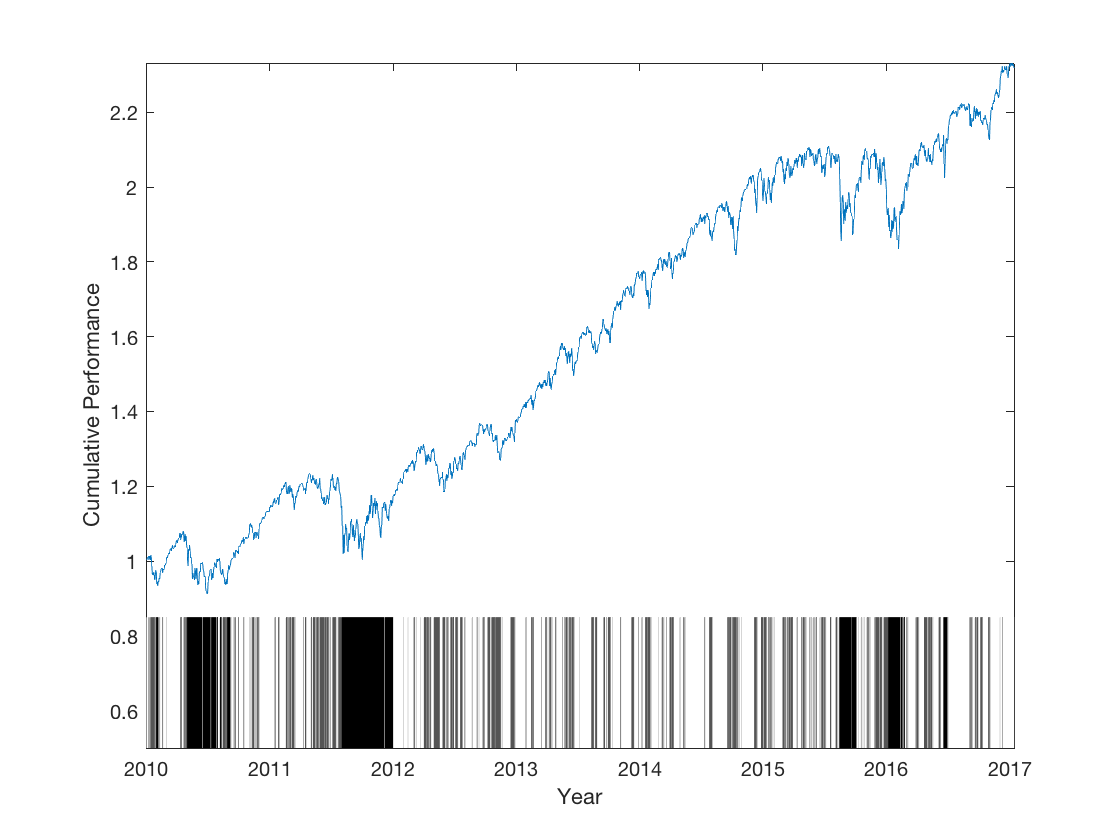}
\end{center}
\caption{Most probable regimes for the four-regime AR(1) model fitted on the S\&P 500
Total Return index from 05/01/2010 to 20/01/2017, together with the cumulative performance
of the index. Darker areas represent higher volatility states.}\label{fig:2010r}
\end{figure}

\begin{figure}
\begin{center}
	\includegraphics[width=10cm]{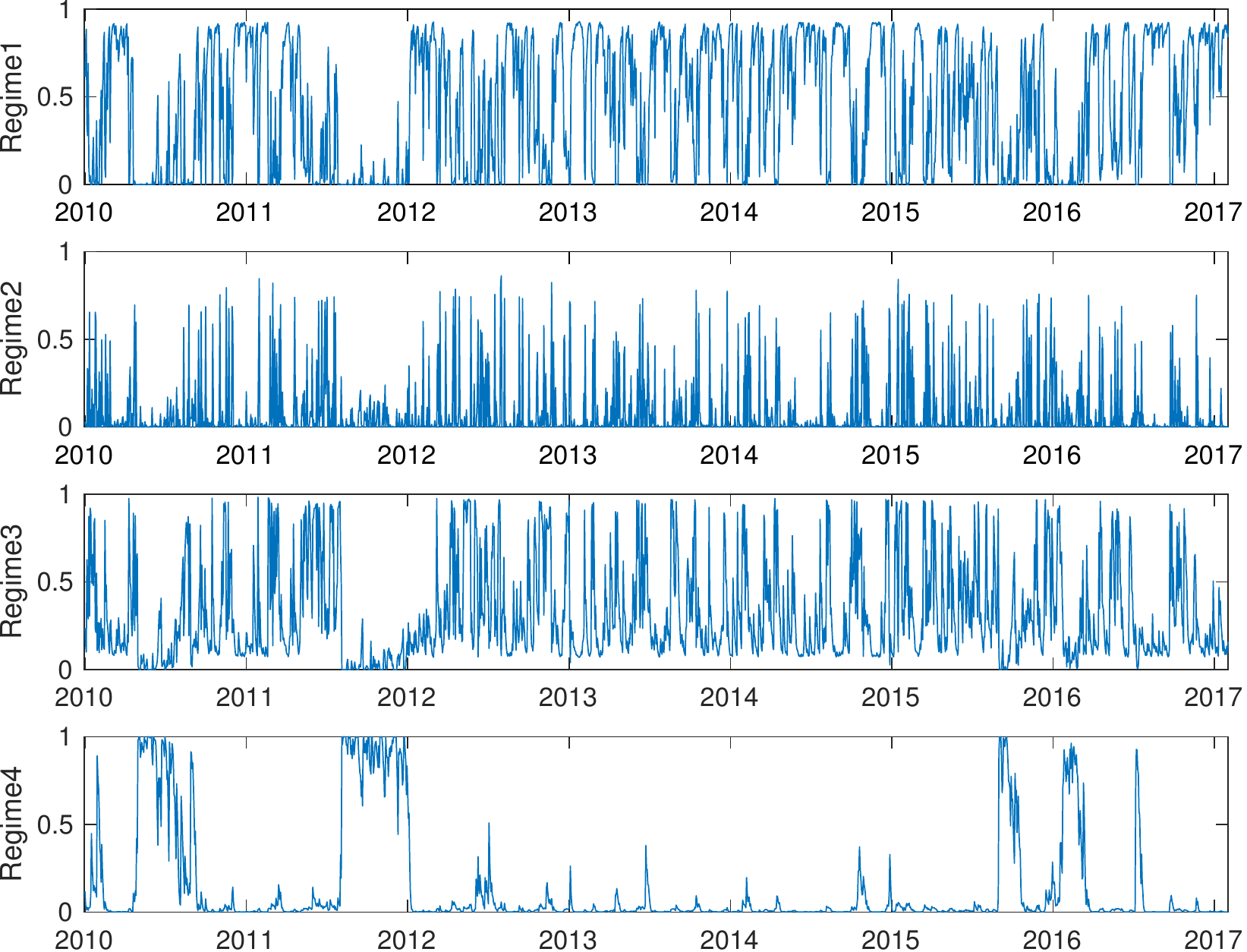}
\end{center}
	\caption{Probability of the regimes, i.e. $\eta_t$, for the four-regime AR(1) model fitted on the S\&P 500
Total Return index from 05/01/2010 to 20/01/2017.}\label{fig:2010r_eta}
\end{figure}
			
			For the last period, we chose four-regimes models. As noted previously, the four-regime HMM did not pass the goodness-of-fit test. We still present the estimated parameters in Table \ref{tab:calib_2010Bull}, as a mean of comparison. The calibration for this period is less intuitive than the previous ones. The inverse correlation between risk and expected premium is not as strong. Also, both models have a non-persistent regime with huge expected returns, (i.e. regime 2). Nevertheless, we still find modest mean-reversion for two regimes (i.e. regime 3 and 4), and the high-risk regime is highly persistent, as highlighted bu $Q_{4,4}=0.97$, as it was in the two previous cases.  The most probable regimes are displayed in Figure \ref{fig:2010r}, while the probability of the regimes  are presented in Figure \ref{fig:2010r_eta}.  Interestingly enough, the crisis in the subprime mortgage market is adequately captured by the high risk regime.. The final part of 2011 was marked by fear of the European sovereign debt crisis spreading to Italy and Spain. Once again, the ARHMM isolated the stock markets fall quite accurately.

\begin{figure}
\begin{center}
	\includegraphics[width=10cm]{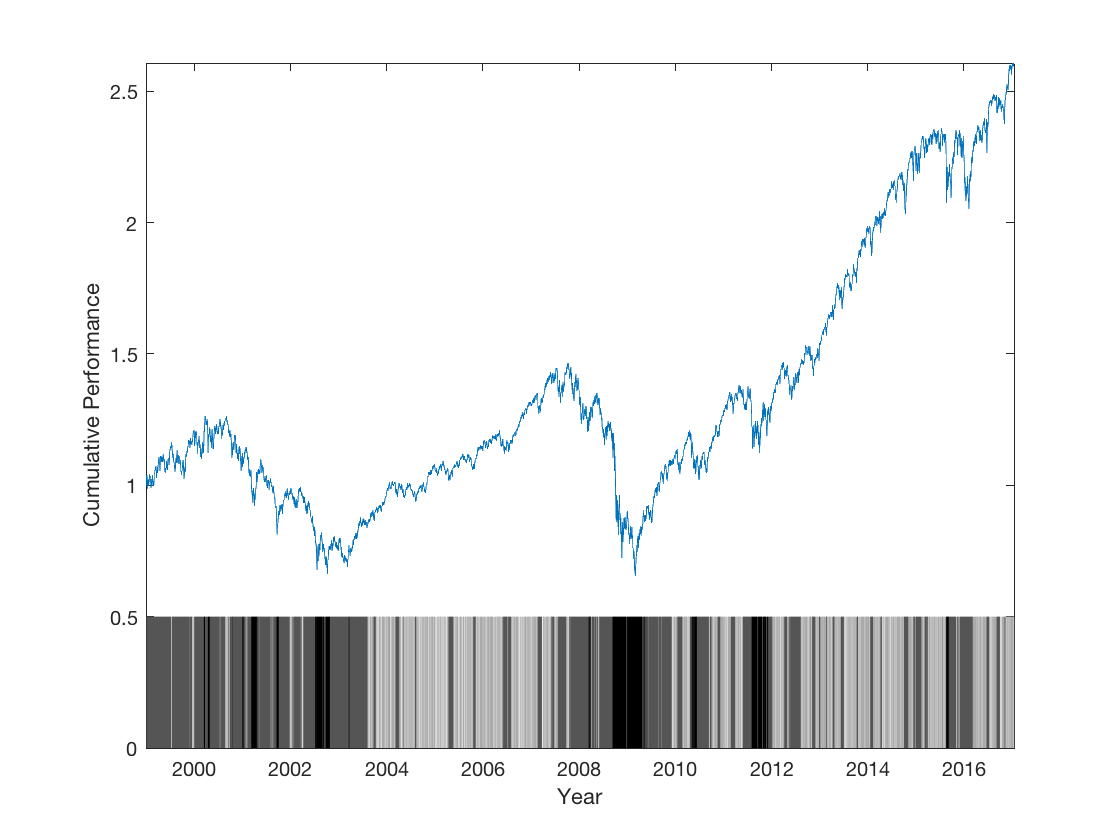}
\end{center}
	\caption{Most probable regimes for the four-regime AR(1) model fitted on the S\&P 500
Total Return index from 01/04/1999 to 01/20/2017, together with the cumulative performance
of the index. Darker areas represent higher volatility states.}\label{fig:sp500}
\end{figure}

			We also estimated the ARHMM on the returns from 01/04/1999 to 01/20/2017. This long period is far from stationary, but it is still interesting to see how the model performs through recessions and recoveries. We chose a four-regime model, as  indicated by the goodness-of-fit tests. We can see on Figure \ref{fig:sp500} that the 2000's bubble burst and the recent financial meltdown (2008-2009) are both correctly captured by the high risk regimes.

\section{Optimal Discrete Time Hedging}\label{sec:opt_hedg}

In what follows, we use the notations and results from \citet{Remillard/Rubenthaler:2013}.

Denote the price process by $S$, i.e., $S_t$ is the value of $d$ underlying assets at period $t$ and let $\mathds{F} =   \{\mathcal{F}_t, \quad t = 0,\ldots,n\}$ a filtration under which $S$ is adapted. Further assume $S$ is square integrable. Set $\Delta_t$ = $\beta_t S_t - \beta_{t-1}S_{t-1}$, where the discounting factors $\beta_t = e^{-r_t}$ are deterministic for $t = 1,\ldots,n$. We are interested in the optimal initial investment amount $V_0$ and the optimal predictable investment strategy $\vec{\varphi}$ = $(\varphi_t)_{t=1}^n$ that minimizes the expected quadratic hedging error for a given payoff, $C$, at time $n$ (e.g a call option). Formally, the problem is stated as
	\begin{equation} \label{eq:opt_hedge}
		\begin{aligned}
			& \underset{\{V_0,\vec{\varphi}\}}{\text{min}}	
			& & E[\{ G(V_0,\vec{\varphi}) \}^2], \\
		\end{aligned}	
	\end{equation}
	where
	\begin{equation*}
		G = G(V_0,\vec{\varphi}) = \beta_n (C-V_n)
	\end{equation*}
	and $V_t$ is the current value of the replicating portfolio at time $t$. In other words, it is the current value of the optimal predictable investment strategy, $\vec{\varphi}$,
	
	\begin{equation*}
		\beta_t V_t = V_0 + \sum_{j=1}^{t} \varphi_j^\top \Delta_j,
	\end{equation*}
	
	for $t = 0,\ldots,n$. \\
	
	To solve (\ref{eq:opt_hedge}), set $P_{n+1}$ = 1, and define
	
	\begin{equation*} \label{offline}
		\begin{aligned}
		    & \gamma_{t+1} = E( P_{t+1} | \mathcal{F}_t), \\
			& \mathfrak{a}_t = E( \Delta_t \Delta_t^\top P_{t+1} | \mathcal{F}_{t-1} )= E( \Delta_t \Delta_t^\top \gamma_{t+1} | \mathcal{F}_{t-1} ), \\
			& \mathfrak{b}_t = E( \Delta_t P_{t+1} | \mathcal{F}_{t-1} ) =  E( \Delta_t \gamma_{t+1} | \mathcal{F}_{t-1} ), \\
			& \rho_t = \mathfrak{a}_t^{-1} \mathfrak{b}_t, \\
			& P_t = \prod_{j=t}^{n} (1-\rho_j^\top \Delta_j),
		\end{aligned}	
	\end{equation*}
	
	for $k = n,\ldots,1.$ \\
	
	We can now state Theorem 1 of \cite{Remillard/Rubenthaler:2013}, which is a multivariate  extension of \cite{Schweizer:1995a}.
	
	\begin{thm}\label{th:opt_hed}
		
		Suppose that $E(P_t|F_{t-1}) \neq 0$ P-a.s., for 1,\ldots,n. This condition is always respected for regime-switching models. Then, the solution  $\left(V_0,\vec{\varphi}\right)$ of the minimization problem \eqref{eq:opt_hedge} is $V_0 = E(\beta_n C P_1) / E(P_1)$, and
		\begin{equation}\label{eq:phi}
			\varphi_t = \alpha_t - \check{V}_{t-1} \rho_t, \quad k=1,\ldots,n.
		\end{equation}
		where
		\begin{equation}\label{eq:alpha}
			\alpha_t = \mathfrak{a}_t^{-1} E(\beta_n C  \Delta_t P_{t+1} | \mathcal{F}_{t-1} ).
		\end{equation}
		and $\check{S}$ and $\check{V}$ are the present values of $S$ and $V$.
	\end{thm}

\begin{rem}
		$V_0$ is chosen such that the expected hedging error, G, is zero.
		\cite{Remillard/Rubenthaler:2013} also showed that $C_t (S_t,\tau_t)$ given by
		\begin{equation}\label{eq:BC}
			\beta_t C_t = \dfrac{E(\beta_n C P_{t+1} | \mathcal{F}_t )}{E(P_{t+1} | \mathcal{F}_t)}
		\end{equation}
		is the optimal investment at period t so that the value of the portfolio at period $n$ is as close as possible to $C$ in term of mean square error $G$, in particular, $V_0=C_0$.

	$C_t$ can be interpreted as the option price at period $t$. By increasing the number of hedging periods, $C_t$ should tend to a price under a risk-neutral measure; see, e.g., \cite{Remillard/Rubenthaler:2016}. For example, when there is only one regime, the density is Gaussian and $\Phi_1$ fixed at 0, $C_t$ tends to the usual Black-Scholes price. The detailed optimal hedging implementation for ARHMM is described in Appendix \ref{app:opt_hedge}. It then follows that
\begin{equation}\label{eq:newC}
	\check C_{t-1} = 		\beta_{t-1} C_{t-1}\gamma_t  = E\{(1-\rho_t^\top \Delta_t)\check C_{t} | \mathcal{F}_{t-1} )
		\end{equation}
\begin{equation}\label{eq:newalpha}
	\alpha_t = \mathfrak{a}_t^{-1} E(\check C_t   \Delta_t | \mathcal{F}_{t-1} ).
\end{equation}
\end{rem}

 To derive the optimal hedging algorithm, we also need the following result, valid for a general ARHMM.

First, write $S_t = D(S_{t-1})e^{Y_t}$, where $e^{Y_t}$ is the vector with components $e^{(Y_t)_j}$, and $D(s)$ is the diagonal matrix with diagonal elements $(s)_j$, $j\in \{1,\ldots,d\}$.

The proof of the following theorem is given in Appendix \ref{pf:thm-functions}.
\begin{thm}\label{thm:functions}
For any $t\in\{1,\ldots,n\}$, $\fa_t =  D(\check S_{t-1} ) a_t(Y_{t-1},\tau_{t-1})D(\check S_{t-1} ) $, $\fb_t= D(\check S_{t-1} )  b_t(Y_{t-1},\tau_{t-1})$, $\rho_t = D^{-1}(\check S_{t-1}) h_t(Y_{t-1},\tau_{t-1})$, and $\gamma_t= g_t(Y_{t-1},\tau_{t-1})$, with $h_t = a_t^{-1} b_t$, where $a_t$, $b_t$, and $g_t$ are deterministic functions given respectively by
\begin{eqnarray}
a_t(y,i) &=& E\left\{  \zeta_t \zeta_t^\top g_{t+1}(Y_{t},\tau_{t})| Y_{t-1}=y,\tau_{t-1}=i\right\},\label{eq:a}\\
b_t(y,i) &=& E\left\{  \zeta_t  g_{t+1}(Y_{t},\tau_{t})| Y_{t-1}=y,\tau_{t-1}=i\right\},\label{eq:b}\\
g_t(y,i) &=& E\left\{   g_{t+1}(Y_{t},\tau_{t}) | Y_{t-1}=y,\tau_{t-1}=i\right\} \label{eq:c}\\
 && \qquad -  b_t^\top(Y_{t-1},\tau_{t-1}) h_t(Y_{t-1},\tau_{t-1}),\nonumber
\end{eqnarray}
with $\zeta_t =  e^{Y_{t}-r_t} -\mathbf{1}$, and $g_{n+1}\equiv 1$.
If in addition
$\beta_n C = \Psi_n(\check S_n)$, then $\check C_t = \Psi_{t}(\check S_{t},Y_{t},\tau_{t})$, where
\begin{equation}\label{eq:Ct}
\Psi_{t-1}(s,y,i) = E\left[ \Psi_t\left\{D(s)e^{Y_t-r_t},Y_{t},\tau_{t}\right\}\left\{ 1- h_{t}(y,i)^\top \zeta_t\right\} |Y_{t-1}=y,\tau_{t-1}=i\right],
\end{equation}
and
\begin{equation}\label{eq:alpha_t}
\alpha_t = D^{-1}(\check S_{t-1})a_t^{-1}(Y_{t-1},\tau_{t-1}) \mathbf{A}_t(\check S_{t-1},Y_{t-1},\tau_{t-1}),
\end{equation}
 where
\begin{equation}\label{eq:At}
 \mathbf{A}_t(s,y,i) = E\left[\Psi_t\left\{D(s)e^{Y_t-r_t},Y_t,\tau_t\right\} \zeta_t|Y_{t-1}=y,\tau_{t-1}=i\right].
 \end{equation}
\end{thm}
For example, for a call option with strike $K$, $\Psi_n(s) = \max(0,s-\beta_n K)$.

	\subsection{Implementation issues}	
	
		There are two main problems related to the implementation of the hedging strategy: $a_t$, $b_t$, $g_t$, $\Psi_t$ and $A_t$ defined in expressions \eqref{eq:a}-\eqref{eq:At} must be approximated and regimes must be predicted. \\
		
		 We discretize $a_t$, $b_t$ $g_t$ functions of the underlying values $y$ with a grid $G$. In a similar manner, we discretize $\Psi_t$ and $A_t$ functions of the underlying values $s$ and $y$. To solve the recursion given by \eqref{eq:Ct}-\eqref{eq:At},  \cite{Remillard/Hocquard/Lamarre/Papageorgiou:2017} interpolate and extrapolate linearly the simulated outcomes on $G$, using a stratified Monte Carlo sampling procedure. Because the simulations are computationally expensive and introduce variability, we propose a novel technique to approximate these integrals using semi-exact calculations, inspired by \cite{Remillard:2013} Chapter 3. The details for the semi-exact calculations are presented in Appendix \ref{app:semi_exact}.
		
		 We also included the Monte Carlo sampling procedure as a mean of comparison. Interestingly, we found that by simply rescaling the Monte Carlo samples to the desired mean and volatility, we achieved results as accurate as the semi-exact calculations, as pointed out in Section \ref{sec:simu_and_hedge}.
		
		 	As for defining the points on the grid, previous literature suggest choosing $10^3$ equidistant points marginally covering at least 3 standard deviations
under the respective highest volatility regimes. Importantly, we found that strategically choosing the points with respect to the percentiles of simulated processes significantly reduces the number of points needed while keeping the accuracy at a reasonable level. \\

		 Next, we need to predict $\tau_1$ based on $(R_1, R_0$ and $\tau_0)$ and so on. The predicted regime is $\hat{\tau}$ is the one having the largest probability given the information on prices up to time $t$, i.e. the $most$ $probable$ $regime$ given by (\ref{eq:most_prob_reg}).  Note that this methodology introduces a bias. We also studied the less biased approach of weighting the regimes proportionally to $\eta_t$, but since the results were comparable and did not lead to any significant improvement, they are omitted from the analysis. For more details on regime predictions, see section \ref{ssec:prediction}.
		
		 Then, according to \eqref{eq:phi} and \eqref{eq:alpha_t}, the optimal hedging weights $\varphi_t$ for period $[t-1,t)$ are approximated by

		 \begin{equation}\label{eq:approx}
			 \hat{\varphi_t} = \alpha_t(\check{S}_{t-1}, Y_{t-1},\hat{\tau}_{t-1}) - D^{-1}(\check{S}_{t-1})\check{V}_{t-1}h_t(Y_{t-1},\hat{\tau}_{t-1}), t = 1,\ldots,n.
		 \end{equation}

		 $V_0$ is approximated by $C_0(S_0,\hat{\tau}_0,0)$
%
%
		 while the remaining monies, $V_0 - \hat{\varphi}_1^\intercal S_0$, are invested in the riskless asset. Next, as $S_1$ is observed, one firsts computes the actual portfolio value $V_1$, then predicts the current regime $\tau_1$ and finally approximates the optimal weights $\varphi_2$. This process is iterated until expiration of the option.

\subsubsection{Using regime predictions}

Here, we obtain option prices and strategies that depend on the unobservable regimes $\tau$, since $(S_t,\tau_t)$ is a Markov chain. However, \citet{Francois/Gauthier/Godin:2014} proposed a very interesting approach: they showed that $(S_t,\eta_t)$ is Markov, so one can obtain prices and hedging strategies depending on  $(S_t,\eta_t)$ instead. This makes sense financially. However, this new Markov chain lives in a $l+d-1$-dimensional space, because the values of $\eta_t$ belong to the simplex $\cS_l=\{x_1,\ldots,x_l; \quad x_i\ge 0,  x-1+\cdots + x_d=1\}$. \citet{Francois/Gauthier/Godin:2014} considered only 2 regimes
and one asset, so the real dimension is $2$. When $l>2$, this becomes numerically intractable.

		\subsection{Global hedging}\label{sec:global}
	
		In practice, an expected hedging error characterized by $V_t - C_t$ will emerge. In other words, the replicating portfolio at period t will not be worth the optimal investment $C_t$. Under the Black-Scholes setting, such error is unaccounted for since derivatives can be replicated perfectly (in continuous time). In contrast, under the proposed optimal hedging protocol, the exposures $\varphi_t$ depend on the replicating portfolio, $V_{t-1}$ (see equation \ref{eq:phi}), which in turn depends on the past strategy path. \\
		Under extreme scenarios, the replication of a call option might lead to optimal exposures $\varphi$ greater than one share. Intuitively, this feature is optimal with respect to closing the gap between $V$ and $C$.
	
		\subsection{Simulated hedging errors}\label{sec:simu_and_hedge}

\begin{figure}[ht]
	\begin{center}
	\includegraphics[width=130mm]{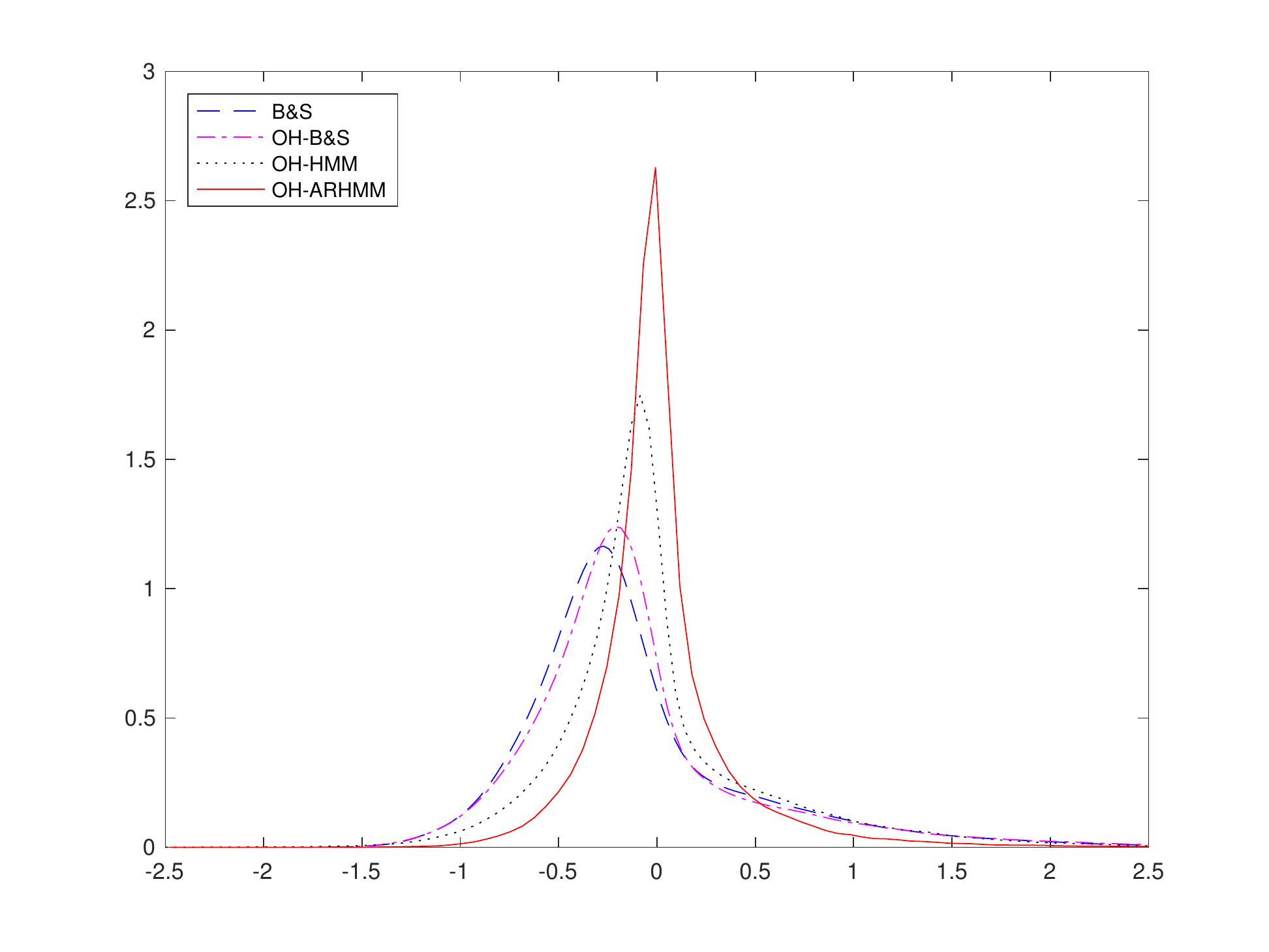}
\end{center}
	\caption{Estimated densities for the hedging errors in an autoregressive hidden Markov model, using 50000 portfolios. Only the semi-exact densities are shown, as they were indiscernible from the Monte Carlo ones.}
				\label{fig:simu}
			\end{figure}

	To assess the proposed strategy's accuracy, we simulated 10000 trajectories under ARHMM and hedge identical options under different hedging strategies. To be realistic, the parameters were taken from Table \ref{tab:calib_2000Bull}. The hedging methodologies are the classical Black-Scholes delta-hedging (B\&S) and optimal hedging under ARHMM (OH-ARHMM), HMM (OH-HMM) and Gaussian (OH-B\&S) returns (i.e., considering only $1$ regime). We also compared semi-exact approximation to Monte-Carlo. The option in question is a call with $S_0$ and $K$ equal to 100, risk-free rate $r=0.01$, 3 month maturity (63 days) with daily hedging.
The main hedging error statistics are given in Table \ref{tab:simu}, while the estimated densities are displayed in Figure \ref{fig:simu}.	
	
	OH-ARHMM achieves a 33\% reduction in RMSE compared to B\&S and OH-B\&S and a 26\% to OH-HMM. The latter is quite impressive, as it highlights how big of an impact the autoregressive dynamic has.

			\begin{table}[ht]
\caption{Statistics for the hedging errors in an autoregressive hidden Markov model, using 10000 simulated portfolios. }\label{tab:simu}
\begin{center}
					\begin{tabular}{ | l | r r r r r r |}
						\hline
											 & B\&S & OH-B\&S & HMM & HMM MC & ARHMM & ARHMM MC \\ \hline
		Average    &  -0.105 &    -0.084 &    0.004 &    \textbf{0.003} &    0.025 &     0.030  \\
		Median     & -0.236 & -0.202 & -0.085 & -0.086 & \textbf{-0.019} & \textbf{-0.019} \\
		Volatility &     0.611 &    0.626 &    0.559 &    0.559 &  \textbf{0.411} & 0.412  \\
		Skewness   & 1.715 &    1.948 &    1.639 &   \textbf{1.629} &  4.644 &   4.738   \\
		Kurtosis   & 7.737 &    9.558 &    9.464 &    9.356 &   \textbf{71.834} & 78.219 \\
		Minimum    & \textbf{-1.658} & \textbf{-1.658} & -4.649 & -4.633 & -2.477 & -2.413 \\
		VaR (1\%)  &  -1.110 &   -1.118 &   -1.087 &   -1.086 &   -0.749 & \textbf{-0.741} \\
		VaR (99\%) & 2.069 &   2.227 &    1.987 &    1.982 &    \textbf{1.526} &  1.531  \\
		Maximum    &  \textbf{4.886} &  6.725 &    6.538 &    6.520 &   12.958 &   14.266  \\
		RMSE       &  0.620 &    0.632 &    0.559 &    0.559 &   \textbf{0.411} &    0.413  \\ \hline	
					\end{tabular}
\end{center}
		\end{table}	
			
\section{Out-of-sample vanilla pricing and hedging}
	
	\subsection{Methodology}\label{sec:hedg_metho}
		
		To exhibit the proposed hedging protocol, we buy and sell vanilla options on the S\&P 500 depending on the how the market prices compare with our theoretical prices. Then, we hedge the positions until expiration. We then assess the impact of model specification on the delta-hedging strategy by examining the statistical properties of the hedging error and of the strategy's returns. All hedging portfolios are re-balanced on a daily basis, as is often assumed in the volatility timing literature; see e.g., \cite{Fleming/Kirby/Ostdiek:2001}.
		
		The market price of an option is defined as the last (i.e. as 4:15 PM EST) midpoint between the bid and the ask. The price of the underlying is its listed close value. For simplicity, we neglect issues related to time-varying discount rates by assuming constant continually compounded daily rates. Risk-free rates, $r$, are linearly interpolated for a given maturity, $n$, from the zero-coupon U.S. yield curve.

\begin{rem}
For the implementation, we chose to present only the results using the most probable regime for the computation of the hedging strategy. These results are  a little bit better than those obtained by weighting the hedging strategy according to the probability of occurrence $\eta_t(1),\ldots,\eta_t(l)$ of the regimes at period $t$.
\end{rem}

\subsubsection{The underlying}
	
			We make the reasonable assumption the spot S\&P 500 is investable and tradable at a minimal cost. The forward rate is retrieved for the maturities of interest directly from the option data at hand, as proposed by \cite{Buraschi/Jackwerth:2001}. From put-call parity, the option implied forward value at $n$, $F_n$, is
			
\begin{equation*}
F_n = (\tilde C(\tilde K, n ) - \tilde P (\tilde K, n)) e^{r_nn} + \tilde K,
\end{equation*}
where $C(\tilde K, T)$ and $P(\tilde K, T)$ are respectively the call and put $market$ values expiring at $T$ with strike $K$ and $\tilde K$ is the at-the-money strike value minimizing $|C(\tilde K, T) -P(\tilde K, T)$ for all strikes offered by the exchange. We use at-the-money options because they are the most liquid and are thus less likely to provide \textit{cash-and-carry} type arbitrage opportunities. We then compute the daily forward rate as $f_n = \dfrac{1}{n} \log(F_n/S_0)$ and the associated daily discounting factor $\beta = e^{-f_n}$, which reflects the current risk-free return on capital net of the implied continuous dividend yield.
			
\subsubsection{Option dataset}
			
			Exchange-traded options on the S\&P 500 are European, heavily traded and have a high number of strikes and maturities. \\
			To assess the accuracy of our model, we will analyze two periods with very different characteristics: the  2008 Financial Crisis, and a chunk of the recent recovery. Dates range from 09/24/2007 to 09/20/2009 and from 09/23/2013 to 07/08/2015, respectively. This will help us discern the impact on hedging and pricing when a dramatic regime change occurs, in the former, and when it does not, in the latter. \\
		    In order to minimize the effect of varying maturities, we will build the dataset of options having a maturity of about $1$ year, more precisely from 231 to 273 trading days till expiration. Also, because \textit{in-the-money} and  \textit{out-of-the-money} are less liquid, we will only include options where  moneyness (strike value divided by the underlying value), is between 0.9 and 1.1. This leaves us with a total of 180 options for the first period, and 478 for the second.
Note that at a given date, more than one option can meet these criteria.
	
\subsubsection{Backtesting}
		
			We apply the AR(1) regime-switching optimal hedging methodology with 3 regimes (ARHMM). We chose 3 regimes because it is the number of regimes that seemed the best  given the time windows studied, which we will describe in the next paragraph. We will compare it to the case with 1 regime and $\Phi$ fixed at 0, corresponding to the optimal hedging under the Black-Scholes  model (OH-B\&S).

			For each option in the dataset, we estimate the ARHMM parameters on the S\&P 500 log-returns with a 500 and 2000 day trailing window. We choose to backtest using 2 estimation windows in order to have a more in depth understanding of model specification on pricing and hedging. The 2000 day trailing window will always include the previous financial meltdown, i.e., dot-com bubble for our first analysis, and the 2008 financial crisis for the second one. The 500 day trailing window won't. Similarly, we applied this methodology to all the hedging protocols included in the analysis, which will be introduced below.
			
			From \cite{Merton:1973}, for a given \textit{moneyness}, the value of an option is homogeneous of degree one with respect to the underlying value. Thus, for each inception date, we normalize the option prices, the strike values and the underlying path at an initial S\&P 500 value of 100. Results can thus be aggregated through time and interpreted as a percentage of S\&P 500. Note that for each inception date, the hedging protocols are applied out-of-sample until maturity.

			To ensure comparability, OH-B\&S assumes the stationary distribution of the ARHMM when the autoregressive parameter $\Phi=0$. The OH-B\&S optimal hedging exposure is derived from an algorithm similar to the one presented in Section \ref{sec:opt_hedg}. Optimal hedging under unconditional distributions is presented in \cite{Remillard:2013}. Both strategies minimize the expected quadratic hedging error under their respective null hypothesis, namely that the returns follow an autoregressive regime-switching model (ARHMM), and a Gaussian  model (OH-B\&S).

			OH-B\&S methodology is not to be confused with the classical Black-Scholes delta hedging protocol. Indeed, the terminology only reflects the fact that we hedge and price under the Black-Scholes framework hypothesis, namely that assets follow geometric Brownian motions. Even though the OH-B\&S prices converge to the usual Black-Scholes prices as the number of hedging periods tends to infinity, the discrete time hedging strategies will not necessarily be the same. For this reason, the classical Black-Scholes delta-hedging methodology (B\&S) is also considered. Similarly to OH-B\&S, the B\&S volatility is calibrated to the stationary volatility of ARHMM. 

			We will add a final benchmark to our analysis, one that reflects how well the market would have hedged the same options, namely the delta-hedging methodology where the volatility is calibrated to the implied volatility at each hedging period (B\&S-M). It will inform us how well the models compare to market's intuition.
The effect of using the implied volatility was discussed  in \citet{Carr:2002}. However, his theoretical analysis cannot be performed here. \\

		To recap, we will buy and sell options depending on their market value compared to the theoretical prices, and hedge the positions until maturity. We will analyse the P\&L of the different methodologies, as well as the hedging errors. Two periods will be studied: the 2008 Financial Crisis and a chunk of the following recovery spanning from mid-2013 to mid-2015.
			
	\subsection{Empirical results}
	
		We define the hedging error as the present value of the liability $\beta_nC$ minus the present value of terminal portfolio $\beta_nV_n$. The options' maturity being set to one year, the annualized root-mean-squared hedging error can be computed by $\sqrt{ \hat{E} (\beta_n V_n - \beta_n C)^2 } $. This realized risk is the empirical counterpart of the quantity we minimized and as such, is the most relevant metric for comparing the different models. Keep in mind that there is a lot of overlap in our dataset, so the hedging error values are not independent, nor identically distributed since the moneyness or other parameters are not constant. Despite these inconveniences, the hedging errors are still useful to compare the models.

		Concerning the trading strategy, if the market is overvalued with respect to the model, we sell the option and hedge our position. Thus, the present value of the return is $(C_0 - V_0) - ( \beta_n C_n - \beta_n V_n )$. If the market is undervalued, we buy the option and hedge our position. The return will be the negative of the former.
	
		\subsubsection{2008-2009 financial crisis}\label{ssec:Crash_exp}

		In this section, we will focus on options with inception dates from September $24^{th}$ 2007 to September $20^{th}$ 2009. This period is really interesting. In the first part, the market experienced a huge increase in volatility and decrease in returns. In the second, the opposite.
		
\begin{figure}[ht!]
				\begin{center}
				(a)	\includegraphics[width=2.24in]{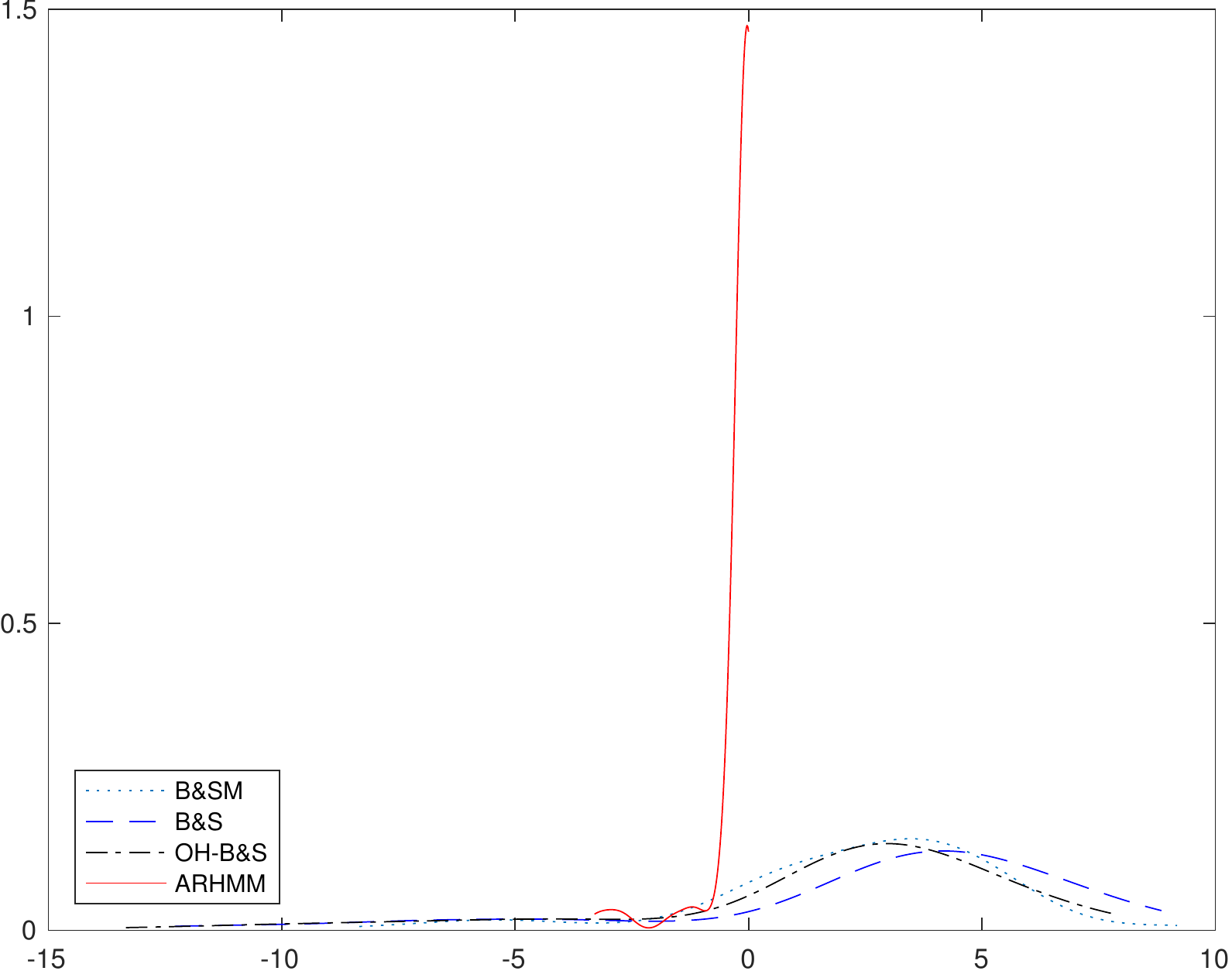}
(b) 	\includegraphics[width=2.24in]{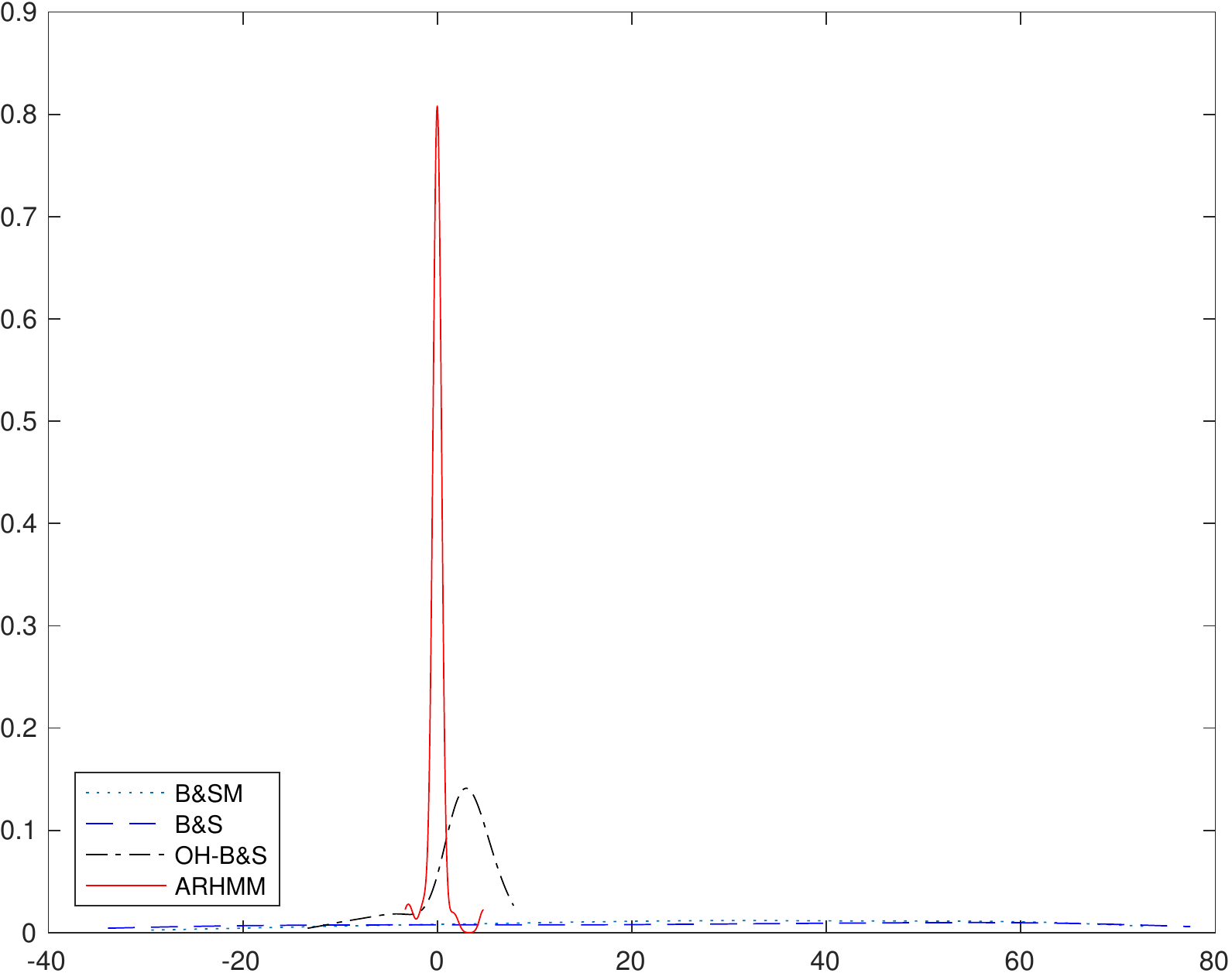}
					\end{center}
			\caption{Hedging error density approximation for the 90 calls (a) and 90 puts (b) traded in the 2008-2009 Financial Crisis with 500 days trailing estimation window.}\label{fig:Crash_t500_he_pdf}
				\end{figure}

\begin{figure}[ht!]
				\begin{center}
				(a)	\includegraphics[width=2.24in]{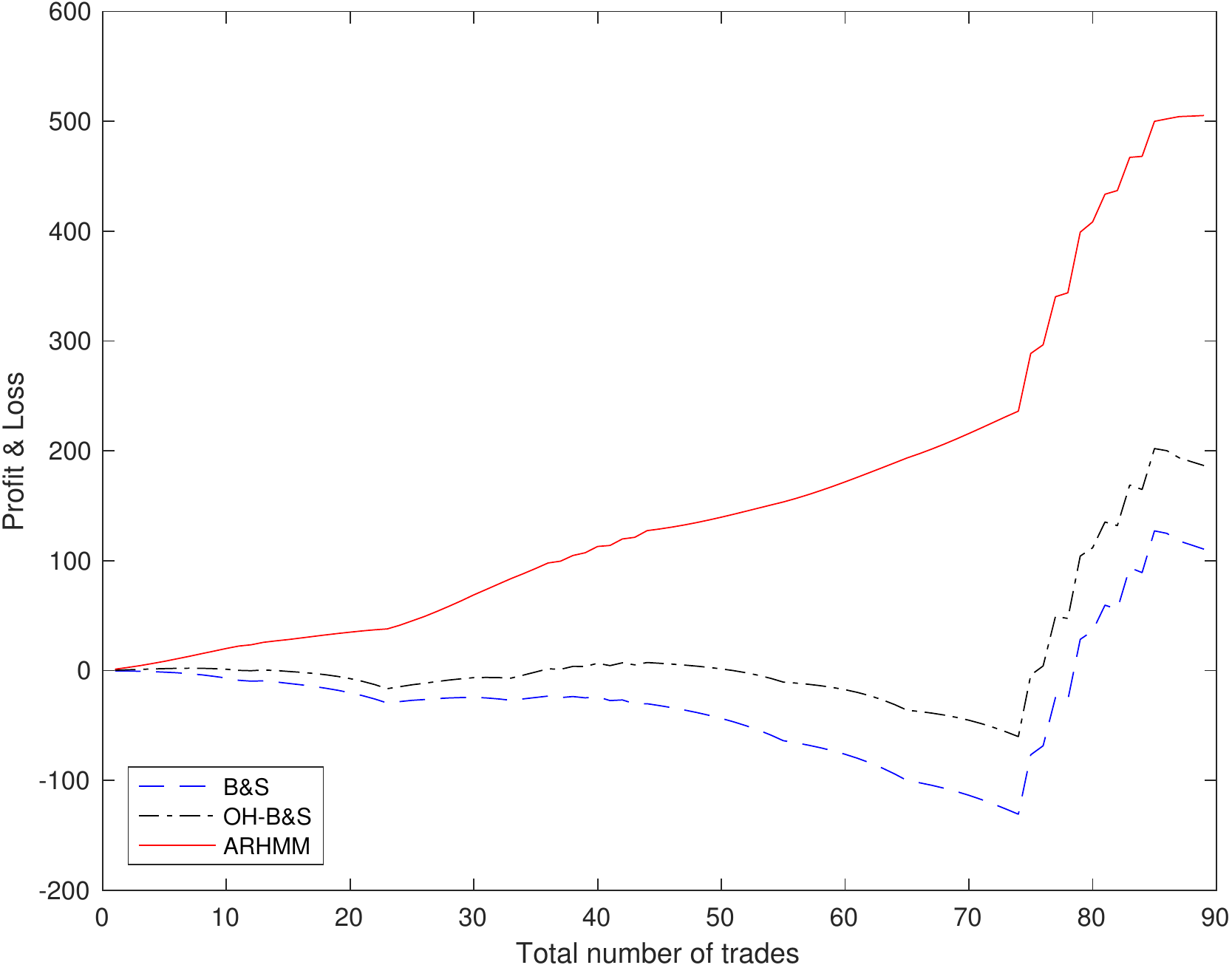}
(b) 	\includegraphics[width=2.24in]{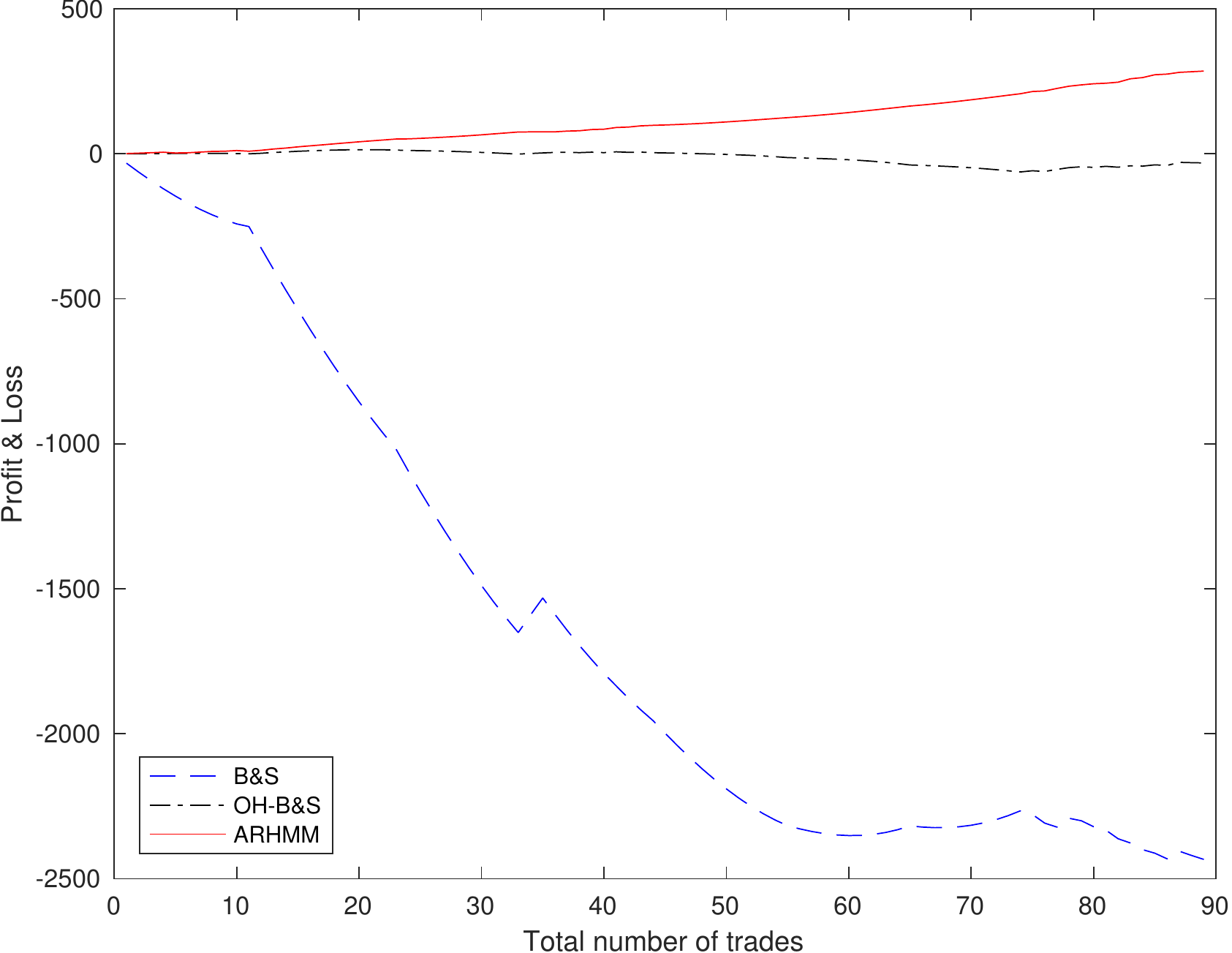}
					\end{center}
			\caption{Profit \& Loss of trading strategy for the 90 calls (a) and 90 puts (b) traded in the 2008-2009 Financial Crisis with 500 days trailing estimation window.}\label{fig:Crash_t500_pnl}
				\end{figure}

		We will first turn our attention to the 500 trailing estimation window case. Table \ref{tbl:Crash_t500_he} and Figure \ref{fig:Crash_t500_he_pdf} present the hedging error's statistics and density approximation. Figure \ref{fig:Crash_t500_pnl} presents the results of the trading strategy, i.e., the cumulative value of a portfolio that traded the 90 options. The x axis is the cumulative number of options traded in chronological order. In this case, ARHMM is by far the superior methodology. It achieved the best hedging error considering all the metrics for both calls and puts. Further, it was the best trading strategy for both type of options, even though the hedging errors are almost entirely negative in the calls case. Note that the statistic ``Bias'' refers to the difference between the market price and the theoretical price. Therefore, it is always $0$ for the BS-M, since the implied volatility is used.
		
		When volatility increases and returns turn negative, the puts' value increase and one needs to be hedge accordingly. B\&S and B\&S-M failed to do so, resulting in huge hedging errors and great losses portfolio wise.

		\begin{table}[ht]
		\caption{Hedging error statistics for the 90 calls and the 90 puts traded in the 2008-2009 Financial Crisis with 500 days trailing estimation window.}\label{tbl:Crash_t500_he}
			\begin{center}
		{\footnotesize
				\begin{tabular}{ | l | c c c c | c c c c |}
					\hline
                     & \multicolumn{4}{c|}{Calls}                            & \multicolumn{4}{c|}{Puts}                                                  \\
					& B\&S-M & B\&S & OH-B\&S & ARHMM                       & B\&S-M & B\&S & OH-B\&S & ARHMM                                          \\ \hline
						RMSE         &3.87&5.27&4.53&\textbf{0.61}            &39.95&42.52&4.53&\textbf{0.98}                                         \\
						Bias   & 0  	&-4.52&-4.37&-5.35                      & 0 &-1.05&-0.91&-1.87                                           \\
						VaR 1\%           & -7.64&-12.02&-12.47&\textbf{-3.17}      & -28.39&-33.24&-12.47&\textbf{-3.18}                                 \\
						Median              &2.9&3.82&2.68&\textbf{-1.65e-04}            &30.78&29.77&2.68&\textbf{0.01}                                        \\
						VaR 99\%            &9.16&8.61&7.57&\textbf{-1.97e-09}	           &72.9&77.22&7.57&\textbf{4.75}                                     \\ \hline
				\end{tabular}
}
			\end{center}
			\end{table}

\begin{figure}[ht!]
		\begin{center}
		(a)	\includegraphics[width=2.24in]{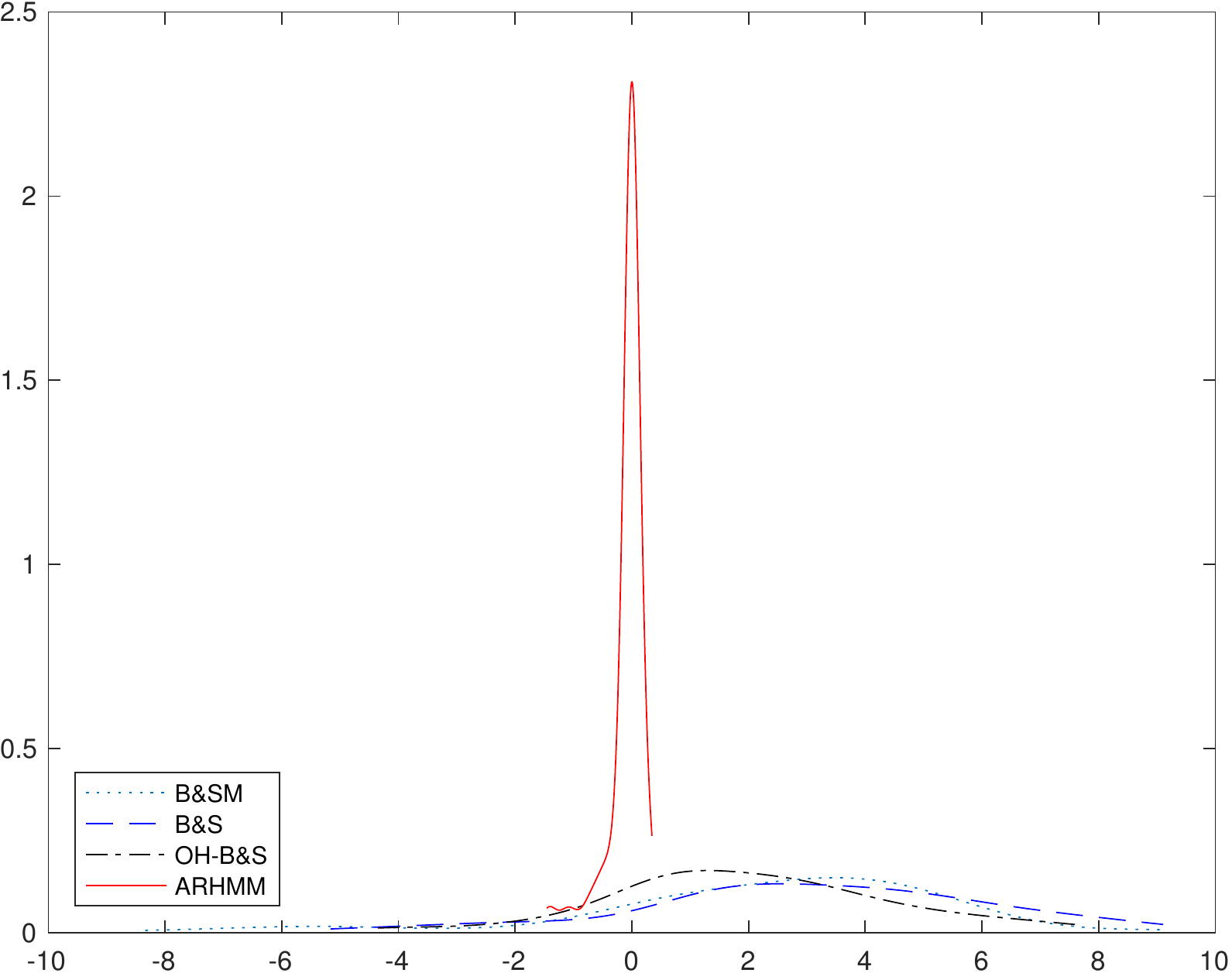}
(b) 	\includegraphics[width=2.24in]{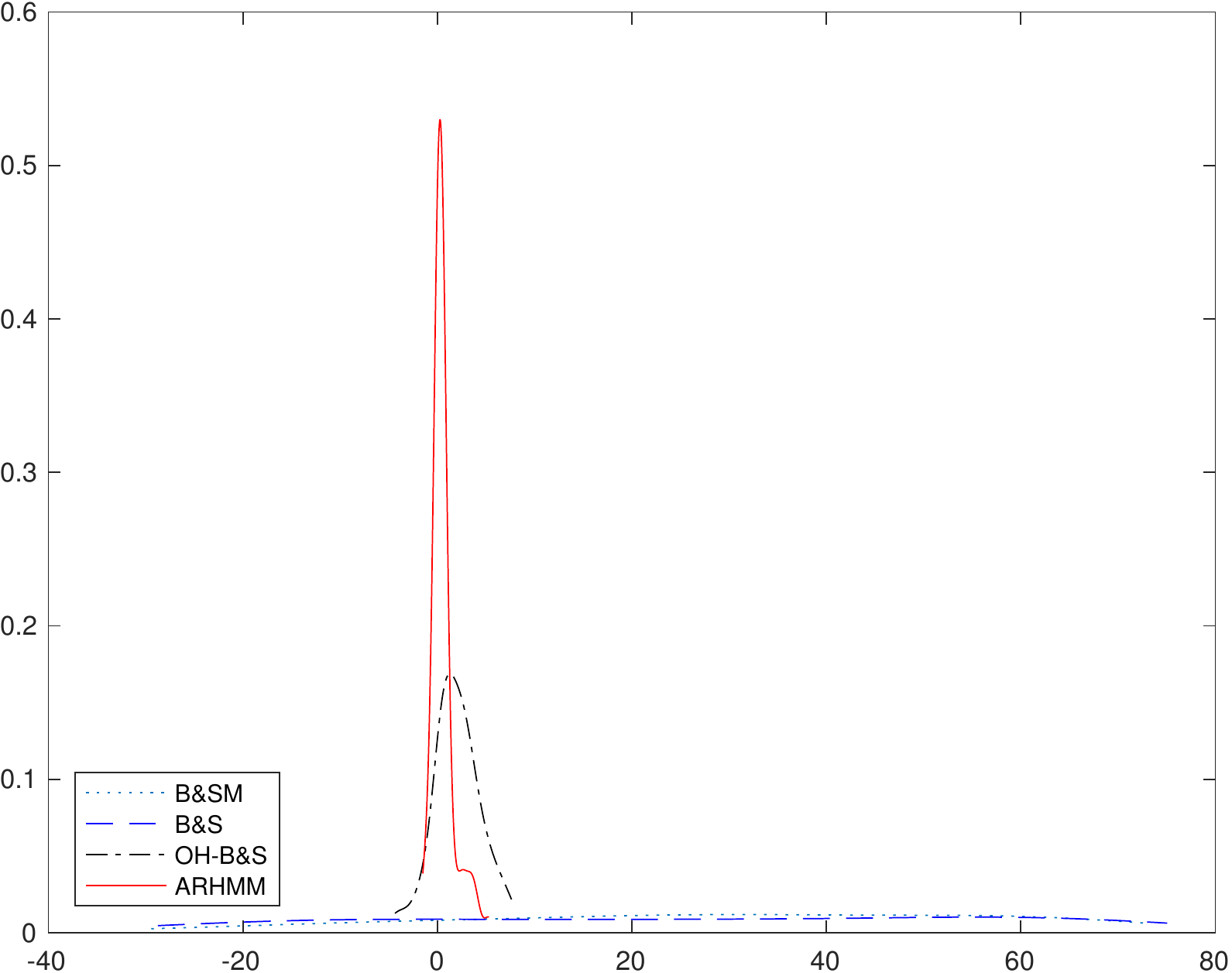}
			\end{center}
			\caption{Hedging error density approximation for the 90 calls (a) and 90 puts (b) traded in the 2008-2009 Financial Crisis with 2000 days trailing estimation window.}\label{fig:Crash_t2000_he_pdf}
		\end{figure}

\begin{figure}[ht!]
		\begin{center}
		(a)	\includegraphics[width=2.24in]{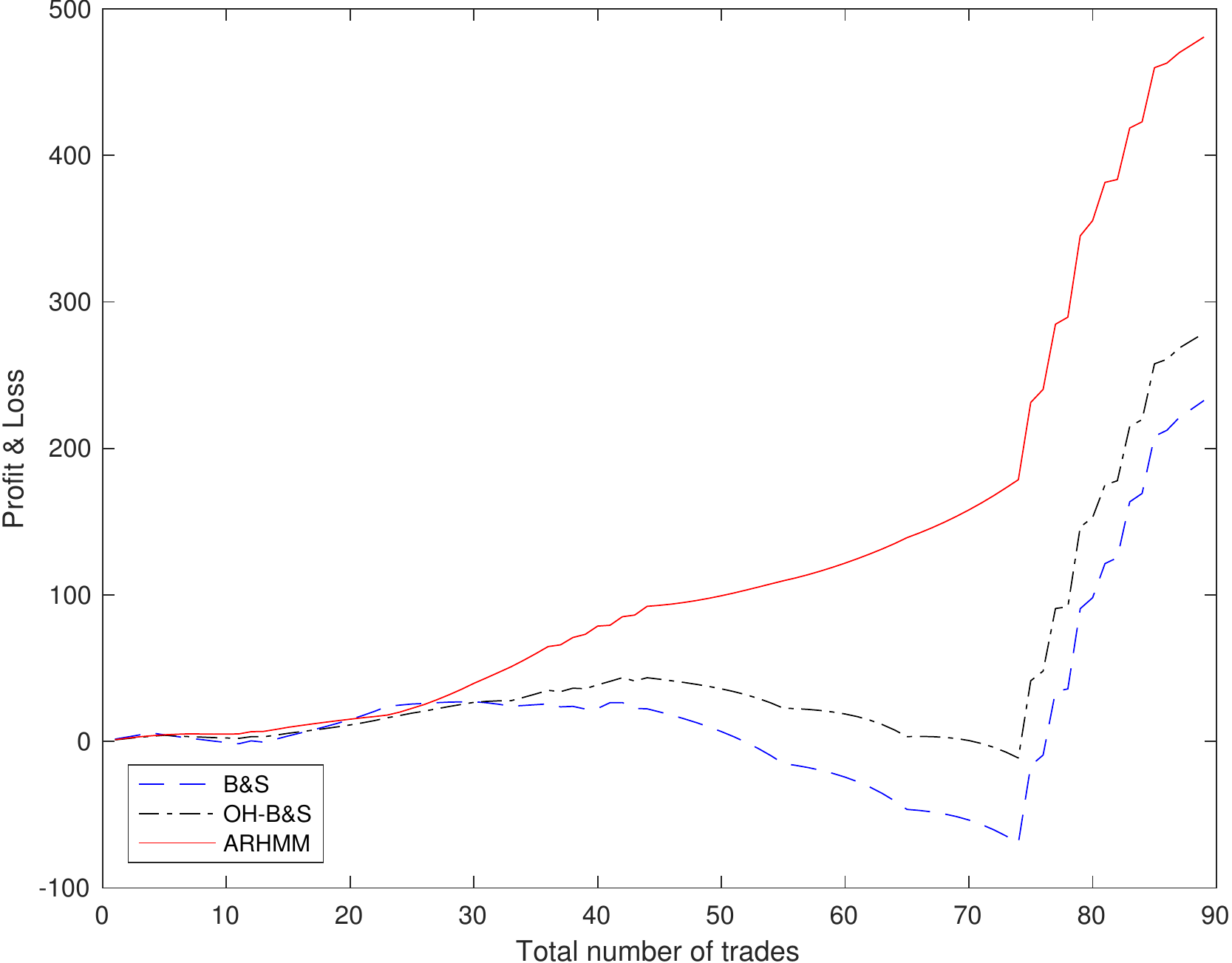}
(b) 	\includegraphics[width=2.24in]{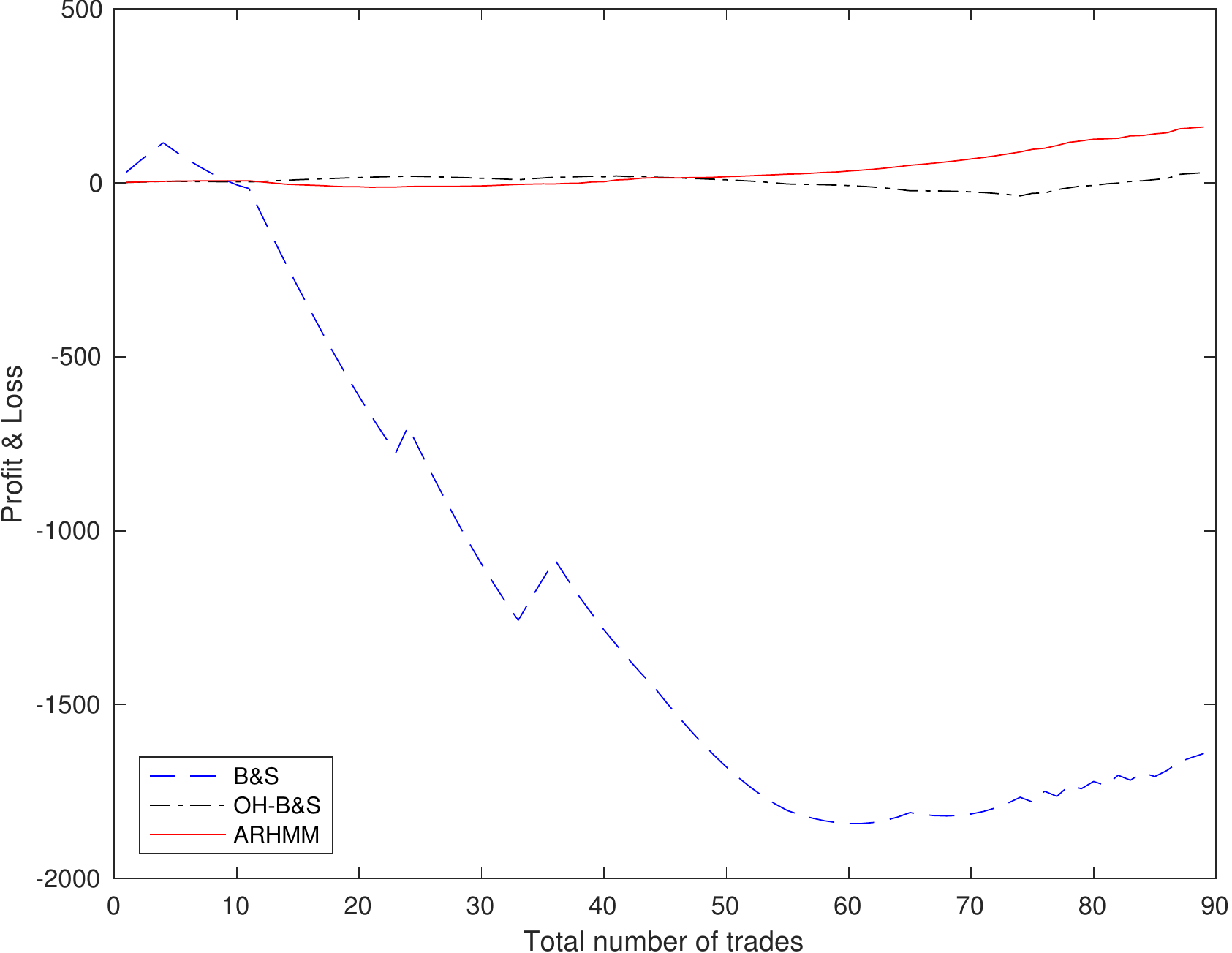}
			\end{center}
			\caption{Profit \& Loss of trading strategy for the 90 calls (a) and 90 puts (b) traded in the 2008-2009 Financial Crisis with 2000 days trailing estimation window.}\label{fig:Crash_t2000_pnl}
		\end{figure}

	Similar results are presented in Table \ref{tbl:Crash_t2000_he} and Figures \ref{fig:Crash_t2000_he_pdf} and \ref{fig:Crash_t2000_pnl}, although the trailing estimation window, previously set to 500 days, is now 2000 days. This estimation window includes  another financial crisis, the Dot-com Bubble. The same conclusion as the previous experience can be drawn.

			\begin{table}[ht]
		\caption{Hedging error statistics for the 90 calls and the 90 puts traded in the 2008-2009 Financial Crisis with 2000 days trailing estimation window.}\label{tbl:Crash_t2000_he}
			\begin{center}
{\footnotesize
				\begin{tabular}{ | l | c c c c | c c c c |}
					\hline
                     & \multicolumn{4}{c|}{Calls}                            & \multicolumn{4}{c|}{Puts}                                                  \\
					& B\&S-M & B\&S & OH-B\&S & ARHMM                       & B\&S-M & B\&S & OH-B\&S & ARHMM                                          \\ \hline
						RMSE         &3.87&4.26&3.15&\textbf{0.33}             & 39.95&40.87&3.15&\textbf{1.25}                                        \\
						Bias   &   	0	&-4.86&-4.69&-4.91                       &0 &-1.4&-1.23&-1.44\                                           \\
						VaR 1\%           & -7.64&-4.76&-4.27&\textbf{-1.43}      & -28.39&-28.68&-4.27&\textbf{-1.4}                                \\
						Median            & 2.9&3.26&1.93&\textbf{0.01}         & 30.78&27.74&1.93&\textbf{0.35}                                        \\
						VaR 99\%          &9.16&8.93&7.62&\textbf{0.33} 	     &72.9&74.76&7.62&\textbf{4.83}                                    \\ \hline
				\end{tabular}
}
			\end{center}
			\end{table}

\subsubsection{2013-2015 Bull markets}\label{ssec:Bull_exp}

	Our second and last analysis focuses on a part of the recent recovery spanning from September $23^{th}$ 2013 to August $7^{th}$ 2015. This period is quite the opposite of a financial crash. It is characterized by steady returns and low volatility. \\
	
\begin{figure}[ht!]
		\begin{center}
		(a)	\includegraphics[width=2.24in]{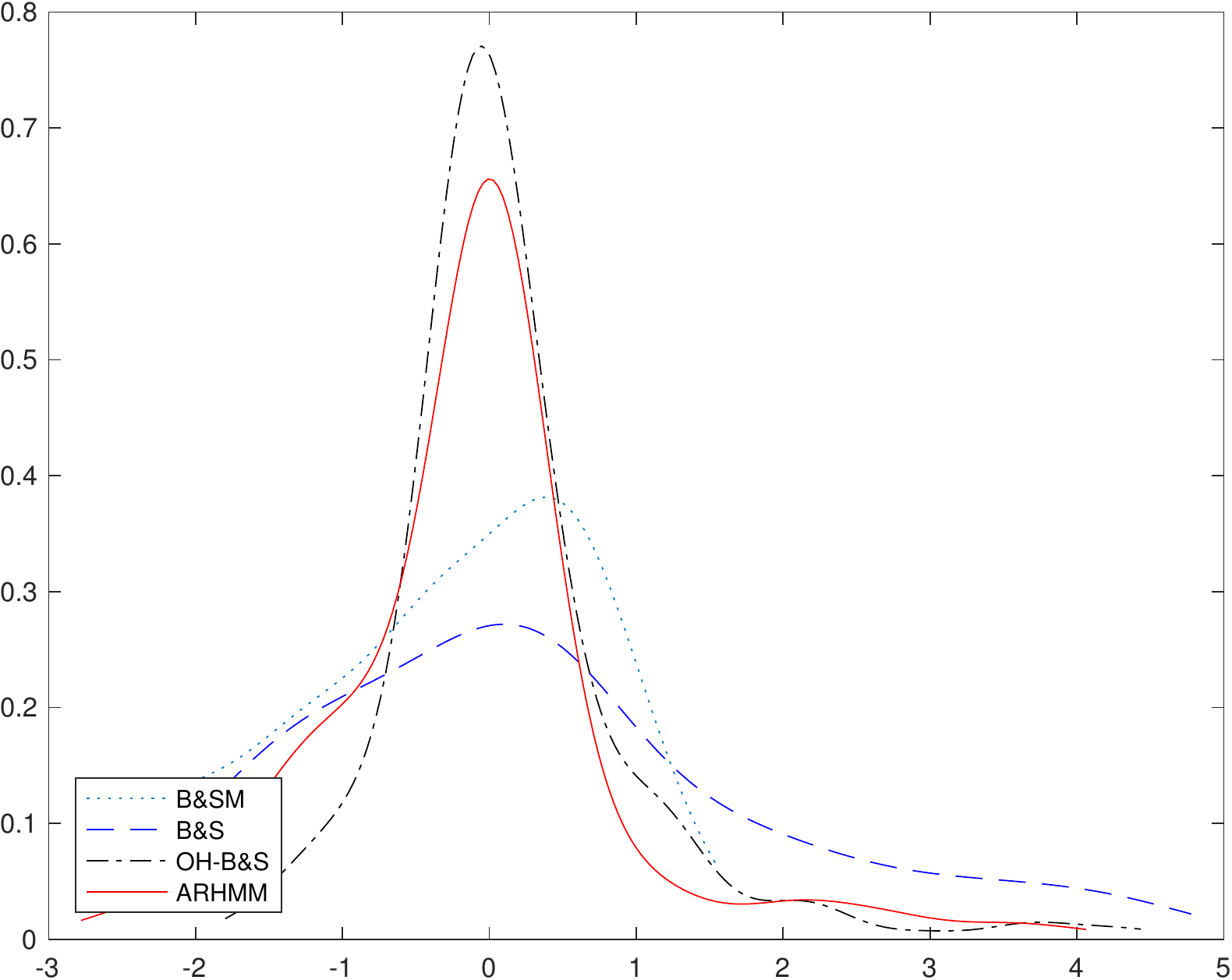}
(b) 	\includegraphics[width=2.24in]{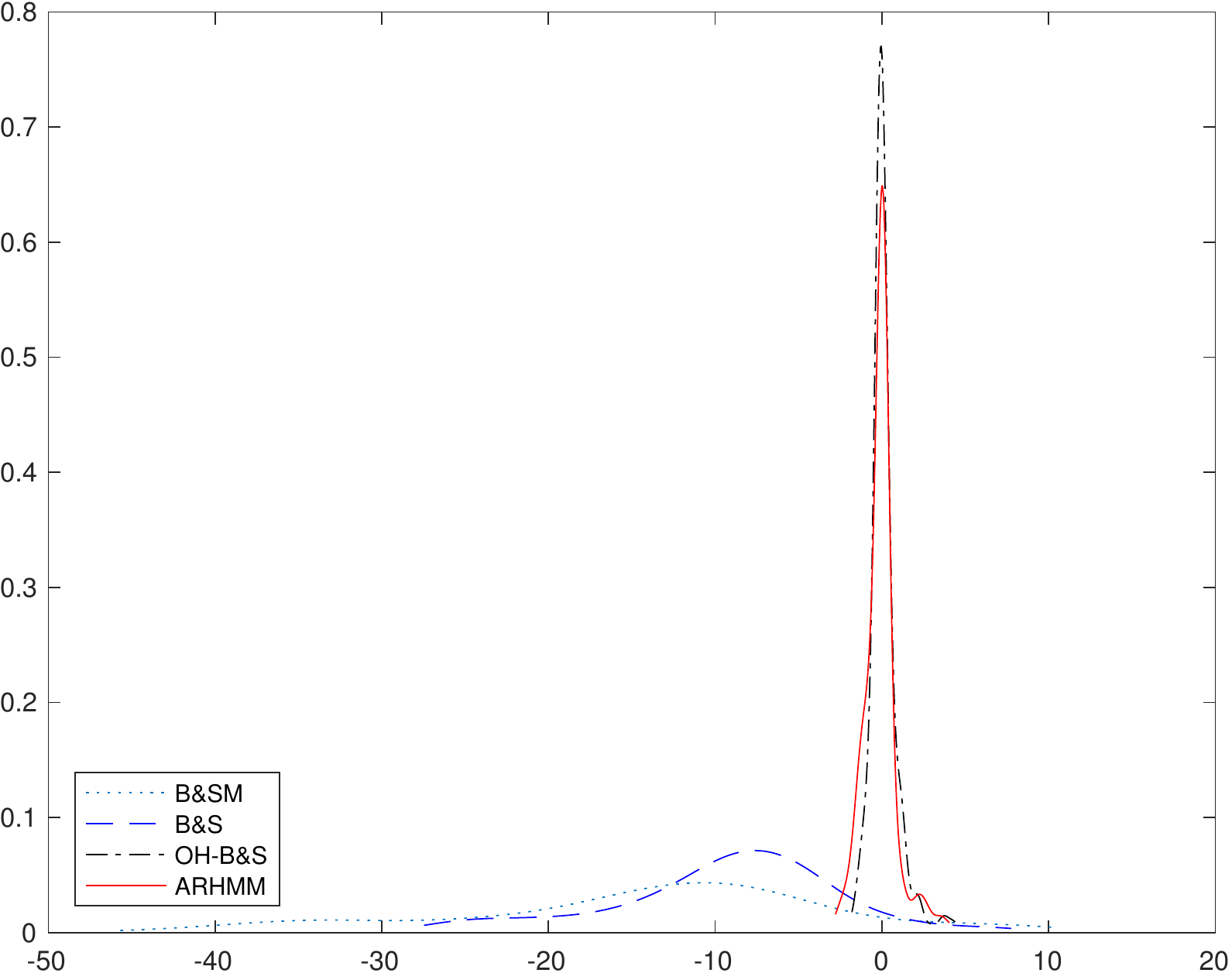}
			\end{center}
			\caption{Hedging error density approximation for the 239 calls (a) and 239 puts (b) traded in the 2013-2015 Bull markets with 500 days trailing estimation window.}\label{fig:Bull_t500_he_pdf}
		\end{figure}

\begin{figure}[ht!]
		\begin{center}
		(a)	\includegraphics[width=2.24in]{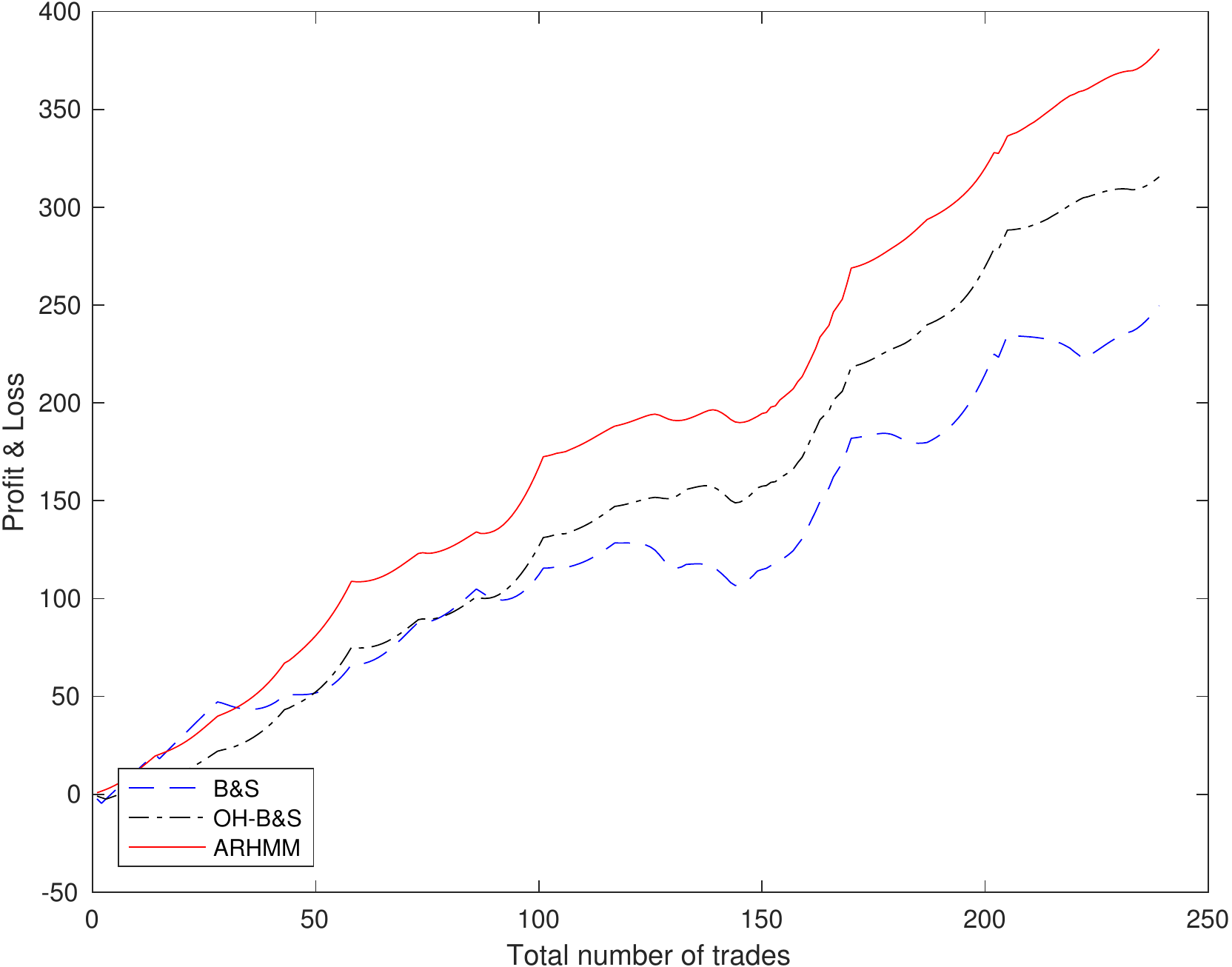}
(b) 	\includegraphics[width=2.24in]{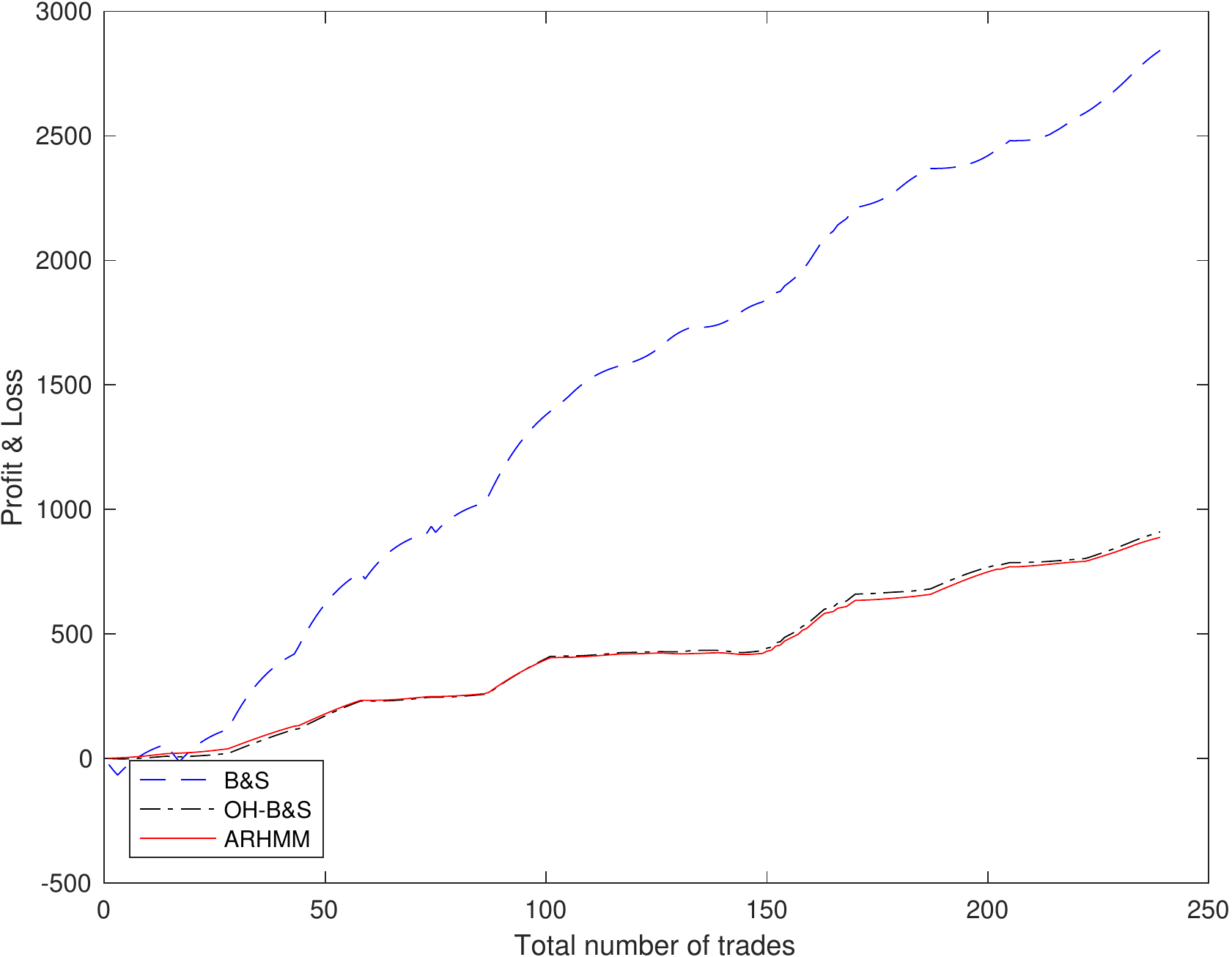}
			\end{center}
			\caption{Profit \& Loss of trading strategy for the 239 calls (a) and 239 puts (b) traded in the 2013-2015 Bull markets with 500 days trailing estimation window.}\label{fig:Bull_t500_pnl}
		\end{figure}

	Again, we start with the small trailing estimation window. We present the results for calls and puts in Table \ref{tbl:Bull_t500_he} and Figures \ref{fig:Bull_t500_he_pdf} and \ref{fig:Bull_t500_pnl}. Considering the hedging errors, OH-B\&S and ARHMM achieved the best and pretty similar statistics for both put and calls. Similarly to the previous experience in Section \ref{ssec:Crash_exp}, B\&S and B\&S-M replicated poorly the put options.

			\begin{table}[ht]
		\caption{Hedging error statistics for the 239 calls and the 239 puts traded in the 2013-2015 Bull markets with 500 days trailing estimation window.}\label{tbl:Bull_t500_he}
			\begin{center}
{\footnotesize
				\begin{tabular}{ | l | c c c c | c c c c |}
					\hline
                     & \multicolumn{4}{c|}{Calls}                            & \multicolumn{4}{c|}{Puts}                                                  \\
					& B\&S-M & B\&S & OH-B\&S & ARHMM                       & B\&S-M & B\&S & OH-B\&S & ARHMM                                          \\ \hline
						RMSE         &1.09&1.64&\textbf{0.84}&0.99         & 18.12&11.6&\textbf{0.84}&0.99                                        \\
						Bias   & 0  &0.12&0.22&0.59                       &0 &-4.08&-3.98&-3.62\                                           \\
						VaR 1\%           & -2.63&-2.53&\textbf{-1.45}&-2.47     &-41.63&-26.59&\textbf{-1.45}&-2.46                                \\
						Median            & -0.2&0.14&-0.02&\textbf{-0.01}       & -12.12&-8.6&-0.02&\textbf{-0.01}                                        \\
						VaR 99\%          & \textbf{1.39}&4.42&3.7&3.54	     &9.42&6.78&3.7&\textbf{3.53}                                    \\ \hline
				\end{tabular}
}
			\end{center}
			\end{table}

\begin{figure}[ht!]
		\begin{center}
		(a)	\includegraphics[width=2.24in]{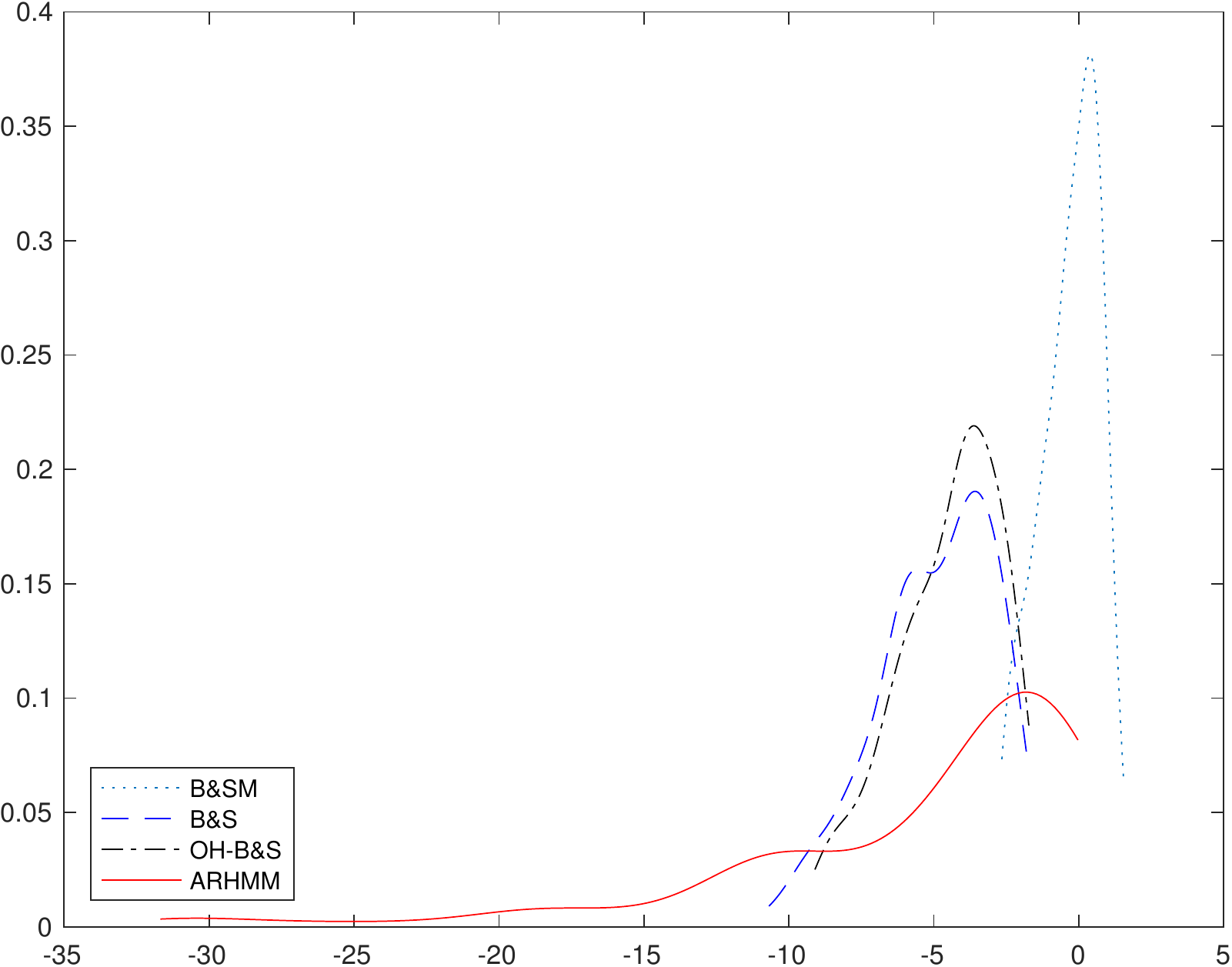}
(b) 	\includegraphics[width=2.24in]{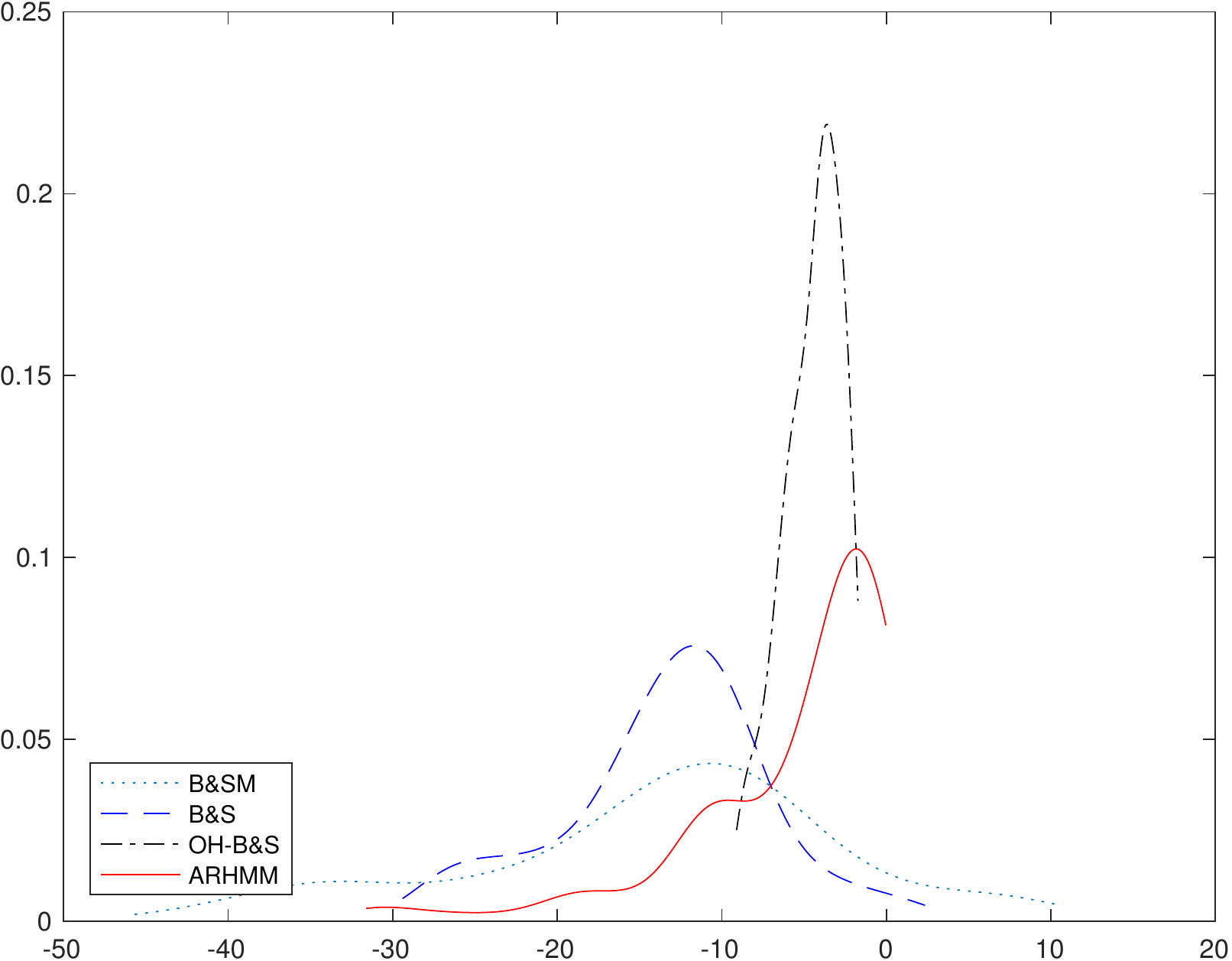}
			\end{center}
			\caption{Hedging error density approximation for the 239 calls (a) and 239 puts (b) traded in the 2013-2015 Bull markets with 2000 days trailing estimation window.}\label{fig:Bull_t2000_he_pdf}
		\end{figure}

\begin{figure}[ht!]
		\begin{center}
		(a)	\includegraphics[width=2.24in]{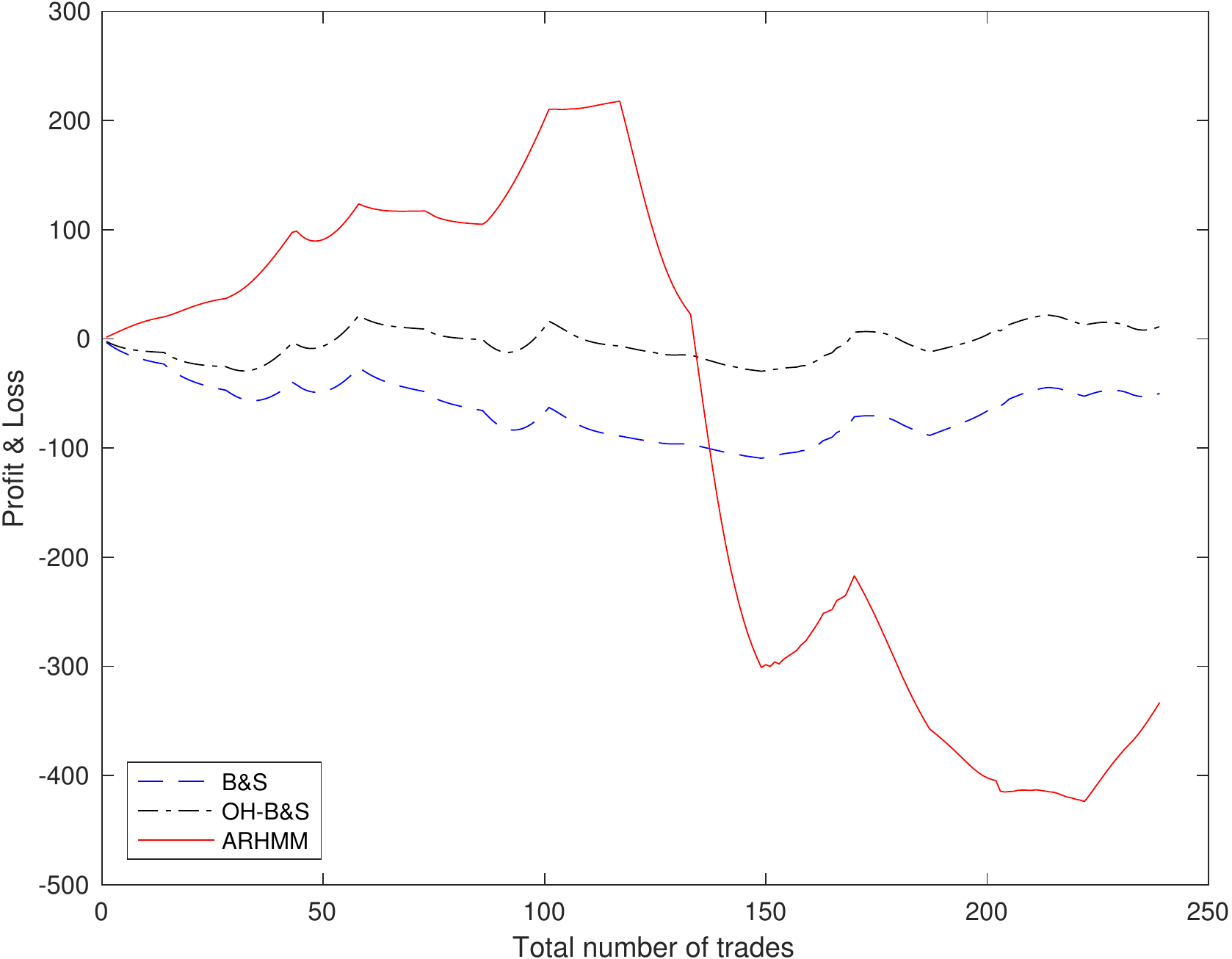}
(b) 	\includegraphics[width=2.24in]{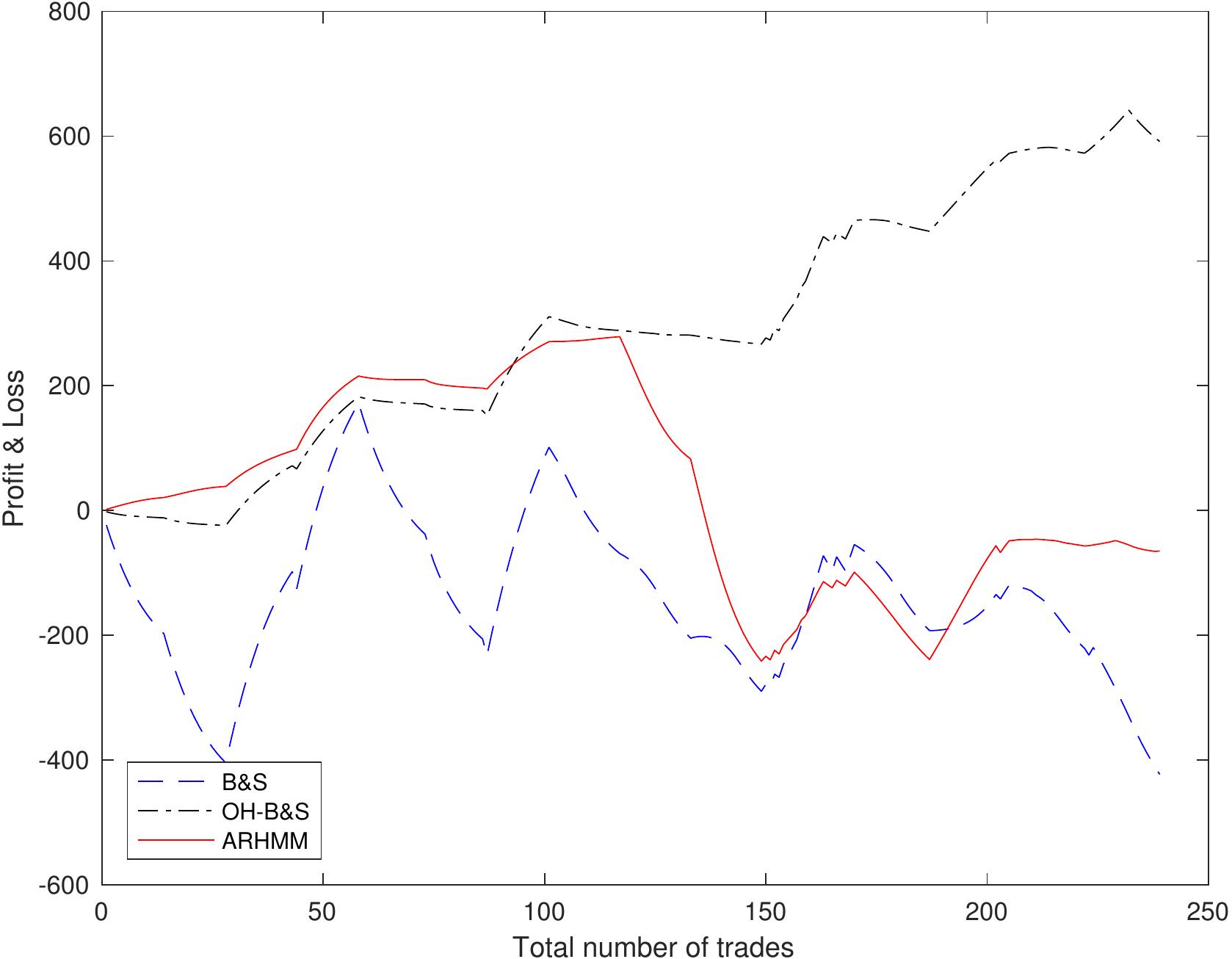}
			\end{center}
			\caption{Profit \& Loss of trading strategy for the 239 calls (a) and 239 puts (b) traded in the 2013-2015 Bull markets with 2000 days trailing estimation window.}\label{fig:Bull_t2000_pnl}
		\end{figure}

	Finally, the results for the longer estimation window case is presented in Table \ref{tbl:Bull_t2000_he} and Figures \ref{fig:Bull_t2000_he_pdf} and \ref{fig:Bull_t2000_pnl}. This is probably the worst environment for the ARHMM, as the estimation window includes a financial crisis (i.e. 2008-2009 Financial Crisis) and the out-of-sample returns are slow and steady. Because our trading strategy takes into account the actual hedging error,  according to \eqref{eq:phi}, the simpler models should perform better. In spite of that, ARHMM managed to perform better than B\&S and B\&-M for the hedging errors of the puts.

		The fact that pricing bias for the calls are strongly positive is noteworthy. In theory, the pricing bias should be negative, to account for the risk premium. In this case, it seems that the market was pretty confident about returns and volatility staying low. In insight, it was right.
		
			\begin{table}[ht!]
			\caption{Hedging error statistics for the 239 calls and the 239 puts traded in the 2013-2015 Bull markets with 2000 days trailing estimation window.}\label{tbl:Bull_t2000_he}
			\begin{center}
{\footnotesize
				\begin{tabular}{ | l | c c c c | c c c c |}
					\hline
                     & \multicolumn{4}{c|}{Calls}                            & \multicolumn{4}{c|}{Puts}                                                  \\
					& B\&S-M & B\&S & OH-B\&S & ARHMM                       & B\&S-M & B\&S & OH-B\&S & ARHMM                                          \\ \hline
						RMSE        &\textbf{1.09}&5.32&4.84&8.57        &18.12&14.75&\textbf{4.84}&8.6                                        \\
						Bias   & 0  &4.71&4.54&4.18                       &0 &0.52&0.34&-0.01 \                                           \\
						VaR 1\%          & \textbf{-2.63}&-10&-8.93&-31.27    &-41.63&-27.66&\textbf{-8.93}&-31.39                               \\
						Median           & \textbf{-0.2}&-4.67&-4.09&-2.82     & -12.12&-12.5&-4.09&\textbf{-2.81}                                         \\
						VaR 99\%         &1.39&-1.79&-1.76&\textbf{-0.04}		  &9.42&1.41&-1.76&\textbf{-0.04}                                   \\ \hline
				\end{tabular}
}
			\end{center}
			\end{table}

	Lastly, we aggregated the P\&L over all the experiences for B\&S, OH-B\&S and ARHMM in Table \ref{tab:final_pnl}. For a fair comparison, we normalized the number of traded options in each cases to 100. Remember that the option prices, strike prices and underlying path are also normalized at an initial S\&P 500 value of 100. Impressively, ARHMM accomplished a 106\% increase in P\&L compared to the second best, OH-B\&S, for the 2-year trailing window, and is only 9\% behind the first for the 8-year case, which is again OH-B\&S.

	\begin{table}[ht!]
	\caption{Total normalized P\&L }
	\label{tab:final_pnl}
\begin{center}
	\begin{tabular}{ | c | c c c |}
					\hline
			    Trailing window (years) & B\&S & OH-B\&S & ARHMM \\ \hline
				2 & -1286.78 & 685.32 & \textbf{1409.54} \\ \hline
				8 & -1761.77 & \textbf{594.56} & 546.38 \\ \hline
	\end{tabular}
\end{center}
	\end{table}

	Overall, by achieving the best Root Mean Square Error (RMSE) two times out of four for both the 2-year and 8-year window, and by being the most profitable strategy three times out of four for the 2-year window and two times out of four for the 8-year window, the ARHMM is the superior hedging protocol.

 However, the practitioners should keep in mind that if the ARHMM is estimated on a window including a financial crisis, they should expect higher hedging errors than the simpler models if returns stay slow and steady. From our results, we strongly suggest to  use a 2-year  trailing window as it consistently achieved an RMSE lower than 1, i.e., the ARHMM can accurately hedge options in a financial crisis without ever seeing one.

\section{Conclusion}

	In this paper, we propose an autoregressive hidden Markov model to fit financial data, and we show how to implement an optimal hedging strategy when the underlying asset returns follow an autoregressive regime-switching random walk.
	
	First, we present estimation and filtering procedures for the ARHMM. In order to determine the optimal number of regimes, we propose a novel goodness-of-fit test for univariate and multivariate ARHMM based on the work of \cite{Bai:2003}, \cite{Genest/Remillard:2008} and \cite{Remillard/Hocquard/Lamarre/Papageorgiou:2017}.
	
	To illustrate the proposed strategy, we model three daily return series of the S\&P 500. Using likelihood test, we show that the ARHMM is a much better fit than the classical HMM, particularly because it has the capacity to model mean-reversion.

	Moreover, we present the implementation of the discrete-time optimal hedging algorithm  minimizing the mean-squared hedging error. Because it further performs pricing, we implemented a trading strategy consisting of selling overpriced and buying underpriced options and hedging the position till maturity. Out of eight cases and compared to three other hedging protocols, our strategy achieves the best root-mean-squared hedging error four times and is the most profitable strategy five times. Furthermore, it realized the best total P\&L.
	
	Because of its ability to model regime switches as well as mean-reversion, it would be interesting to see this model applied to multivariate time series. The hedging algorithm can also be applied to multivariate or American options.

\bibliography{all2005}
\bibliographystyle{apalike}

\appendix

\section{Extension of Baum-Welch Algorithm}\label{app:BWA}

For $i\in \{1,\ldots,l\}$ and $1\le t\le n$, define
$$
\lambda_t(i)=  P(\tau_t=i|Y_1,\ldots, Y_n).
$$
Also, for $i,j\in \{1,\ldots,l\}$ and $1\le t\le n-1$, define
$$
\Lambda_t(i,j)=  P(\tau_t=i,\tau_{t+1}=j|Y_1,\ldots, Y_n),
$$
and let $\bar\eta_t(i)$ be the conditional density of
$(Y_{t+1},\ldots, Y_n)$, given $Y_t$ and $\tau_t=i$. Further set
$\bar\eta_n\equiv 1$. Note that $\Lambda_n(i,j) =
\lambda_n(i)Q_{ij}$,  for any $i,j\in\{1,\ldots,l\}$.

The proof of the following proposition is given in Appendix
\ref{app:algo}.

\begin{prop}\label{prop:filt}
	For all $i,j\in\{1,\ldots,l\}$,
	\begin{eqnarray}
	\eta_{t+1}(i) &=&  \frac{f_i(Y_{t+1}|Y_t)\sum_{\beta=1}^l
		\eta_{t}(\beta)Q_{\beta i} }{\sum_{\alpha=1}^l \sum_{\beta=1}^l
		f_\alpha(Y_{t+1}|Y_t) \eta_{t}(\beta)Q_{\beta \alpha} }, \quad t=0,\ldots, n-1, \label{eq:eta}\\
	\bar \eta_t(i) &=&  \sum_{\beta=1}^l   Q_{i\beta} \bar
	\eta_{t+1}(\beta)f_\beta(Y_{t+1}|Y_t), \quad t=0,\ldots, n-1, \label{eq:etabar}\\
	\lambda_t(i) &= & \frac{\eta_t(i)\bar \eta_t(i)}{\sum_{\alpha=1}^l \eta_t(\alpha)\bar \eta_t(\alpha)}, \quad t=0,\ldots, n,\label{eq:lambda}\\
	\Lambda_t(i,j) &=& \frac{\eta_t(i)Q_{ij}\bar
		\eta_{t+1}(j)f_j(Y_{t+1}|Y_t)}{\sum_{\alpha=1}^l
		\eta_t(\alpha)  \bar \eta_{t}(\alpha)}, \quad  t=0,\ldots, n-1.\label{eq:Lambda}
	\end{eqnarray}
	In particular,
	\begin{equation}\label{eq:sumlambda}
	\sum_{\beta=1}^l \Lambda_t(i,\beta)=\lambda_t(i), \quad t=0,\ldots,
	n.
	\end{equation}
\end{prop}

\section{Proof of Proposition \ref{prop:filt}}\label{app:algo}

Let $i\in \{1,\ldots,l\}$ and $t \in \{1,\ldots,n\}$ be given. Set
$X=(Y_1,\ldots,Y_{t-1})$, $\zeta=Y_t$ and $W=(Y_{t+1},\ldots,Y_n)$. Let
$f$ denotes the density of $X$. If follows from the definition of
conditional expectations that for any bounded measurable functions
$F$, $G$, and $H$,
\begin{eqnarray*}
	E\{F(X)G(\zeta)\eta_t(i)\}&=& E\{F(X)G(\zeta)\I(\tau_t=i)\}
	\\
	&=& \sum_{\beta=1}^l Q_{\beta i} \int F(x)G(z)\eta_{t-1}(\beta)
	f(x)f_i(z|x)dz dx.
\end{eqnarray*}
As a by-product, one gets
$$
E\{F(X)G(\zeta)\eta_t(i)\}=  \sum_{\alpha=1}^l \sum_{\beta=1}^l Q_{\beta
	\alpha} \int F(x)G(z)\eta_{t}(i)
\eta_{t-1}(\alpha)f(x)f_\alpha(z|x)dz dx.
$$
Since the last equation holds for any $F$ and $G$, it follows that
\eqref{eq:eta} holds true.

Next,
\begin{eqnarray*}
	E\{G(\zeta)H(W)|Y_t=y,\tau_{t-1}=i\} &= &\int
	G(z)H(w)\bar\eta_{t}(i)dzdw\\
	&=&\sum_{\beta=1}^l Q_{i\beta}\int
	f_\beta(z|y)\bar\eta_{t+1}(\beta)G(z)H(w)dzdw,
\end{eqnarray*}
proving that \eqref{eq:etabar} holds.

Next, let $\tilde f(x,z)$ be the density of $(X,\zeta) =
(Y_1,\ldots,Y_t)$. Then
\begin{eqnarray*}
	E\{F(X)G(\zeta)H(W)\lambda_t(i)\}&=& E\{F(X)G(\zeta)H(W)\I(\tau_t=i)\}
	\\
	&=& \int F(x)G(z)H(w)\eta_{t-1}(i)Q_{ij}\bar\eta_t(j) \tilde
	f(x,z)dw dz dx.
\end{eqnarray*}
As a by-product,
$$
E\{F(X)G(\zeta)H(W)\lambda_t(i)\}=\sum_{\alpha=1}^l
\int F(x)G(z)H(w)\lambda_t(i) \eta_{t}(\alpha)\bar\eta_t(\alpha)
\tilde f(x,z)dw dz dx,
$$
proving \eqref{eq:lambda}. It is easy to extend the last argument to
the case $t=0$.

Finally,
\begin{eqnarray*}
	E\{F(X)G(\zeta)H(W)\Lambda_{t-1}(i,j)\}&=&
	E\{F(X)G(\zeta)H(W)\I(\tau_{t-1}=i,\tau_t=j)\}
	\\
	&=& Q_{ij}\int F(x)G(z)H(w)\eta_{t}(i)\bar\eta_t(j)  f(x)f_i(z|x) dw
	dz dx.
\end{eqnarray*}
As a by-product, the latter can also be written as
$$
\sum_{\alpha=1}^l \sum_{\beta=1}^l
\int F(x)G(z)H(w)\Lambda_{t-1}(i,j) Q_{\alpha\beta} \eta_{t}(\alpha)\bar\eta_t(\beta)
f(x)f_\beta(z|x)dw dz dx,
$$
proving \eqref{eq:Lambda}. It is easy to extend the last argument to
the case $t=0$. This completes the proof.
\qed

\section{Estimation of regime-switching models}\label{app:em}

To describe the EM algorithm for the estimation, suppose that at
step $k\ge 0$, one has the parameters $Q$, $\mu_i$, $\Phi_i$, $A_i$,
$ i\in\{1,\ldots,l\}$.

Let $w_t(i) = \lambda_t(i)\Big{/}\sum_{k=1}^n  \lambda_k(i)$, and
set $\bar y_i = \sum_{t=1}^n w_t(i) y_{t}$ and $\underline{y}_i =
\sum_{t=1}^n w_t(i) y_{t-1}$, $i\in \{1,\ldots,l\}$, where
$\lambda_t$ and $\Lambda_t$ are given in Proposition
\ref{prop:filt}.

Then, at step $k+1$,  for $ i,j\in\{1,\ldots,l\}$, one has
\begin{eqnarray}
\label{eq:newQ} Q_{ij}^{(k+1)} &=& \frac{\sum_{t=1}^n
\Lambda_{t-1}(i,j) }{\sum_{\beta=1}^l \sum_{t=1}^n
\Lambda_{t-1}(i,\beta)} = \frac{\sum_{t=1}^n \Lambda_{t-1}(i,j) }{
\sum_{t=1}^n \lambda_{t-1}(i)},\\
\label{eq:newmu} \mu_i^{(k+1)} &=&
\left(I-\Phi_i^{(k+1)}\right)^{-1}\left(\bar
y_i-\Phi_i^{(k+1)}\underline{y}_i \right),\\
\Phi_i^{(k+1)} &=& \left\{\sum_{t=1}^n
w_t(i)\left(y_{t-1}-\underline{y}_i\right)\left(y_{t-1}-\underline{y}_i\right)^\top\right\}^{-1}\label{eq:newphi}\\
&& \qquad \qquad \times  \left\{\sum_{t=1}^n w_t(i)\left(y_{t}-\bar
y_i\right)\left(y_{t-1}-\underline{y}_i\right)^\top
\right\},\nonumber
\\
 A_i^{(k+1)} &=&  \sum_{t=1}^n w_t(i)e_{ti} e_{ti}^\top,\label{eq:newA}
\end{eqnarray}
where $ e_{ti} = y_{t}-\bar
y_i-\Phi_i^{(k+1)}\left(y_{t-1}-\underline{y}_i\right)$, $
t=1,\ldots,n$.

The proof is given in the next section.

\subsection{Proof of the EM algorithm for the estimation}

The EM algorithm for estimating parameters consists of two steps, expectation and maximization:

\noindent \textbf{E-Step:} Compute the conditional probabilities.
$$
\lambda_t(i) = P(\tau_t=i|Y_1,\ldots, Y_n) \quad \mbox{and}  \quad
\Lambda_t(i,j)=  P(\tau_t=i,\tau_{t+1}=j|Y_1,\ldots, Y_n),
$$
for all $1\le t\le n$ and $i,j\in\{1,\ldots,l\}$.\\

\noindent \textbf{M-Step:}

Let $\cQ$ be the set of $l\times l$ transition matrices with
positive entries. Suppose that $Q\in \cQ$,  and  $\theta \in \Theta
$. Then the log-likelihood is
$$
L(Y_1,\ldots. Y_n, \tau_1,\ldots,\tau_n,Q,\theta) = \sum_{t=1}^n
\log Q_{\tau_{t-1},\tau_t} +\sum_{t=1}^n \log
f_{\tau_t}(y_t|y_{t-1},\theta).
$$
It then follows that
\begin{eqnarray*}
	\cL(\tilde Q,\tilde \theta; Q,\theta) &=&
	E_{Q,\theta}\left\{L\left(Y_1,\ldots, Y_n, \tau_1,\ldots,\tau_n,
	\tilde
	Q,\tilde \theta\right)|Y_1=y_1,\ldots, Y_n=y_n\right\}\\
	&=& \sum_{t=1}^n \sum_{i=1}^l\sum_{j=1}^l \Lambda_{t-1}(i,j)\log
	\tilde Q_{ij} +\sum_{t=1}^n  \sum_{i=1}^l \lambda_t(i) \log
	f_{i}(y_t|y_{t-1},\tilde\theta).
\end{eqnarray*}

If $Q,\theta$ are the parameters at step $k$, then the parameters
$Q^{(k+1)},\theta^{(k+1)}$ at step $k+1$ are
$$
\left(Q^{(k+1)},\theta^{(k+1)}\right) = \arg\max_{\tilde Q \in
	\cQ,\tilde \theta \in \Theta}\cL(\tilde Q,\tilde \theta; Q,\theta).
$$

It is easy to check that
$$
Q^{(k+1)} = \arg\max_{\tilde Q \in \cQ} \sum_{t=1}^n
\sum_{i=1}^l\sum_{j=1}^l \Lambda_{t-1}(i,j)\log \tilde Q_{ij}
$$
satisfies
$$
Q_{ij}^{(k+1)} = \frac{\sum_{t=1}^n \Lambda_{t-1}(i,j)
}{\sum_{\beta=1}^l \sum_{t=1}^n \Lambda_{t-1}(i,\beta)} =
\frac{\sum_{t=1}^n \Lambda_{t-1}(i,j) }{ \sum_{t=1}^n
	\lambda_{t-1}(i)},\quad i,j\in \{1,\ldots,l\},
$$
proving \eqref{eq:newQ}.
Also,
$$
\theta^{(k+1)} = \arg\max_{\tilde \theta \in \Theta} \sum_{t=1}^n
\sum_{i=1}^l \lambda_t(i) \log f_{i}(y_t|y_{t-1},\tilde\theta).
$$

	\subsubsection{Estimation for Gaussian AR(1) regime-switching models (M-Step)}\label{app:emgaussian}
	
	For the estimation procedure, we assume the densities $f_1,\ldots,
	f_l$ are given by \eqref{eq:VAR1}, so $\theta = (\mu_1,\ldots,\mu_l,
	\Phi_1, \ldots,\Phi_l, A_1,\ldots, A_l) \in \Theta = \dR^{p \otimes
		l} \times \cB_d^{\otimes l} \times S_d^{\otimes l}$.
	
	In this case, the function $L(\tilde\theta)= -\sum_{t=1}^n
	\sum_{i=1}^l \lambda_t(i) \log f_{i}(y_t|y_{t-1},\tilde\theta)$ to
	minimize is given by
	\begin{eqnarray*}
		L(\tilde\theta) &=& \frac{1}{2}\sum_{t=1}^n \sum_{i=1}^l
		\lambda_t(i)\left\{y_{t}-\tilde\mu_i-\tilde\Phi_i(y_{t-1}-\tilde\mu_i)\right\}^\top
		{\tilde
			A}_i^{-1}\left\{y_t-\tilde\mu_i-\tilde\Phi_i(y_{t-1}-\tilde\mu_i)\right\}\\
		&& \qquad +\frac{nd}{2}\log{2\pi} + \frac{1}{2}\sum_{t=1}^n
		\sum_{i=1}^l \lambda_t(i) \log{ |\tilde A_i|}.
	\end{eqnarray*}
	
	Let $w_t(i) = \lambda_t(i)\Big{/}\sum_{k=1}^n  \lambda_k(i)$. It
	then follows that for any $i\in\{1,\ldots,l\}$,
	\begin{equation}\label{eq:gradmu}
	\sum_{t=1}^n
	w_t(i)\left\{y_{t}-\mu_i^{(k+1)}-\Phi_i^{(k+1)}\left(y_{t-1}-\mu_i^{(k+1)}\right)\right\}=0,
	\end{equation}
	\begin{equation}\label{eq:gradphi}
	\sum_{t=1}^n
	w_t(i)\left\{y_{t}-\mu_i^{(k+1)}-\Phi_i^{(k+1)}\left(y_{t-1}-\mu_i^{(k+1)}\right)\right\}\left(y_{t-1}-\mu_i^{(k+1)}\right)^\top=0,
	\end{equation}
	and
	\begin{equation}\label{eq:Aopt}
	A_i^{(k+1)} =  \sum_{t=1}^n w_t(i)z_{ti} z_{ti}^\top,
	\end{equation}
	where  $z_{ti} =
	y_{t}-\mu_i^{(k+1)}-\Phi_i^{(k+1)}\left(y_{t-1}-\mu_i^{(k+1)}\right)$,
	$t=1,\ldots,n$, since
	$$
	\sum_{t=1}^n \sum_{i=1}^l \lambda_t(i)z_{ti}^\top {\tilde
		A}_i^{-1}z_{ti} = n \nu_i^{(k+1)} {Tr} \left({\tilde A}_i^{-1}
	A_i^{(k+1)}\right),
	$$
	where $\bar\nu_i^{(k+1)} = \sum_{t=1}^n  \lambda_t(i)/n$, and for
	any non singular $d\times d$ matrix $B$,
	$$
	Tr(B)-\log{|B|}\ge d.
	$$
	The latter is true because $f(x) = x-\log(x) \ge f(1)=1$ for any
	$x>0$.

	Next, set $\bar y_i = \sum_{t=1}^n w_t(i) y_{t}$ and
	$\underline{y}_i = \sum_{t=1}^n w_t(i) y_{t-1}$. Then it follows
	from \eqref{eq:gradmu}  that
	$$
	\mu_i^{(k+1)} = \left(I-\Phi_i^{(k+1)}\right)^{-1}\left(\bar
	y_i-\Phi_i^{(k+1)}\underline{y}_i \right),\quad i\in\{1,\ldots,l\},
	$$
	proving \eqref{eq:newmu}.
	Now, for $i\in\{1,\ldots,l\}$,
	\begin{eqnarray*}
		\sum_{t=1}^n
		w_t(i)\left(y_{t}-\mu_i^{(k+1)}\right)\left(y_{t-1}-\mu_i^{(k+1)}\right)^\top
		&=& \sum_{t=1}^n w_t(i)\left(y_{t}-\bar
		y_i\right)\left(y_{t-1}-\underline{y}_i\right)^\top \\
		&& \qquad  + \left(\mu_i^{(k+1)}-\bar
		y_i\right)\left(\mu_i^{(k+1)}-\underline{y}_i\right)^\top\\
		&=& \sum_{t=1}^n w_t(i)\left(y_{t}-\bar
		y_i\right)\left(y_{t-1}-\underline{y}_i\right)^\top \\
		&& \qquad  +
		\Phi_i^{(k+1)}\left(\mu_i^{(k+1)}-\underline{y}_i\right)\left(\mu_i^{(k+1)}-\underline{y}_i\right)^\top,
	\end{eqnarray*}
	using \eqref{eq:newmu}, and
	\begin{eqnarray*}
		\sum_{t=1}^n
		w_t(i)\left(y_{t-1}-\mu_i^{(k+1)}\right)\left(y_{t-1}-\mu_i^{(k+1)}\right)^\top
		&=& \sum_{t=1}^n w_t(i)\left(y_{t-1}-\underline{y}_i\right)\left(y_{t-1}-\underline{y}_i\right)^\top \\
		&& \qquad  +
		\left(\mu_i^{(k+1)}-\underline{y}_i\right)\left(\mu_i^{(k+1)}-\underline{y}_i\right)^\top
		.
	\end{eqnarray*}
	
	As a result, for $i\in \{1,\ldots,l\}$,
	$$
	\Phi_i^{(k+1)} = \left\{\sum_{t=1}^n
	w_t(i)\left(y_{t-1}-\underline{y}_i\right)\left(y_{t-1}-\underline{y}_i\right)^\top\right\}^{-1}
	\left\{\sum_{t=1}^n w_t(i)\left(y_{t}-\bar
	y_i\right)\left(y_{t-1}-\underline{y}_i\right)^\top \right\},
	$$
	proving \eqref{eq:newphi}
	
	It also follows from \eqref{eq:newmu} that \eqref{eq:Aopt} can be
	written as
	\begin{equation}\label{eq:Aoptnew} A_i^{(k+1)} =  \sum_{t=1}^n
	w_t(i)e_{ti} e_{ti}^\top,\quad i\in \{1,\ldots,l\},
	\end{equation}
	where
	$$
	e_{ti} = y_{t}-\bar
	y_i-\Phi_i^{(k+1)}\left(y_{t-1}-\underline{y}_i\right) ,\quad i\in
	\{1,\ldots,l\}, t=1,\ldots,n.
	$$


\section{Goodness-of-fit Test for Autoregressive Hidden Markov Model} \label{app:gof}

	In this Appendix, we state the goodness-of-fit test, which can be performed to asses the suitability of a Gaussian AR(1) regime-switching models as well as to select the optimal number of regimes, $l^*$. The proposed test, based on the work of \cite{Diebold/Gunther/Tay:1998}, \cite{Genest/Remillard:2008} and \cite{Remillard:2011}, uses the Rosenblatt's transform. For conciseness, we detail the implementation for two dimensional Gaussian AR(1) regime-switching models, but the approach can be easily generalized.
	
	\subsection{Conditional distribution functions and the Rosenblatt's transform.}
	
		Let $ i \in \{1,\ldots,l\}$ be fixed an $R_i$ be a random vector with density $f_i$. For any $ q \in \{1,\ldots,d\}$, denote by $f_{i,1:q}$ the density of $(R_i^{(1)},\ldots,R_i^{(q)})$, and by $f_{i,q}$ the density of $R_i^{(q)}$ given $(R_i^{(1)},\ldots,R_i^{(q-1)})$. Further denote by $F_{i,q}$ the distribution function associated with density $f_{i,q}$. By convention, $f_{i,1}$ denotes the $unconditional$ density of $R_i^{(1)}$. Then, the Rosenblatt's transform
		
		\begin{equation*}
			x \mapsto T_i(x) = \big(F_{i,1}(x^{(1)}), F_{i,2}(x^{(1)},x^{(2)}),\ldots,F_{i,d}(x^{(1)},\ldots,x^{(d)})\big)^\intercal
		\end{equation*}
	
		is such that $T_i(R_i)$ is uniformly distributed in $[0,1]^d$. \\
		For example, if $f_i$ is the density of a bivariate Gaussian distribution with mean $u_i$ and covariance matrix
		
		\begin{equation*}
			\Sigma_i = \left(\begin{array}{cc} v_i^{(1)} & \rho_i \sqrt{v_i^{(1)}v_i^{(2)}} \\
			 \rho_i \sqrt{v_i^{(1)}v_i^{(2)}} & v_i^{(2)}  \end{array}\right),
		\end{equation*}
		
		$f_{i,2}$ is the density of a Gaussian distribution with mean $\mu_i^{(2)}+ \beta_i(y_i^{(1)}-\mu_i^{(1)})$ and variance $v_i^{(2)}(1-\rho_i^2)$, with $\beta_i = \rho_i \sqrt{v_i^{(2)} / v_i^{(1)}}$. These results can easily be extended to the Gaussian AR(1) distribution. \\
		However, for regime-switching random walks models, past returns must also be included in the conditioning information set. For any $x^{(1)},\ldots,x^{(d)} \in \mathbb{R}$, the ($d$-dimensional) Rosenblatt's transform $\Psi_t$ corresponding to the density (\ref{eq:cond_dens}) conditional on $x_1,\ldots,x_{t-1} \in \mathbb{R}^d$ is given by
		\begin{equation*}
			\Psi_t^{(1)}(x_t^{(1)}) = \Psi_t^{(1)}(x_1,\ldots,x_{t-1},x_t^{(1)}) = \sum_{i=1}^{l}W_{t-1}(i)F_{i,1}(x_t^{(1)})
		\end{equation*}
		and
		\begin{equation*}
			\begin{aligned}
			\Psi_t^{(q)}(x_1^{(1)},\ldots,x_t^{(q)}) = \Psi_t^{(q)}(x_1,\ldots,x_{t-1},x_t^{(1)},\ldots,x_t^{(q)}) \\
											= \dfrac{
															\sum{i=1}^{l} W_{t-1}(i)f_{i,1:q-1}(x_t^{(1)},\ldots,x_t^{(q-1)})F_{i,q}(x_t^{(q)})
														}{
															\sum_{i=1}^{l} W_{t-1}(i) f_{i,1:q-1} (x_t^{(1)},\ldots,x_t^{(q-1)})
														}
			\end{aligned}									
		\end{equation*}
		for $q \in \{2,\ldots,d\}$. \\
		Suppose $R_1,\ldots,R_n$ is a size n sample of $d$-dimensional vectors drawn from a joint (continuous) distribution $P$. Also, let $\mathcal{P}$ be the parametric family of Gaussian AR(1) regime-switching models with $l$ regimes. Formally, the hypothesis to be tested is
		\begin{equation*}
			\mathcal{H}_0: P \in \mathcal{P} = \{ P_{\theta}; \theta \in \Theta  \} \quad  vs \quad \mathcal{H}_1:P \notin \mathcal{P}
		\end{equation*}
		Under the null, it follows that $\big( U_1 = \Psi_1(R_1,\theta), U_2 = \Psi_2(R_1,R_2,\theta),\ldots,U_n = \Psi(R_1,\ldots,R_n,\theta) \big) $ are independent and uniformly distributed over $[0,1]^d$, where $\Psi_1(\cdot,\theta),\ldots,\Psi_n(\cdot,\theta)$ are the Rosenblatt's transforms conditional on the set of parameters $\theta \in \Theta$. \\
		Since $\theta$ is unknown, it must be estimated by some $\theta_n$. Then, the $pseudo$-$observations$, $\big( \hat{U}_1 = \Psi_1(R_1,\theta_n),\ldots,\hat{U}_n = \Psi_n(R_1,\ldots,R_n,\theta_n)\big)$ are $approximately$ uniformly distributed over $[0,1]^d$ and $approximately$ independent. We next propose a test statistic based on these pseudo observations.
		
	\subsection{Test statistic}
	
		The test statistic builds from the following empirical process:
		\begin{equation*}
			D_n(u) = \dfrac{1}{n} \sum_{t=1}^{n} \prod_{q=1}^{d} \mathbb{I} \big(\hat{U}_t^{(q)} \leq u^{(q)} \big), \quad
			u \equiv (u^{(1)},\ldots,u^{(d)}) \in [0,1]^d.
		\end{equation*}
		To test $\mathcal{H}_0$ against $\mathcal{H}_1$ we propose a Cram\'er-von Mises type statistic:
		\begin{align*}
				S_n & \equiv B_n(\hat{U}_1,\ldots,\hat{U}_n) = n \int_{[0,1]^d} \Bigg\{ D_n(u) - \prod_{q=1}^{d} u^{(q)} \Bigg\}^2 du \\
				& = \dfrac{1}{n} \sum_{t=1}^{n} \sum_{k=1}^{n} \prod_{q=1}^{d} \bigg\{1 - \max{\hat{U}_t^{(q)},\hat{U}_k^{(q)}} \Bigg\} - \dfrac{1}{2^{d-1}} \sum_{t=1}^{n} \prod_{q=1}^{d} (1 - \hat{U}^{(q)2}) + \dfrac{n}{3^d}
		\end{align*}
		Since $\hat{U}_i$ is almost uniformly distributed on $[0,1]^d$ under the null hypothesis, large values of $S_n$ should lead to rejection of the null hypothesis. Unfortunately, the limiting distribution of the test statistic will depend on the unknown parameter set, $\theta$. Since it is impossible to construct tables, we use a different methodology, namely parametric bootstrap, to compute $P$-values. The validity of the parametric bootstrap approach has been shown for a wide range of assumptions in \cite{Genest/Remillard:2008}. These results were recently extended to dynamic models \citep{Remillard:2011a}, including regime-switching random walks. In this paper, we generalized the procedure to AR(1) Gaussian regime-switching model by conditioning the Rosenblatt's transform on the previous return.
		
\subsection{Parametric bootstrap algorithm} 
	
	\begin{enumerate}[a)]
		\item For a given number of regimes, estimate parameters with $\theta_n$ computed from the EM algorithm applied to $(R_1,\ldots,R_n)$

		 \item Compute the test statistic,
		\begin{equation*}
			S_n = B_n(\hat{U},\ldots,\hat{U}_n),
		\end{equation*}
		from the estimated pseudo observations, $\hat{U}_i = \Psi(R_1,\ldots,R_n,\theta_n)$, for $i \in \{1,\ldots,n\}$. \\
		\item For some large integer N (say 1000), repeat the following steps for every $k \in {1,\ldots,N}$:
		
		\begin{enumerate}[i)]
	
			\item Generate a random sample $\{ R_1^k,\ldots,R_n^k,\theta_n^k\}$ from distribution $P_{\theta_n}$
			\item Compute $\theta_n^k$ by applying the EM algorithm to the simulated sample,  $ R_1^k,\ldots,R_n^k$.
			\item Let $\hat{U}_i^k = \Psi_i(R_1^k,\ldots,R_n^k,\theta_n^k) $ for $i \in {1,\ldots,n}$, and finally compute
			\begin{equation*}
				S_n^k = B_n\left(\hat{U}_1^k,\ldots,\hat{U}_n^k\right).
			\end{equation*}
		\end{enumerate}	
	\end{enumerate}		

	Then, the approximate $P$-value for the test based on the Cram\'er von Mises statistic $S_n$ is given by
	\begin{equation*}
		\dfrac{1}{N} \sum_{k=1}^{N} \mathbb{I}\left(S_n^k>S_n\right).
	\end{equation*}

\section{Optimal Hedging} \label{app:opt_hedge}

\subsection{Proof of Theorem \ref{thm:functions}}\label{pf:thm-functions}

\begin{proof}
First, we show this is true for the off-line random sequences $\fa,\fb, \gamma, \rho$ and $P$.  The result is clearly true for $P_{n+1}=1$. Now,
\begin{eqnarray*}
\fa_n &=& E\left(\Delta_n\Delta_n^\top |\cF_{n-1}\right) \\
 &= &  D(\check S_{n-1} ) E\left\{  \left( e^{Y_n-r_n} -\mathbf{1}\right) \left( e^{Y_n-r_n} -\mathbf{1}\right)^\top | \cF_{n-1}\right\}  D(\check S_{n-1})  \\
 &=&  D(\check S_{n-1} ) a_n(Y_{n-1},\tau_{n-1})D(\check S_{n-1} ),
 \end{eqnarray*}
since $(Y,\tau)$ is a Markov process. For the same reason,
\begin{eqnarray*}
\fb_n &=& E\left(\Delta_n |\cF_{n-1}\right) \\
 &= & D(\check S_{n-1} ) E\left\{  \left( e^{Y_n-r_n} -\mathbf{1}\right)  | \cF_{n-1}\right\}  \\
 &=& D(\check S_{n-1} )  b_n(Y_{n-1},\tau_{n-1}).
 \end{eqnarray*}
 As a result, $\rho_n = D^{-1}(\check S_{n-1}) a_n^{-1}(Y_{n-1},\tau_{n-1}) b_n((Y_{n-1},\tau_{n-1})$, so
 $$
 \rho_n^\top \Delta_n = b_n((Y_{n-1},\tau_{n-1})^\top a_n^{-1}(Y_{n-1},\tau_{n-1}) \left( e^{Y_n-r_n} -\mathbf{1}\right).
 $$
 Therefore, $\gamma_n = g_n(Y_{n-1},\tau_{n-1})$. The rest of the proof is done by induction. Assume this is true for $t+1$, then we have to prove this is true for $t$.
 Now
 \begin{eqnarray*}
\fa_t &=& E\left(\Delta_t\Delta_t^\top \gamma_{t+1} |\cF_{t-1}\right) \\
 &= &  D(\check S_{t-1} ) E\left\{  \left( e^{Y_{t}-r_t} -\mathbf{1}\right) \left( e^{Y_{t}-r_t} -\mathbf{1}\right)^\top g_{t+1}(Y_{t},\tau_{t})| \cF_{t-1}\right\}  D(\check S_{t-1}) \\
 &=& D(\check S_{t-1} ) a_t(Y_{t-1},\tau_{t-1})D(\check S_{t-1} ),
 \end{eqnarray*}
since $(Y,\tau)$ is a Markov process. Similarly,
\begin{eqnarray*}
\fb_t &=& E\left(\Delta_t \gamma_{t+1}|\cF_{t-1}\right) \\
 &= &  D(\check S_{t-1} ) E\left\{  \left( e^{Y_{t}-r_t} -\mathbf{1}\right) g_{t+1}(Y_{t},\tau_{t}) | \cF_{t-1}\right\} \\
 &=& D(\check S_{t-1} )  b_t(Y_{t-1},\tau_{t-1}).
 \end{eqnarray*}
 Hence, $\rho_t = D^{-1}(\check S_{t-1}) a_t^{-1}(Y_{t-1},\tau_{t-1}) b_t(Y_{t-1},\tau_{t-1})$, so
 $$
 \rho_t^\top \Delta_t = b_t((Y_{t-1},\tau_{t-1})^\top a_t^{-1}(Y_{t-1},\tau_{t-1}) \left( e^{Y_{t}-r_t} -\mathbf{1}\right).
 $$
 Therefore,
\begin{eqnarray*}
\gamma_t &=& E\left\{(1-\rho_t^\top \Delta_t) \gamma_{t+1}|\cF_{t-1}\right\} \\
 &= &  E\left\{   g_{t+1}(Y_{t},\tau_{t}) | \cF_{t-1}\right\} \\
 && \qquad -  b_t(Y_{t-1},\tau_{t-1})^\top a_t^{-1}(Y_{t-1},\tau_{t-1})E\left\{  \left( e^{Y_{t}-r_t} -\mathbf{1}\right) g_{t+1}(Y_{t},\tau_{t}) | \cF_{t-1}\right\} \\
 &=&  E\left\{   g_{t+1}(Y_{t},\tau_{t}) | \cF_{t-1}\right\}  -  b_t(Y_{t-1},\tau_{t-1})^\top a_t^{-1}(Y_{t-1},\tau_{t-1})b_t(Y_{t-1},\tau_{t-1}) \\
 &=& g_t(Y_{t-1},\tau_{t-1}).
 \end{eqnarray*}
 This proves that \eqref{eq:a}--\eqref{eq:c} hold true for any $t\in \{1,\ldots,n\}$.

 Suppose now that $\beta_n C_n = \Psi_n(\check S_n)$.
 Using induction together with  \eqref{eq:newC}, one gets
  \begin{eqnarray*}
 \check C_{t-1} = \beta_{t-1}C_{t-1}\gamma_t &=& E[(1-\rho_t^\top \Delta_t) \check C_t |\cF_{t-1}] \\
 &= & E[ \{1-h_t(Y_{t-1},\tau_{t-1})^\top \left( e^{Y_{t}-r_t} -\mathbf{1}\right) \} \Psi_t(\check S_{t},Y_{t},\tau_{t}) |\cF_{t-1}] \\
 &=& \Psi_{t-1}(\check S_{t-1},Y_{t-1},\tau_{t-1}).
 \end{eqnarray*}

 Finally, it follows from \eqref{eq:alpha} and \eqref{eq:newC} that
 $$
 \alpha_t = \mathfrak{a}_t^{-1} E(\beta_n C  \Delta_t P_{t+1} | \mathcal{F}_{t-1} ) =
 \mathfrak{a}_t^{-1} E(\beta_t C_t  \Delta_t \gamma_{t+1} | \mathcal{F}_{t-1} ) =  \mathfrak{a}_t^{-1} E(\check C_t  \Delta_t  | \mathcal{F}_{t-1} ).
 $$
 As a result, using \eqref{eq:a}--\eqref{eq:Ct}, one gets
  $$\alpha_t = D^{-1}(\check S_{t-1})a_t^{-1}(Y_{t-1},\tau_{t-1}) \mathbf{A}_t(\check S_{t-1},Y_{t-1},\tau_{t-1}),
 $$
 where
 $$
 \mathbf{A}_t(s,y,i) = E\left[\Psi_t\left\{D(s)e^{Y_t-r_t},Y_t,\tau_t\right\} \left( e^{Y_{t}-r_t} -\mathbf{1}\right)|Y_{t-1}=y,\tau_{t-1}=i\right].
 $$
 \end{proof}

	\subsection{Optimal hedging algorithm}

		We now need to evaluate the \eqref{eq:a}--\eqref{eq:At}. First,		
		\begin{eqnarray*}
		 a_t(y,i) &=& E\left\{  \left( e^{Y_t-r_t} -\mathbf{1}\right) \left( e^{Y_t-r_t} -\mathbf{1}\right)^\top g_{t+1}(Y_t,\tau_t) | Y_{t-1}=y, \tau_{t-1}=i\right\} \\
&=&  \sum_{j=1}^{l} Q_{i j} \int \left( e^{z-r_t} -\mathbf{1}\right) \left( e^{z-r_t} -\mathbf{1}\right)^\top g_{t+1}(z,j) f_j(z|y)dz, \\
		 b_t(y,i) &=& E\left\{  \left( e^{Y_t-r_t} -\mathbf{1}\right) g_{t+1}(Y_t,\tau_t) | Y_{t-1}=y, \tau_{t-1}=i\right\} \\
		& = & \sum_{j=1}^{l} Q_{i j} \int \left( e^{z-r_t} -\mathbf{1}\right) g_{t+1}(z,j)f_j(z|y)dz,  \\
		 g_t(y,i) &=& E\left\{ g_{t+1}(Y_t,\tau_t) | Y_{t-1}=y, \tau_{t-1}=i\right\} - b_t^\top(y,i) h_t(y,i) \\
		 & = & \sum_{j=1}^{l} Q_{i j} \int g_{t+1}(z,j) f_j(z|y)dz - b_t^\top(y,i) h_t(y,i) ,\\
\Psi_{t-1}(s,y,i) &=& E\left[ \Psi_t\left\{D(s)e^{Y_t-r_t},Y_{t},\tau_{t}\right\}\left\{ 1- h_{t}(y,i)^\top \left(e^{Y_t-r_t}-\mathbf{1}\right)\right\} |Y_{t-1}=y,\tau_{t-1}=i\right] \\
& = & \sum_{j=1}^{l} Q_{i j} \int \Psi_t\left\{D(s)e^{z-r_t},z,j\right\} \left\{ 1- h_{t}(y,i)^\top \left(e^{z-r_t}-\mathbf{1}\right)\right\} f_j(z|y)dz \\
		\mathbf{A}_t(s,y,i) & = &  E\left\{\Psi_t\left\{D(s)e^{Y_t-r_t},Y_t,\tau_t\right\} \left(e^{Y_t-r_t}-\mathbf{1}\right)|Y_{t-1}=y,\tau_{t-1}=i\right \} \\
		 & = &  \sum_{j=1}^{l} Q_{i j} \int \Psi_t\left\{D(s)e^{z-r_t},z,j \right\} \left(e^{z-r_t}-\mathbf{1}\right)f_j(z|y)dz .
		 \end{eqnarray*}
		
		 Since these variables are weighted expectations they can be approximated by using Monte Carlo simulations, coupled with interpolations. This method was proposed in \cite{Papageorgiou/Remillard/Hocquard:2008}. Because the simulations are computationally expensive and introduce variance, we propose a novel technique to approximate these expectations using semi-exact calculations, based on \cite{Remillard:2013}.

	\subsection{Semi-exact calculations.}\label{app:semi_exact}	
	
		By defining a grid, i.e. $x_0<x_1<...<x_m<x_{m+1}$, one can approximate a function $F$ by the continuous piecewise linear function
		
		\begin{equation}\label{eq:iterp1}
		\hat{f}(x) = \sum_{q=0}^{m} \I( x_q \le x < x_{q+1})\left\{A^f(q)+x B^f(q)\right\}),
		\end{equation}
		where	
		\begin{eqnarray}
			\label{eq:iterp1-B}
			B^f(q) =  \frac{ f(x_{q+1}) -  f(x_{q})  }{ x_{q+1}-x_q}, \\
			\label{eq:iterp1-A}
			A^f(q) =  f(x_q) - x_q B^f(q).
		\end{eqnarray}

		Now, for simplicity, take $d=1$. Using \eqref{eq:iterp1}-\eqref{eq:iterp1-A}, we can approximate $g_t$ the following way
		$$\hat{g}_t(y,i) = \sum_{v=0}^{k} \I( y_v \le y < y_{v+1})(A_t^g(v,i)+yB_t^g(v,i))$$
		
		where	
		$$
		B_t^g(v,i) =  \frac{ \hat{g}_t(y_{v+1},i) - \hat{g}_t(y_v,i)  }{y_{v+1} - y_v}
		$$	
		$$
		A_t^g(v,i) =  \hat{g}_t(y_v,i) - y_v B_t(v,i).
		$$

		Remember that $g_{n+1}=1$. The formula for $g_t$ is shown later.\\

Now, let $\mathcal{N}(x)$ be the cumulative distribution function of the standard normal distribution and let $\mathcal{N}'(z) = \frac{e^{-z^2/2}}{\sqrt{2\pi}}$ be the associated density. Also,  $\mathcal{N}''(z) = -z \mathcal{N}'(z)$. If $X \sim N( \mu,\,\sigma^{2})$, and if $a<b$ are given, then  for any $\theta\in \mathbb{R}$,  one has
		 \begin{eqnarray*}
			M(\theta) &=& E\left[e^{\theta X}\I( a < X < b) \right] = e^{\theta \mu + \theta^2 \sigma^2/2} \big[ \mathcal{N}\{ \kappa(b)\} - \mathcal{N}\{ \kappa(a) \}\big],
		\end{eqnarray*}	
		\begin{eqnarray*}
			M'(\theta) &=& 		E\left[X e^{\theta X}\I( a < X < b) \right]   \\
			 	&=& (\mu+\theta\sigma^2)M(\theta)-\sigma e^{\theta \mu + \theta^2 \sigma^2/2}\big[ \mathcal{N}'\{ \kappa(b)\} - \mathcal{N}'\{ \kappa(a) \}\big],
		\end{eqnarray*}
		and
	
		\begin{eqnarray*}
			M''(\theta) & = & E\left[X^2 e^{\theta X}\I( a < X < b) \right]   \\
			 	&=& \sigma^2 M(\theta) + (\mu+\theta\sigma^2)M'(\theta) \\
			 	& & \quad +\sigma e^{\theta \mu + \theta^2 \sigma^2/2}\bigg\{ \sigma \big[ \mathcal{N}''\{ \kappa(b)\} - \mathcal{N}''\{ \kappa(a) \}\big]\\
 && \qquad  -(\mu +\theta \sigma^2)\big[\mathcal{N}'\{ \kappa(b)\}- \mathcal{N}'\{ \kappa(a) \} \big] \bigg\},
		\end{eqnarray*}
	where for any $x\in\dR$, 		
		\begin{equation*}
						\kappa(x) = \frac{x -\theta \sigma^2 - \mu}{\sigma}.
		\end{equation*}
In particular,
	\begin{eqnarray*}
M'(0) &= & E\left[X \I( a < X < b) \right] \\
&=& \mu\left[ \mathcal{N}\left\{ \frac{b-\mu}{\sigma}\right\} - \mathcal{N}\left\{  \frac{a-\mu}{\sigma} \right\}\right] -\sigma \left[ \mathcal{N}'\left\{ \frac{b-\mu}{\sigma}\right\} - \mathcal{N}'\left\{ \frac{a-\mu}{\sigma} \right\}\right].
	\end{eqnarray*}
	This way, we can approximate $a$, $b$ and $g$ as follows.\\

For any $u\in\{1,\ldots,k\}$ and any $i \in \{1\ldots,l\}$, set
		\begin{eqnarray*}
		 \hat{a}_t(y_u,i) &=& \sum_{j=1}^{l} Q_{i j} \sum_{v=0}^{k} \int_{y_v}^{y_{v+1}}  \left( e^{z-r_t} -1\right) \left( e^{z-r_t} -1\right) \\
		  && \quad \times (A_{t+1}^g(v,j)+ z B_{t+1}^g(v,j)) f_j(z|y_u) dz  \\
		  & = & \sum_{j=1}^{l} Q_{i j} \sum_{v=0}^{k} \int_{y_v}^{y_{v+1}} \bigg\{ A_{t+1}^g(v,j) + z B_{t+1}^g(v,j) - 2 e^{-r_t} e^z A_{t+1}^g(v,j) \\
		  & & \quad- 2 e^{-r_t}  z e^z B_{t+1}^g(v,j) + e^{-2 r_t} e^{2z} A_{t+1}^g(v,j) + e^{-2 r_t} z e^{2z} B_{t+1}^g(v,j) \bigg\} \\
&& \qquad \qquad \qquad  f_j(z|y_u) dz,
		 \end{eqnarray*}
\begin{eqnarray*}
		 \hat{b}_t(y_u,i) &=& \sum_{j=1}^{l} Q_{i j} \sum_{v=0}^{k} \int_{y_v}^{y_{v+1}}  \left( e^{z-r_t} -1\right)  (A_{t+1}^g(v,j)+ z B_{t+1}^g(v,j)) f_j(z|y_u) dz  \\
		  & = & \sum_{j=1}^{l} Q_{i j} \sum_{v=0}^{k} \int_{y_v}^{y_{v+1}} \bigg\{ - A_{t+1}^g(v,j) - z B_{t+1}^g(v,j) + e^{-r_t} e^z A_{t+1}^g(v,j) \\
		  & & \quad+ e^{-r_t}  z e^z B_{t+1}^g(v,j)  \bigg\} f_j(z|y_u) dz.
		 \end{eqnarray*}
		and
	\begin{eqnarray*}
		\hat{g}_t(y_u,i) &=&
		 \sum_{j=1}^{l} Q_{i j} \sum_{v=0}^{k} \int_{y_v}^{y_{v+1}}\left\{ A_t^g(v,j)  		  + zB_t^g(v,j)   \right\} f_j(z|y_u) dz - \hat{b}_t(y_u,i) \hat{h}_t(y_u,i),
	\end{eqnarray*}
	where $ \hat{h}_t(y_u,i) = \hat{a}_t^{-1}(y_u,i) \hat{b}_t(y_u,i)$.

	For values of $y$ not on the grid, we can also interpolate $\hat{a}$, $\hat{b}$, and $\hat g$  using \eqref{eq:iterp1}-\eqref{eq:iterp1-A}.
	\\
	
		To approximate $\Psi$ and $\alpha$, we need to define a second grid, defined by $0 = s_0<s_1<...<s_m<s_{m+1}= \infty $. $\Psi_t$ can be also be approximated by a product of continuous piecewise linear functions viz.
		
		\begin{eqnarray*}
			\hat{\Psi}_t(s,y,i) &=& \sum_{q=0}^{m} \sum_{v=0}^k \I( s_q \le s < s_{q+1})
				\I( y_v \le y < y_{v+1}) \\
			&& \qquad \times \left\{A_t^{\Psi}(q,v,i)+s B_t^{1,\Psi}(q,v,i)+ y B_t^{2,\Psi}(q,v,i)+sy B_t^{3,\Psi}(q,v,i)\right\},
		\end{eqnarray*}
		where		
		\begin{eqnarray*}
		B_t^{3,\Psi}(q,v,i) & = & \frac{\Psi(s_{q+1},y_{v+1},i) - \Psi(s_{q+1},y_{v},i) - \Psi(s_{q},y_{v+1},i) + \Psi(s_{q},y_{v},i)}{(s_{q+1}-s_q)(y_{v+1}-y_v)}, \\
B_t^{2,\Psi}(q,v,i) & = & \frac{\Psi(s_{q},y_{v+1},i) - \Psi(s_{q},y_{v},i)}{y_{v+1}-y_v} - B_t^{3,\Psi}(q,v,i) s_q, \\
		B_t^{1,\Psi}(q,v,i) & = & \frac{\Psi(s_{q+1},y_{v},i) - \Psi(s_{q},y_{v},i)}{s_{q+1}-s_q}  - B_t^{3,\Psi}(q,v,i)  y_v ,\\
				A_t^{\Psi}(q,v,i) & = & \Psi(s_{q},y_{v},i) - \frac{\Psi(s_{q+1},y_{v},i) - \Psi(s_{q},y_{v},i)}{s_{q+1}-s_q} s_q \\
		& & \quad - \frac{\Psi(s_{q},y_{v+1},i) - \Psi(s_{q},y_{v},i)}{y_{v+1}-y_v} y_v + B_t^{3,\Psi}(q,v,i) s_q y_v,
		\end{eqnarray*}
	
        using the results in Appendix \ref{app:interpol}.

	As a result, for $q\in \{1,\ldots,m\}$ and for $v \in \{1,\ldots,k\}$,
\begin{eqnarray*}
\hat \Psi_{t-1}(s_q,y_v,i) & = &  \sum_{j=1}^l Q_{ij} \int  \hat\Psi_t\left(s_q e^{z-r_t},z,j\right)  \\
&& \qquad \qquad \times  \left\{ 1- h_{t}(y_v,i) \left(e^{z-r_t}-1\right)\right\} f_j(z|y_v) dz  \\
		& = & \sum_{j=1}^l Q_{ij} \sum_{q'=1}^m \sum_{v'=1}^k  \int_{y_{v'}}^{y_{v'+1}}  \I( s_{q'} \le s_q e^{z-r_t} < s_{q'+1})
 \\
&& \quad \times \Big{\{}\{A_t^{\Psi}(q',v',j)+s_q e^{z-r_t} B_t^{1,\Psi}(q',v',j) \\
 && \qquad \qquad + z B_t^{2,\Psi}(q',v',j)+zs_q e^{z-r_t} B_t^{3,\Psi}(q',v',j)\Big{\}} \\
&& \quad \qquad \qquad \times \left\{ 1- h_{t}(y_v,i) \left(e^{z-r_t}-1\right)\right\} f_j(z|y_v) dz\\
& = &  \sum_{j=1}^l Q_{ij}\sum_{q'=1}^m \sum_{v'=1}^k  \int_{\max\{y_{v'}, \log(s_{q'}/s_q)\}}^{\min\{y_{v'+1},\log(s_{q'+1}/s_q)\}}
 \\
&& \quad \times \Big{\{}A_t^{\Psi}(q',v',j)+s_q e^{z-r_t} B_t^{1,\Psi}(q',v',j)\\
&& \qquad \qquad + z B_t^{2,\Psi}(q',v',j)+zs_q e^{z-r_t} B_t^{3,\Psi}(q',v',j)\Big{\}} \\
&& \qquad \qquad \quad \times \left\{ 1- h_{t}(y_v,i) \left(e^{z-r_t}-1\right)\right\} f_j(z|y_v) dz,
\end{eqnarray*}
with  the convention that the integral is $0$ if
$$\max\{y_{v'}, \log(s_{q'}/s_q)\} \ge \min\{y_{v'+1},\log(s_{q'+1}/s_q)\}.
$$

	There is a similar expression for $\hat{A}_t(s_q,y_v,i)$, namely
	
	\begin{eqnarray*}
	\hat{A}_t(s_q,y_v,i) & = &
		   \sum_{j=1}^{l} Q_{i j} \int \hat\Psi_t\left(s_q e^{z},z,j\right) \left(e^{z-r_t}-1\right) f_j(z|y_v) dz \\
		   & = &  \sum_{j=1}^{l} Q_{i j} \sum_{q'=1}^m \sum_{v'=1}^k  \int_{y_{v'}}^{y_{v'+1}}  \I( s_{q'} \le s_q e^{z-r_t} < s_{q'+1}) \\
		   && \quad \times \Big{\{}A_t^{\Psi}(q',v',j)+s_q e^{z-r_t} B_t^{1,\Psi}(q',v',j) \\
&& \qquad \qquad + z B_t^{2,\Psi}(q',v',j)+zs_q e^{z-r_t} B_t^{3,\Psi}(q',v',j)\Big{\} }\\
		   &&  \qquad \qquad \quad \times \left(e^{z-r_t}-1\right) f_j(z|y_v) dz \\
		   & = &  \sum_{j=1}^{l} Q_{i j} \sum_{q'=1}^m \sum_{v'=1}^k  \int_{\max\{y_{v'}, \log(s_{q'}/s_q)\}}^{\min\{y_{v'+1},\log(s_{q'+1}/s_q)\}} \\
		   && \quad \times \Big{\{}A_t^{\Psi}(q',v',j)+s_q e^{z-r_t} B_t^{1,\Psi}(q',v',j) \\
&& \qquad \qquad + z B_t^{2,\Psi}(q',v',j)+zs_q e^{z-r_t} B_t^{3,\Psi}(q',v',j)\Big{\}} \\
		   &&  \qquad \qquad \quad \times \left(e^{z-r_t}-1\right) f_j(z|y_v) dz.
	\end{eqnarray*}

	Note that $\Psi_{t-1}(0,y,i)$ = $\Psi_n(0,y,i)$ and $A_{t}(0,y,i)$ = $h_t(y,i) \Psi_n(0,y,i)$ for all $t=1,\ldots,n$. For example, $\Psi_t(0,y,i) \equiv 0$ for all $t$, while for a put option with strike $K$, $\Psi_t(0,y,i) = \beta_n K$, for all $t \in \{0,\ldots,n\}$.

	\subsection{Bi-linear interpolation}\label{app:interpol}

One wants to interpolate $f$ by $\hat f$ over $[x_0,x_1]\times [y_0,y_1]$. Set $f_{i,j} = f(x_i,y_j)$, $i,j\in \{0,1\}$.

Then $\hat f(x,y) = \frac{(x-x_0)}{(x_1-x_0)} \frac{(y-y_0)}{(y_1-y_0)} B_{11}+ \frac{(x-x_0)}{(x_1-x_0)}B_{10}+ \frac{(y-y_0)}{(y_1-y_0)} B_{01}+ B_{00}$, where
\begin{eqnarray*}
B_{00} &=& f_{00},\\
B_{10} &=& f_{10}-f_{00},\\
B_{01} &=& f_{01}-f_{00},\\
B_{11} &=& f_{11}-f_{10}-f_{01}+f_{00}.\\
\end{eqnarray*}
		\subsection{Simulation of Gaussian ARHMM(1)}\label{app:simulation}

If $(Y_t,\tau_t)=(y,i)$, then with $\tau_{t+1}=j$ with probability $Q_{ij}$ and then $Y_{t+1} = \mu_j + \Phi_j(y-\mu_j)+ B_j \varepsilon$, where $B_j^\top B_j = \Sigma_j$ and $\varepsilon \sim N(0,I)$.

Also, when $d=1$, a random sample of size $N$ from $f_j(\cdot|y)$ is given by $z_{ij}  = \mu_i+\Phi_j(y-\mu_j)+\sigma_j \epsilon_i$, with $\epsilon_i\sim N(0,1)$. So the sequence of innovations $\epsilon$ are independent of $j$.

In this case, every integral $I_{j}(h) = \int h(z)f_j(z|y)dz$ is approximated by
$$
\hat I_j(h) = \frac{1}{N}\sum_{i=1}^N h(z_{ij}).
$$
In particular, this means that for any $j_1,j_2\in \{1,\ldots,l\}$,
$$
\hat I_{j_1}(h_1) =  \frac{1}{N}\sum_{i=1}^N h_1(z_{ij_1}),\quad
\hat I_{j_2}(h_2) =  \frac{1}{N}\sum_{i=1}^N h_2(z_{ij_2}).
$$
\end{document}